\documentclass[a4paper,11pt]{article}
\usepackage[text={16.8cm,22.4cm}]{geometry}
\usepackage{amsmath,amsfonts,slashed,amssymb,tikz,bm,psfrag,graphicx,color,dsfont}
\usepackage{multicol,multirow}
\usepackage{float}
\usepackage{authblk}
\usepackage[nottoc]{tocbibind}
\usepackage[english]{babel}
\usepackage{array} 

\usepackage{graphicx}
\usepackage{dcolumn}
\usepackage{bm}
\usepackage{xcolor}
\usepackage{colortbl,booktabs}
\usepackage{float}
\usepackage{slashed}
\usepackage{multirow}
\usepackage{multicol}
\usepackage{subfigure}
\usepackage[citecolor=blue]{hyperref} 
\allowdisplaybreaks 
\usepackage[figuresright]{rotating}
\usepackage{cancel}
\renewcommand{\arraystretch}{1.2}

\def\lrpartial{\buildrel\leftrightarrow\over\partial}

\def\slash#1{#1 \hskip-0.53em /}
\def\nn{\nonumber\\}

\RequirePackage[sort&compress,square,comma,numbers]{natbib}
\allowdisplaybreaks
\renewcommand{\arraystretch}{1.2}

\title{\bf An improved calculation of the $D_{(s)}^*D_{(s)}V$ and $B_{(s)}^*B_{(s)}V$ couplings from light-cone sum rules}

\vspace{0.5cm}
\author[a]{Su-Ping Jin \thanks{jinsuping@nankai.edu.cn}}
\author[a]{Hua-Yu Jiang \thanks{huayujiang@nankai.edu.cn}}

\vspace{0.7cm}
\affil[a]{School of Physics, Nankai University, Weijin Road 94, 300071 Tianjin, China \,}

\begin{document}

\maketitle

\medskip

\vspace{0.2cm}
\begin{abstract}

We present an improved calculation of the $D_{(s)}^*D_{(s)}V$ and $B_{(s)}^*B_{(s)}V$ coupling constants, where $V$ denotes $\rho$, $K^\ast$, $\omega$, and $\phi$ meson. These couplings govern the QCD long-distance dynamics in interactions between heavy pseudoscalar/vector mesons and light vector mesons. Our analysis is conducted within the framework of QCD light-cone sum rules (LCSRs) by utilizing the light-cone distribution amplitudes (LCDAs) of light vector mesons. By systematically incorporating the subleading power corrections and higher-twist contributions at leading order (LO) and including the next-to-leading order (NLO) corrections at leading power, we achieve enhanced accuracy in the light-cone operator product expansion (OPE) for the underlying correlation function. In assessing the reliability of the established LCSRs, we consider uncertainties arising from the choice of the quark-hadron duality region in the double dispersion relation. Building upon these improvements, we accomplish an optimized computation and analysis for the strong coupling constants $g_{H^\ast HV}$ which are used to extract the effective coupling $\lambda$ in the heavy meson chiral perturbation theory (HM$\chi$PT). Furthermore, we investigate the $SU(3)$ flavour symmetry breaking effects in detail and compare our sum rule calculations with previous studies in an exploratory way.




\end{abstract}
\vfil

\newpage


\newpage

\section{Introduction}
The strong couplings among the heavy vector mesons ($D_{(s)}^\ast$ or $B_{(s)}^\ast$ meson), the heavy pseudoscalar mesons ($D_{(s)}$ or $B_{(s)}$) and the light vector mesons ($V$ involving $\rho,\,K^\ast,\,\omega$ and $\phi$), represented by $g_{D^*DV}$ and $g_{B^*BV}$, offer valuable insights into the non-perturbative aspects of the long-distance dynamics within quantum chromodynamics (QCD). 
By combining the chiral and heavy quark symmetry of QCD, the heavy meson chiral perturbative theory (HM$\chi$PT) \cite{Wise:1992hn,Burdman:1992gh,Casalbuoni:1996pg,Yan:1992gz} provides an effective framework to describe interactions between heavy and light mesons at low energy. 
The $D_{(s)}^*D_{(s)}V$ and $B_{(s)}^*B_{(s)}V$ couplings are essential ingredients for the extraction of the effective coupling parameter $\lambda$ which governing the $H^\ast H V$ interactions in the HM$\chi$PT Lagrangian \cite{Casalbuoni:1996pg,Cheng:2004ru}.


Furthermore, the $D_{(s)}^*D_{(s)}V$ and $B_{(s)}^*B_{(s)}V$ couplings play a crucial role in determining the residue of the vector and tensor form factors for $D \to V$ and $B\to V$ transitions at the $D^\ast$ and $B^\ast$ poles, as dictated by the dispersion relation \cite{SentitemsuImsong:2014plu,Li:2010rh,Wang:2020yvi,Khodjamirian:2020mlb}.
This insight enhances our understandings of the behavior of these form factors, which are vital for extracting the CKM parameters and the lepton flavour universality (LFU) parameters in the $D \to V\ell\nu$ and $B\to V\ell\nu$ semileptonic decays.
Moreover, the $D_{(s)}^*D_{(s)}V$ couplings are crucial for understanding the final-state interactions (FSIs), which can make significant contributions to the long-distance dynamics and may contribute strong phases, potentially leading to direct CP violation in $B$ meson non-leptonic decays \cite{Cheng:2004ru}.


Due to the phase space limit, direct experimental measurements of these couplings are not feasible. Consequently, some couplings, particularly $g_{B^*B\rho}$ and $g_{D^*DV}$, have been determined through calculations employing QCD techniques such as SVZ sum rules (QCDSRs) \cite{Shifman:1978bx,Shifman:1978by,Colangelo:2000dp} and light-cone sum rules (LCSRs) \cite{Balitsky:1989ry,Braun:1988qv,Chernyak:1990ag}. However, these calculations, mainly conducted at leading order \cite{Aliev:1996xb,Li:2002pp,Li:2007dv,Wang:2007mc,Wang:2007zm,Azizi:2010jj,OsorioRodrigues:2010fen,Khosravi:2014rwa}, often come with relatively large uncertainties. 
Recently, significant progress has been made in calculating the correlation functions using QCD (SCET) factorization approach in light-cone sum rules \cite{Khodjamirian:2023wol}. 
Incorporating next-to-leading order (NLO) ($\alpha_s$) corrections and subleading power ($\Lambda/m_Q$) contributions has granted us the ability to more precisely determine the spectral density of the dispersion relation for building up the LCSRs. 
This approach has been widely used to calculate some essential non-perturbative physical quantities, for instance, the heavy-to-light ($B\to \pi$, etc.) and heavy-to-heavy ($B \to D$, etc.) transition form factors. 
The calculation of the correlation function based on QCD (SCET) factorization is carried out from two-distinct aspects: the hard-collinear factorization by using the light-cone distribution amplitudes (LCDAs) of light hadrons, such as in  \cite{Ball:2004ye,Duplancic:2008ix,Ball:2004rg,Bharucha:2015bzk,Li:2020rcg}, and a slightly more complex factorization procedure by separating the hard, hard-collinear, collinear and soft scales with the using of the heavy-light hadrons LCDAs, e.g., the calculations in \cite{Wang:2015vgv,Cui:2022zwm,Gao:2019lta,Feldmann:2011xf,Wang:2015ndk,Gao:2021sav,Cui:2023jiw,Wang:2018wfj,Cui:2023yuq}.   



The QCD (SCET) factorization-based LCSRs has emerged as a robust technique for investigating strong couplings beyond leading order and power like $g_{H^*H\rho}$ \cite{Wang:2020yvi}, and $g_{H^*H\pi}$ \cite{Khodjamirian:2020mlb}. The higher-order/power contributions can significantly affect the magnitude of strong couplings, potentially altering the overall results by approximately $10\%$ to $20\%$ for NLO and $20\%$ to $30\%$ for subleading-power effects as shown in \cite{Wang:2020yvi}. However, discrepancies between calculation results in LCSRs \cite{Aliev:1996xb,Wang:2020yvi,Li:2002pp,Li:2007dv} and the extrapolated coupling $g_{B^*B\rho}$ from $B\to \rho$ form factors \cite{Bharucha:2015bzk,Gao:2019lta}, using dispersion relation, have been observed. These discrepancies may stem from incomplete analyses of subleading power contributions at leading order (LO) and next-to-leading order (NLO) contributions at higher-twist, along with uncertainties arising from different shapes of the duality region at NLO \cite{Khodjamirian:2020mlb}. Another potential explanation for this inconsistency is the inclusion of contributions from excited heavy meson states to the dispersion relation \cite{Becirevic:2002vp}, which introduces considerable model dependence and raises questions about effectively isolating ground-state contributions within the duality region.

Given the importance of these strong couplings for our understanding of QCD long-distance dynamics, as introduced previously, this paper aims to present a comprehensive investigation of the $D_{(s)}^*D_{(s)}V$ and $B_{(s)}^*B_{(s)}V$ strong couplings using LCSRs. The essential new ingredients are summarized as follows.

\begin{itemize}
\item{We improve the calculation accuracy of the established light-cone sum rules from two perspectives. Firstly, we incorporate more complete subleading power corrections\footnote{We define $\delta_V\equiv m_V/m_Q,\,Q=(b,c)$ and $\delta_s\equiv m_s/m_V$ ($m_V$ denotes the mass of vector mesons, $m_Q$ and $m_s$ are respectively the heavy and strange quark mass). A detailed discussion on the effectiveness of our power expansion is presented in Table \ref{tab:power}.} (up to $\delta_V^4$, $\delta_s^1$, and $(\delta_V\delta_s)^1$) and higher-twist corrections (up to two-particle twist-5 and three-particle twist-4) of the LCDAs of light vector mesons at leading order and the NLO contributions at leading-power. Secondly, we consider the uncertainties due to the choice of the quark-hadron duality region in the double dispersion relation, an aspect which was not addressed in the previous study \cite{Wang:2020yvi}. Based on these enhancements, we present the optimized and comprehensive computations for the $D_{(s)}^*D_{(s)}V$ and $B_{(s)}^*B_{(s)}V$ couplings to date. This enables us to obtain a more accurate prediction for the effective coupling $\lambda$ of the HM$\chi$PT, which still exhibits significant deviation from the estimation provided by the vector meson dominance (VMD) model \cite{Isola:2003fh}.}


\item{We conduct a detailed investigation into the impact of $SU(3)$ flavour ($SU(3)_F$) symmetry breaking effects, which stem from modifications in the light quark mass, the decay constants and mass of the vector mesons and heavy mesons, and the parameters related to the vector meson LCDAs. 
Our analysis indicates that the $SU(3)_F$ breaking effects play a non-negligible role in accurately determining the coupling constants $g_{H_{s}^\ast H K^\ast}$, $g_{H^\ast H_s K^\ast}$, $g_{H^\ast H \omega}$ and $g_{H_s^\ast H_s \phi}$. As expected, the most significant breaking effects are observed in the coupling $g_{H^*_s H_s\phi}$, primarily due to the involvement of the maximum number of strange quarks in this channel.}


\item{We investigate the dispersion relation, which establishes a connection between the residue of various $B_{(s)} \to V$ transition form factors (for $V$ and $T_1$) at the $B_{(s)}^\ast$ pole and the corresponding $B^*_{(s)} B_{(s)} V$ strong coupling. The couplings $g_{B^*_{(s)} B_{(s)} V}$, accompanied by systematic uncertainties, are computed from the dispersion relation using form factors obtained from two distinct LCSRs methods \cite{Bharucha:2015bzk,Gao:2019lta}. 
}

\end{itemize}

We update all input parameters utilized in LCSRs. Specifically, we employ the RunDec package \cite{Chetyrkin:2000yt} to adopt the four-loop evolution values for the QCD coupling and quark mass within the $\overline{\text{MS}}$ scheme. The decay constants of vector and pseudoscalar heavy-light mesons are determined using two distinct methods: QCD two-point sum rules \cite{Khodjamirian:2020mlb,Gelhausen:2013wia} with the same NLO accuracy as the LCSRs, and lattice QCD (LQCD) \cite{Lubicz:2017asp,Colquhoun:2015oha}.


The structure of this paper is organized as follows: In section \ref{sec-2}, we introduce the conventions and definitions for the $D_{(s)}^*D_{(s)}V$ and $B_{(s)}^*B_{(s)}V$ couplings and establish the corresponding LCSRs using the double dispersion relation. In section \ref{sec-3}, we present the calculation details for the LO and NLO double spectral density of the correlation function. Section \ref{sec-4} is dedicated to discussing the uncertainties arising from the quark-hadron duality ansatz for the double dispersion relation and constructing the finalized form of the LCSRs.  We focus on determining the various input parameters and analyzing the numerical results in section \ref{sec-5}. The final section concludes the work. Essential details regarding the vector meson LCDAs are provided in Appendices A and B, while Appendix C contains expressions for the double spectral densities at NLO.

\section{Light-cone Sum Rules for the $D_{(s)}^{\ast}D_{(s)}V$ and $B_{(s)}^{\ast}B_{(s)}V$ couplings \label{sec-2}}
Our study begins with the introduction of the phenomenological effective Lagrangian \cite{Yan:1992gz,Cheng:2004ru,Casalbuoni:1996pg} governing the strong interactions between heavy mesons and light vector resonances,
\label{section: Tree-level LCSR}
\begin{equation}
{\cal L}_{eff} = -\frac{1}{2} g_{H^*HV} \epsilon_{\mu\nu\alpha\beta}
 (\partial^\mu V^\nu)^i_j
 \Big( H_i^\dagger\lrpartial{}^{\!\alpha} H^{*\beta j}-H_i^{*\beta\dagger}\lrpartial{}{\!^\alpha} H^j \Big) \,,
\label{eq:LDDV}
\end{equation}
with the convention $\epsilon_{0123}=-1$, $g_{H^*HV}$ represents the strong couplings, and $V$ stands for the nonet matrix of vector mesons,
\begin{eqnarray}\setlength{\tabcolsep}{3pt}
V=\left(\begin{array}{ccc}
\frac{\rho^{0}}{\sqrt{2}}+\frac{\omega}{\sqrt{2}} & \rho^{+} & K^{*+}\\
\rho^{-} & ~~ -\frac{\rho^{0}}{\sqrt{2}}+\frac{\omega}{\sqrt{2}} ~~~~~& K^{*0}\\
K^{*-} &\bar{K}^{*0} &\phi
\end{array}\right) ,\, 
\end{eqnarray}
 and $H^{(\ast)} = \big( D^{(\ast)0},D^{(\ast)+},D_s^{(\ast)+} \big)$ or $\big( B^{(\ast)-},\bar{B}^{(\ast)0},\bar{B}_s^{(\ast)0} \big)$. 
The hadronic matrix element of the effective Lagrangian, describing the coupling between specific initial and final states, plays a crucial role in determining the strength of the interaction between particles in a given process,
\begin{equation}
\langle\, V(p,\eta^*)\,H^*(p+q,\varepsilon)\,|{\cal L}_{eff}|\,H(q)\, \rangle= -g_{H^*HV}\,\epsilon_{\alpha\beta\rho\sigma}\,\eta^{*\alpha}\,\varepsilon^{*\beta}\,p^\rho\,q^\sigma \,,
\end{equation}
where the vector meson $V(p,\eta)$ is characterized by its mass $m_V$, 4-momentum $p$, and polarization vector $\eta$. These couplings are determined by the charge and flavour quantum numbers of the initial and final hadrons. In this work, we focus on the couplings 
$g_{D^{*+}D^0\rho^+}$, $g_{D^{*+}D^+\omega}$, $g_{D^{*+}_s D^0  K^{*+}}$, $g_{D^{*+} D^+_s {\bar K}^{*0}}$ and $g_{D^{*+}_s D^{+}_s \phi}$ in the charm sector, as well as
$ g_{\bar{B}^{*0}{B}^-\rho^+}$, $g_{B^{*0}D^0\omega}$, $g_{\bar{B}^{*0}_s {B}^-  K^{*+}}$, $g_{\bar{B}^{*0} \bar{B}^0_s \bar{K}^{*0}}$ and $g_{\bar{B}^{*0}_s {B}^{0}_s \phi}$  in the bottom sector.
The others can be related to the aforementioned couplings through isospin symmetry, which dictates certain relationships between couplings involving particles with different isospin quantum numbers,
\begin{eqnarray}
{\rm D~ case}:\qquad&&g_{D^*D\rho} \equiv g_{D^{*+}D^0\rho^+}=
-\sqrt{2}g_{D^{*+}D^{+}\rho^0}=\sqrt{2}g_{D^{*0}D^0\rho^0}
=-g_{D^{*0}D^+\rho^-}~,\nn
&&g_{D^*_s D K^{*}} \equiv g_{D^{*+}_s D^0  K^{*+}}=
-g_{D^{*+}_s D^{+} K^{*0}}~,\,g_{D^{*} D_s  K^{*}} \equiv g_{D^{*+} D^+_s {\bar K}^{*0}}=
g_{D^{*0} D^+_s K^{*-}}~;\label{eq:coupling-def1}\\
{\rm B~ case}:\qquad && g_{B^*B\rho} \equiv g_{\bar{B}^{*0}{B}^-\rho^+}=
-\sqrt{2}g_{\bar{B}^{*0}\bar{B}^{0}\rho^0}=\sqrt{2}g_{{B}^{*-}{B}^-\rho^0}
=-g_{{B}^{*-}\bar{B}^0 \rho^-}~,\nn
&&g_{B^*_s B K^{*}} \equiv g_{\bar{B}^{*0}_s {B}^-  K^{*+}}=
-g_{\bar{B}^{*0}_s \bar{B}^{0} K^{*0}}~,\,g_{B^*B_s K^{*}} \equiv g_{\bar{B}^{*0} \bar{B}^0_s  \bar{K}^{*0}}=
g_{{B}^{*-}_s \bar{B}^{0}_s K^{*-}}~.\label{eq:coupling-def2}
\end{eqnarray}

In the more general $SU(3)$ flavour symmetry framework, specific relations exist for the $D$ and $B$ cases. These relations govern certain patterns or constraints on the interactions between particles within the $D$ and $B$ meson sectors, based on their shared properties under $SU(3)$ transformations,
\begin{eqnarray}
{\rm D~ case}:\qquad && g_{D^*D\rho}=g_{D^*_s D K^{*}}=g_{D^{*} D_s  K^{*}}=g_{D^*_s D_s \phi}=\sqrt{2} g_{D^*D\omega} ;\\
{\rm B~ case}:\qquad && g_{B^*B\rho} =g_{B^*_s B K^{*}}=g_{B^*B_s K^{*}} =g_{B^*_s B_s \phi}=\sqrt{2} g_{B^*B\omega}.
\end{eqnarray}

To establish the sum rules for the coupling $g_{H^*HV}$, we adopt the approach outlined in Ref.\cite{Wang:2018wfj}. This involves investigating the correlation function between the vacuum and $V$-meson, utilizing two distinct local interpolating currents for vector and pseudoscalar heavy mesons, denoted by $j_{\mu}$ and $j_5$. These currents are employed to effectively probe the relevant properties and dynamics of the heavy meson sector,
\begin{align}
F_\mu(p,q)= i\int d^4x\,e^{-i(p+q)\cdot x}\langle\, V(p,\eta^*)\,|\,T\big\{j_{\mu}(x),\,j_{5}(0)\big\}\,|\,0\rangle\,=\epsilon_{\mu\nu\rho\sigma}\eta^{*\nu}\,p^\rho\,q^{\sigma}F((p+q)^2,q^2).
\label{eq:corr}
\end{align}
Here,  $j_{\mu}=\bar{q_1}\,\gamma_{\mu}\,Q$ and $j_{5}=(m_Q+m_{q_2})\bar{Q}\,i\,\gamma_5\,q_2$, where the heavy quark, denoted as $Q$, corresponds to either $c$ or $b$, while $q_{1,2}$ represent one of the light quarks, $u$, $d$, or $s$.

Following \cite{Belyaev:1994zk}, one can utilize the analytic properties of the amplitude in two distinct invariant variables:  $(p+q)^2$ and $q^2$, corresponding to the squared momenta of interpolating currents in the two channels. By incorporating the complete sets of intermediate states with quantum numbers of $H$ and $H^*$ into these channels, we derive the double dispersion relation,
\begin{eqnarray}
F_{\mu}(p,q)= \frac{\langle\, V(p,\eta)H^*(p\!+\!q, \varepsilon)\,|H(q)  \rangle\langle H(q)|\,j_{5}|0\rangle\langle H^*(p\!+\!q, \varepsilon)|j_{\mu}|0\rangle }{\left[(p+q)^{2}-m_{H^{*}}^{2}\right]\left[q^{2}-m_{H}^{2}\right]} + \cdots \,.
\end{eqnarray}
 It's worth noting that we have omitted the continuum part in this derivation.

To establish the LCSRs for the strong coupling constant, we begin by substituting its definition and expressing the matrix elements of interpolating currents in terms of the decay constants $f_{H}$ and $f_{H^*}$. Subsequently, we derive the double dispersion relations for the amplitude $F((p+q)^2,q^2)$,
\begin{eqnarray}
F((p+q)^2,q^2)=-\frac{g_{H^{*} H  V} f_{H^{*}} f_{H} m_{H}^{2}m_{H^{*}}}{\left[m_{H^{*}}^{2}-(p+q)^{2}\right]\left[m_{H}^{2}-q^{2}\right]}\,
+\iint\limits_{\hspace{4mm}{\Sigma}} \frac{ \rho_{cont}\left(s_1, s_2\right) d s_1 d s_2}{\left[s_1-(p+q)^{2}\right]\left(s_2-q^{2}\right)} + \cdots\,.
\label{eq:double_disp}
\end{eqnarray}
The relevant matrix elements used to define the heavy meson decay constants are outlined as follows:
\begin{align}
&\left\langle 0\left|j_5\right| H(p)\right\rangle= f_{H} {m_{H}^{2}}, \qquad
\left\langle H^{*}(p+q,  \varepsilon) \left|j_{\mu}\right|0\right\rangle=f_{H^{*}} m_{H^{*}} \varepsilon_{ \mu}.
\label{eq:HB_decay_constant}
\end{align}

The leading term in Eq. (\ref{eq:double_disp}) represents the contribution from the ground-state double-pole, which involves the product of the desired strong coupling $g_{H^*HV}$ and the corresponding decay constants. The term $\rho_{cont}{(s_1,s_2)}$ captures the combined spectral density of excited and continuum states, and $\Sigma$ denotes the duality region occupied by these states in the $(s_1,s_2)$-plane. Additional terms resulting from subtractions, represented by ellipses, vanish after the application of the double Borel transformation.

In QCD, the invariant function $F((p+q)^2,q^2)$ can be calculated at momentum transfers $q^2$ and $(p+q)^2$ much smaller than the heavy quark mass $m_Q^2$ through the light-cone OPE applied to the correlation function (\ref{eq:corr}) near the light-cone $x^2\sim 0$. The result can be factorized in terms of the convolution of a hard kernel and the LCDAs of vector mesons, classified by their twist. The former represents the short-distance perturbative contributions, while the latter parameterizes the long-distance non-perturbative effects. The OPE calculations can be represented in the form of a double dispersion relation:
\begin{align}
F^{\rm (OPE)}((p+q)^2,q^2)=
\iint ds_1\,ds_2
\frac{\rho^{\rm (OPE)}(s_1, s_2)}{(s_1-(p+q)^2)(s_2-q^2)}\,,
\label{eq:ddispOPE}
\end{align}
where the involved dual spectral density $\rho^{\rm (OPE)} (s_1, s_2)$ refers to,
\begin{equation}
\rho^{\rm (OPE)} (s_1, s_2) \equiv \frac{1}{\pi^2}
\mbox{Im}_{s_1}\mbox{Im}_{s_2} F^{\rm (OPE)}(s_1,s_2) \,,
\label{eq:rhoope}
\end{equation}
where ${\rm Im}_{s_1}$ and ${\rm Im}_{s_2}$ correspond to the sequential extraction of the imaginary part of the corrleation function $F^{\rm (OPE)}(s_1,s_2)$ in the variable $s_1$ and $s_2$.

After performing the double Borel transformation with respect to the variables $(p+q)^2 \to M_1^2$ and $ q^2 \to M_2^2$, which is a technique used to enhance the convergence and suppress the contributions from higher-dimensional operators in the operator product expansion (OPE), we arrive at the sum rules for the coupling constant $g_{H^*HV}$,
\begin{eqnarray}
f_H f_{H^*}\,g_{H^*HV}= -\frac{1}{m_H^2 m_{H^*}}\iint \limits^{\hspace{4mm}\widetilde{\Sigma}} ds_1 \, ds_s\, \exp\left(\frac{m^2_{H^*}-s_1}{M_1^2}+\frac{m^2_{H}-s_2}{M_2^2}\right)
\rho^{\rm (OPE)} (s_1, s_2)\,.
\label{eq:SR2}
\end{eqnarray}
The inclusion of the integration boundary $\widetilde{\Sigma}$, which is dual to the ground-state contribution to \eqref{eq:double_disp}, arises from subtracting the continuum contributions using the parton-hadron duality ansatz.

\section{Double spectral density of the correlation function \label{sec-3}}
\subsection{Double  spectral density at LO}

\begin{figure}[tb]
\begin{center}
\includegraphics[width=0.35\columnwidth]{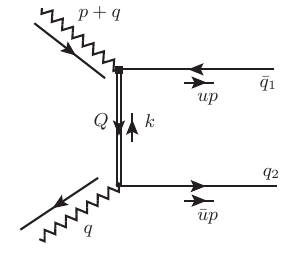} \\
\vspace*{0.1cm}
\caption{Diagrammatical representation of the leading-order (LO) contribution. }
\label{fig:twist-2-LO}
\end{center}
\end{figure}

Near the light-cone\ref{fig:twist-2-LO}, characterized by $x^2 \sim 0$, the OPE for the correlation function given in \eqref{eq:corr} remains applicable under the condition that both external momenta squared, $(p+q)^2$ and $q^2$, are significantly below the heavy quark threshold $m_Q^2$. Specifically, to ensure the validity of power counting within the OPE, it suffices to have:
\begin{eqnarray}
m_Q^2-q^2 \sim m_Q^2-(p+q)^2\sim {\cal O}(m_Q\tau) \,,n \cdot p \sim {\cal O}(m_Q).
\end{eqnarray}

Here, $\tau \gg \Lambda_{\text{QCD}}$ is a parameter independent of $m_Q$. The heavy quark propagating in the correlation function is consequently highly virtual and can be expanded near the light-cone. The initial expression in Eq. (\ref{eq:corr}) undergoes transformation to:
\begin{equation}
F_\mu((p+q)^2,q^2)=-(m_Q+m_{q_2})\!\int \! d^4x e^{-i(p+q)x}\langle V(p,\eta^*)| \bar{q}_{1}(x)\gamma_\mu
S(x,0) \gamma_5 q_{2}(0)  |0\rangle\,,
\label{eq:corr1}
\end{equation}
where, $S_Q(x,0)$ consists of the free-quark propagator and a term representing one-gluon emission, to our level of accuracy. Subsequently, we obtain the two-particle (2p) and three-particle (3p) contributions to the correlation function (\ref{eq:corr1}),
\begin{eqnarray}
&&F^{\rm (LO)}_{\mu,2p}((p+q)^2,q^2)=\epsilon_{\mu\nu\rho\sigma}\eta^{\nu}\,p^\rho\,q^{\sigma}F^{\rm (LO)}_{2p}((p+q)^2,q^2)\nn
&& \qquad =-m_Q(1+{\Hat{m}_{q_2}})\int d^{4}x  \int \frac{d^{4} k}{(2 \pi)^{4}}e^{-i(p+q+k) \cdot x}  \langle V(p,\eta^*)|\bar{q}_1(x) \gamma_{\mu} \frac{i\left(\slash{k}+m_{Q}\right)}{k^{2}-m_{Q}^{2}} \gamma_{5} q_2(0)| 0\rangle ,\nn
&&F^{\rm (LO)}_{\mu,3p}((p+q)^2,q^2)=\epsilon_{\mu\nu\rho\sigma}\eta^{\nu}\,p^\rho\,q^{\sigma}F^{\rm (LO)}_{3p}((p+q)^2,q^2)\nn
&& \qquad = i g_{s}m_Q(1+{\Hat{m}_{q_2}}) \int d^{4} x \int \frac{d^{4} k}{(2 \pi)^{4}} e^{-i(k+p+q) \cdot x}  \int_{0}^{1} d v\left\langle V(p,\eta^*)\right| \bar{q}_1(x) \gamma_{\mu}\nn
&&\qquad\quad\times\left[{{\bar v}}(k\llap{/}+m_Q)\sigma_{\mu\nu}+v\sigma_{\mu\nu}(k\llap{/}+m_Q)\over 2(k^2-m^2)^2\right]G^{\mu\nu}(v x)\gamma_{5} q_2(0)|0\rangle.
\end{eqnarray}

Upon inserting the DAs into the correlation function, we focus on the key factors discussed below,
\begin{eqnarray}
&&\text{\rm for\, two-particle\, case}:\int_{0}^{1} d u \int \frac{d^{4} k}{(2 \pi)^{4}} \int d^{4} x e^{-i(q+\bar{u} p+k) \cdot x},\nn
&&\text{\rm for\, three-particle\, case}:\int_{0}^{1} d u \int \frac{d^{4} k}{(2 \pi)^{4}} \int d^{4} x e^{-i\left(q+\bar{\alpha} p+k\right) \cdot x},
\end{eqnarray}
 which reveal that for the two-particle case, $k\equiv-(q+\bar{u}p)$, with $\bar{u}=1-u$ and for the three-particle case, $k\equiv-(q+\bar{\alpha}p)$, with $\bar{\alpha}=1-\alpha=1-(\alpha_1+v \alpha_3)$. 
The removal of the $1/px$ factors is accomplished by applying the replacement rules, as outlined below:
\begin{eqnarray}
&&\text{\rm for\, two-particle\, case}: \frac{1}{p \cdot x} \phi_{2 p}(u) \rightarrow-i \int_{0}^{u} d z \phi_{2 p}(z),\nn
&&\text{\rm for\, three-particle\, case}:\frac{1}{p\cdot x}\Phi_{3p}(\underline{\alpha}) \rightarrow -i \widehat{\Phi}_{3p}(\underline{\alpha}) ,
\end{eqnarray}
where the three-particle DAs with hat are defined as: 
\begin{align}
& \widehat{\Phi}_{3p}(\underline{\alpha}) \equiv \int_0^{\alpha_1} d\alpha_1^{\prime}\int_0^{1-\alpha_1^{\prime}} d\alpha_3^\prime \;\delta(\alpha_3^\prime - \alpha_3) \Phi_{3p}( \alpha_1^{\prime}, 1-\alpha_1^{\prime}-\alpha_3,  \alpha_3 ).
\label{3p-redefine}
\end{align}
The LO results for the invariant amplitude are obtained by summing the contributions from individual twists up to twist-5 accuracy,
\begin{eqnarray}
F^{\rm (LO)}((p+q)^2,q^2)=\big[F_{2p,tw2}^{\rm (LO)}
+F_{2p,tw3}^{\rm (LO)}+F_{2p,tw4}^{\rm (LO)}+F_{2p,tw5}^{\rm (LO) }+F_{3p,tw4}^{\rm (LO)}\big]
((p+q)^2,q^2)\,.
\label{eq:LO-OPE}
\end{eqnarray}

For the three-particle case, a parallel expression is obtained by replacing $u(\bar{u})$ with $\alpha(\bar{\alpha})$.
Consequently, we arrive at the following compact form through the power expansion:
\begin{eqnarray}
F^{\rm (LO)}((p+q)^2,q^2) \,&=&~~\, \frac{1+{\hat{m}_{q_2}}}{m_Q} \sum_{i=0}^4 \sum_{j=1}^4 \bigg\{ \int_0^1 du \, \delta_V^i \, \mathcal{C}^j(r_1,r_2,u) \, \mathcal{A}_{\mathrm{2p},\,ij}^{\rm LO}(u) \nn
&& +\int_0^1 dv \int_0^1 d\alpha_1 \int_0^{1-\alpha_1} d\alpha_3 \, \delta_V^i \, \mathcal{C}^j(r_1,r_2,\alpha)\, \mathcal{A}_{\mathrm{3p},\,ij}^{\rm LO}(v,\underline{\alpha})   \bigg\} \,,
\label{eq:sum}
\end{eqnarray}
where $\delta_V = {m_V/m_Q}$, $\underline{\alpha} = (\alpha_1, \alpha_3)$, $\Hat{m}_{q_2}=m^2_{q_2}/m^2_Q$, 
\begin{align}
  \mathcal{C}(r_1,r_2,u) & =  \frac{1}{\bar{u}r_1 + u r_2 - 1} \,, 
\end{align}
and $r_1 = (p+q)^2/m_Q^2$, $r_2 = q^2/m^2_Q$. The notations include:
\begin{align}
& \mathcal{A}_{2\mathrm{p},\,01}^{\rm (LO)}(u) = f_V^{\perp} \phi^{\perp}_{2;V} (u)\,, \quad
\mathcal{A}_{2\mathrm{p},\,12}^{\rm (LO)}(u) = -{1 \over 2} f_V^{\|} {\psi}^{\perp}_{3;V}(u)\,, \\ 
& \mathcal{A}_{2\mathrm{p},\,22}^{\rm (LO)}(u) = {f_V^{\perp}\over 4} \big[ 4 u\bar{u} \phi^{\perp}_{2;V} (u) + \phi^{\perp}_{4;V}(u) \big]\,, \quad \mathcal{A}_{2\mathrm{p},\,23}^{\rm (LO)}(u) =- {1 \over 2} f_V^{\perp} \phi^{\perp}_{4;V}(u)\,, \\
& \mathcal{A}_{2\mathrm{p},\,33}^{\rm (LO)}(u) =-f_V^{\|} u \bar{u} {\psi}^{\perp}_{3;V}(u)\, , \quad 
\mathcal{A}_{2\mathrm{p},\,34}^{\rm (LO)}(u) =  {3 \over 4} f_V^{\|}  {\psi}^{\perp}_{5;V}(u), \\
& \mathcal{A}_{2\mathrm{p},\,43}^{\rm (LO)}(u) = {f_V^{\perp}\over 2} u\bar{u} \big[ 2 u\bar{u} \phi^{\perp}_{2;V} (u) + \phi^{\perp}_{4;V}(u) \big]\,, \quad \mathcal{A}_{2\mathrm{p},\,44}^{\rm (LO)}(u) = - {3 \over 2} f_V^{\perp} u\bar{u} \phi^{\perp}_{4;V}(u)\,, \\
& \mathcal{A}_{3\mathrm{p},\,22}^{\rm (LO)}(v,\underline{\alpha}) = f_V^\perp \Big\{ 2 \bar{v} \big[\Phi^{\perp(2)}_{4;V}(\underline{\alpha}) -\Phi^{\perp(1)}_{4;V}(\underline{\alpha}) \big]  - {\Psi}^{\perp}_{4;V}(\underline{\alpha}) - (v-\bar{v})\widetilde{\Psi}^{\perp}_{4;V}(\underline{\alpha}) \nn
&\qquad\qquad\qquad~ - {2 (v - \bar{v}) \over \bar{\alpha}} \big[ \widehat{\Phi}^{\perp(2)}_{4;V}(\underline{\alpha}) - \widehat{\Phi}^{\perp(1)}_{4;V}(\underline{\alpha}) + \widehat{\Phi}^{\perp(3)}_{4;V}(\underline{\alpha}) - \widehat{\Phi}^{\perp(4)}_{4;V}(\underline{\alpha}) \big] \Big\}\,,\label{eq:3p-22} \\
& \mathcal{A}_{3\mathrm{p},\,23}^{\rm (LO)}(v, \underline{\alpha}) =  f_V^\perp {2 (v - \bar{v}) (r_2-1) \over \bar{\alpha}} \big[\widehat{\Phi}^{\perp(2)}_{4;V}(\underline{\alpha}) - \widehat{\Phi}^{\perp(1)}_{4;V}(\underline{\alpha}) + \widehat{\Phi}^{\perp(3)}_{4;V}(\underline{\alpha}) - \widehat{\Phi}^{\perp(4)}_{4;V}(\underline{\alpha}) \big]\,.\label{eq:3p-23}
\end{align}
Our next step is to derive the LO double spectral density, which is given by,
\begin{eqnarray}
\rho^{\rm LO}(s_1,s_2) ={1 \over \pi^2} \, {\rm Im}_{s_1} \, {\rm Im}_{s_2} F^{\rm (LO)}((p+q)^2,q^2). 
\end{eqnarray}
We extend the derivation of the dispersion representation within the framework of the tree-level factorization formula \eqref{eq:LO-OPE} by considering the general expression for double spectral densities governing invariant amplitudes in the two-particle case.  The expression is given by,
\begin{eqnarray}
\rho_{j,2p}(s_1,s_2)\equiv~&& {1 \over \pi^2} \, {\rm Im}_{s_1} \, {\rm Im}_{s_2}  \,
\int_0^1 d u \, {\phi_{2p}(u)  \over [\bar u \, s_1 + u \, s_2 - m_Q^2 + i \, 0]^{j}}  \nn
&& = {1 \over \Gamma(j)} \, {d^{j - 1} \over (d \, m_Q^2)^{j -1}}  \,
\sum_k \, {(-1)^{k+1} \, c^{(\phi_{2p})}_k\over  \Gamma(k+1) } \, (s_1 - m_Q^2)^k  \, \delta^{(k)}(s_1-s_2) \,\nn
&&=\sum_k c^{(\phi_{2p})}_k\rho_{j k}(s_1,s_2),
\label{eq:rho-2p}
\end{eqnarray}
where the  function $\phi_{2p}(u)$ is expressed through a Taylor expansion at $u=0$
\begin{eqnarray}
\phi_{2p}(u) = \sum_k \, c^{(\phi_{2p})}_k \, u^k \,.
\label{eq:expansion}
\end{eqnarray}
The three-particle scenario follows a framework similar to that of the two-particle case,
\begin{eqnarray}
&&\{\rho_{j,3p}(s_1,s_2),\widehat{\rho}_{j,3p}(s_1,s_2)\}~\nn
&&= {1 \over \pi^2} \, {\rm Im}_{s_1} \, {\rm Im}_{s_2}  \,
\int_0^1 d v \, \int {\cal D} \alpha \,\,
\frac{v^\ell\,\{\Phi_{3p}(\underline{\alpha}),\widehat\Phi_{3p}(\underline{\alpha})/\bar{\alpha} \}} {\left [\bar{\alpha} \, s_1 + \alpha \, s_2 - m_Q^2 + i \, 0 \right ]^j}\nn
 &&={1 \over \Gamma(j)} \, {d^{j - 1} \over (d \, m_Q^2)^{j -1}}  {1 \over \pi^2} \, {\rm Im}_{s_1} \, {\rm Im}_{s_2}  \,
\int_0^1 d \alpha \, 
\frac{\{\overline{\Phi}_{3p}(\alpha,\ell),\overline{\widehat\Phi}_{3p}(\alpha,\ell) \}} {\bar{\alpha} \, s_1 + \alpha \, s_2 - m_Q^2 + i \, 0 }
 \nn
&& = {1 \over \Gamma(j)} \, {d^{j - 1} \over (d \, m_Q^2)^{j -1}}  \,
\sum_k \, {(-1)^{k+1} \, \{c^{{\Phi}_{3p}}_{\ell, k},c^{\widehat{\Phi}_{3p}}_{\ell, k}\}\over  \Gamma(k+1) } \, (s_1 - m_Q^2)^k  \, \delta^{(k)}(s_1-s_2) \,\nn
&&=\sum_k \{c^{{\Phi}_{3p}}_{\ell, k} \rho_{jk,3p}(s_1,s_2),c^{\widehat{\Phi}_{3p}}_{\ell, k} \widehat{\rho}_{jk,3p}(s_1,s_2)\},
\label{eq:3p-rho}
\end{eqnarray}
where the coefficient $c^{{\Phi}{3p}}_{\ell, k}$ and $c^{\widehat{\Phi}{3p}}_{\ell, k}$ originate from the series expansion of two ``effective" DAs, namely ${\overline{\Phi}_{3p}(\alpha,\ell),\overline{\widehat\Phi}_{3p}(\alpha,\ell) }$,
\begin{align}
\{\overline{\Phi}_{3p}(\alpha,\ell),\overline{\widehat\Phi}_{3p}(\alpha,\ell) \} & = \int_0^{\alpha} \, d \alpha_1 \, \int_{\alpha-\alpha_1}^{1-\alpha_1} \, d \alpha_3 \,
{ (\alpha - \alpha_1)^\ell \over\alpha_3^{\ell+1} } \, \{\Phi_{3p}(\underline{\alpha}) ,\widehat{\Phi}_{3p}(\underline{\alpha})/\bar{\alpha} \} \nonumber \\
& = \sum_k \,\{ c^{{\Phi}_{3p}}_{\ell, k},c^{{\widehat{\Phi}}_{3p}}_{\ell, k}\} \, \alpha^k \,, \quad \text{with}~ k\geq 2 \,.
\end{align}

To account for the terms proportional to $q^2$ in Eq.\eqref{eq:3p-23}, we introduce supplementary functions\\ $\widetilde{\widehat{\rho}}_{jk,3p}(s_1,s_2)$ as described by Eq.\eqref{eq:3p-rho}:
\begin{equation}
    \widetilde{\widehat{\rho}}_{jk,3p}(s_1,s_2)=s_2 \widehat{\rho}_{jk,3p}(s_1,s_2).
\end{equation}
Performing a variable substitution by replacing $\alpha$ with $u$,
\begin{eqnarray}
\{\overline{\Phi}_{3p}(\alpha,\ell),\overline{\widehat\Phi}_{3p}(\alpha,\ell) \} \leftrightarrow
\{\overline{\Phi}_{3p}(u,\ell),\overline{\widehat\Phi}_{3p}(u,\ell) \} \,, 
\end{eqnarray}
 facilitates the  combination of contributions from three-particle processes with those from the previously considered two-particle contributions. With the replacement of each twist and multiplicity component in the sum (\ref{eq:sum}) with its corresponding double dispersion form, we derive the LO double spectral density of the correlation function,
\begin{eqnarray}
\rho^{\rm (LO)}(s_1,s_2) 
= ~~ && m_Q (1+{\Hat{m}_{q_2}}) \Bigg\{
\rho^{(\phi^{\perp}_{2;V})}_1
+ { m^2_Q \delta_V} \, \rho^{({\psi}^{\perp}_{3;V})}_1  \nn
&& + \, { m^2_Q \delta^2_V} \left(\rho^{(\phi^{\perp}_{2;V})}_2+\rho^{(\phi^{\perp}_{4;V})}_1+\sum^2_{i=1}\rho^{(\overline{\Phi}^{\perp(i)}_{4;V})}+m^2_Q\rho^{({\phi}^{\perp}_{4;V})}_2\right)\nn
&& + \, { m^2_Q \delta^2_V} \Bigg[\rho^{({\overline{\Psi}}^{\perp}_{4;V})}+\rho^{(\overline{\widetilde{\Psi}}^{\perp}_{4;V})}+ m^2_Q\sum^4_{i=1}\left(\rho^{(\overline{\widehat{\Phi}}^{\perp(i)}_{4;V})}+\widetilde{\rho}
^{(\overline{\widehat{\Phi}}^{\perp(i)}_{4;V})}\right)\Bigg] \nn
&& + \, m^4_Q \delta^3_V \left(\rho^{({\psi}^{\perp}_{3;V})}_2+ m^2_Q \rho^{({\psi}^{\perp}_{5;V})}\right) \nn
&& + \, m^4_Q\delta^4_V\Bigg[\left(\rho^{(\phi^{\perp}_{2;V})}_3+\rho^{(\phi^{\perp}_{4;V})}_3\right)+m^2_Q\rho^{(\phi^{\perp}_{4;V})}_4\Bigg]\Bigg \}(s_1,s_2)\,.
\label{eq:rhoLO}
\end{eqnarray}
This expression features an expansion form (\ref{eq:expansion}) for each term, where the coefficients $c_k^{(\phi)}$ are readily determined from the polynomial form of the DAs, as explicitly presented in Appendix \ref{app:lcda}.
The resulting expression (\ref{eq:rhoLO}) contributes to a new and comprehensive understanding. For instance, we illustrate the contributions to $\rho^{\rm (LO)}$ from the twist-2 and twist-3 DAs, emphasizing their asymptotic behavior, with $\delta^{(k)}(x)\equiv d^k/dx^k[\delta(x)]$,
\begin{align}
&\rho_1^{(\phi^{\perp}_{2;V})} (s_1,s_2)=6 f^{\perp}_V
\left [ (s_1- m_Q^2)\delta^{(1)}(s_1-s_2)+ \frac{1}{2}
(s_1 -m_Q^2)^2\delta^{(2)}(s_1-s_2) \right ] \, \theta(s_2-m_{Q}^2)
\,,\nn
&\rho_1^{(\psi^{\perp}_{3;V})} (s_1,s_2)=-3 f^{\parallel}_V\frac{d}{d m^2_Q}
\left [ (s_1- m_Q^2)\delta^{(1)}(s_1-s_2)+ \frac{1}{2}
(s_1 -m_Q^2)^2\delta^{(2)}(s_1-s_2) \right ] \, \theta(s_2-m_{Q}^2),\nn
&\rho_2^{(\phi^{\perp}_{2;V})} (s_1,s_2)=-f^{\perp}_V\frac{d}{d m^2_Q}
\bigg[ 3(s_1- m_Q^2)^2\delta^{(2)}(s_1-s_2)+ 2
(s_1 -m_Q^2)^3\delta^{(3)}(s_1-s_2)\nn
&\hspace*{2.5cm}+\frac1{20}(s_1 -m_Q^2)^4\delta^{(4)}(s_1-s_2)\bigg] \, \theta(s_2-m_{Q}^2),\nn
&\rho_2^{(\psi^{\perp}_{3;V})} (s_1,s_2)=\frac1{2}f^{\parallel}_V\frac{d^2}{d (m^2_Q)^2}
\bigg[ 3(s_1- m_Q^2)^2\delta^{(2)}(s_1-s_2)+ 2
(s_1 -m_Q^2)^3\delta^{(3)}(s_1-s_2)\nn
&\hspace*{2.5cm}+\frac1{20}(s_1 -m_Q^2)^4\delta^{(4)}(s_1-s_2)\bigg] \, \theta(s_2-m_{Q}^2).
\label{eq:rhophiV}
\end{align}

\subsection{Double  spectral density at NLO}

To achieve next-to-leading order (NLO) precision, we express the invariant amplitude associated with the correlation function (\ref{eq:corr}) in the following form,
\begin{equation}
F^{\rm (OPE)}((p+q)^2,q^2)=F^{\rm (LO)}((p+q)^2,q^2)+
\frac{\alpha_s C_F}{4\pi}F^{\rm (NLO)}((p+q)^2,q^2).
\end{equation}

\begin{figure}[tb]
\begin{center}
\includegraphics[width=1.0 \columnwidth]{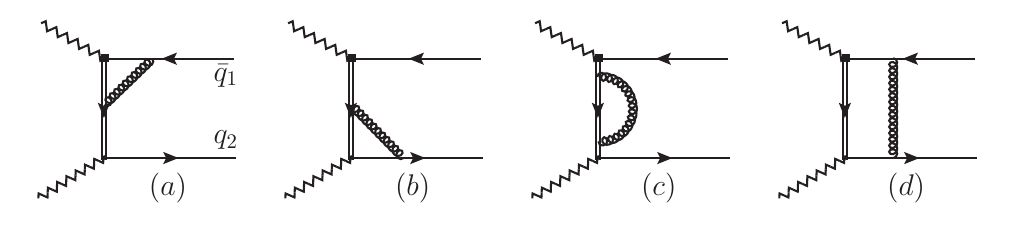} \\
\vspace*{0.1cm}
\caption{NLO QCD corrections of leading-twist contributions.}
\label{twist-2-NLO}
\end{center}
\end{figure}
The loop computations in this context closely follow the procedures detailed in \cite{Li:2020rcg} for the leading twist. The one-loop diagrams depicted in Figure \ref{twist-2-NLO} are computed using the method of regions, initially introduced in Ref. \cite{Beneke:1997zp}. It is evident that only the hard region contributes significantly, while the collinear region yields scaleless integrals within dimensional regularization, resulting in negligible contributions. The one-loop diagrams yield the following results:
\begin{align}
&F_{H^*}^{(1), h}((p+q)^2,q^2)=  \frac{\alpha_s C_F}{4 \pi}\left\{\Big[\frac{2\left(1-r_1\right)}{r_1-r_3} \ln \frac{1-r_3}{1-r_1}-1\Big]\Big[\frac{1}{\epsilon}+\ln \frac{\mu^2}{m_Q^2}-\ln \left[\left(1-r_3\right)\left(1-r_1\right)\Big]\right]\right. \nn
&\qquad -\ln \left[\left(1-r_3\right)\left(1-r_1\right)\right]+\frac{1}{r_3-r_1}\Big[2\left(1-r_1\right)\left(\operatorname{Li}_2\left(r_3\right)-\operatorname{Li}_2\left(r_1\right)\right) \nn
&\qquad \left.+\frac{\left(1-r_3\right)\left(1-r_3-2 r_1\right)}{r_3} \ln \left(1-r_3\right)-\frac{\left(1-r_1\right)\left(1-3 r_1\right)}{r_1} \ln \left(1-r_1\right)\Big]-4\right\} F^{(0)}((p+q)^2,q^2)\,, \nn
&F_{ H}^{(1), h}((p+q)^2,q^2)=  -\frac{\alpha_s C_F}{4 \pi}\left\{2\Big[\frac{1-r_2}{r_3-r_2} \ln \frac{1-r_3}{1-r_2}-1\Big]\Big[\frac{1}{\epsilon}+\ln \frac{\mu^2}{m_Q^2}-\ln \Big[\left(1-r_3\right)\left(1-r_2\right)\Big]\Big]\right. \nn
&\qquad -2 \ln \left[\left(1-r_3\right)\left(1-r_2\right)\right]-\frac{2}{r_3-r_2}\Big[\left(1-r_2\right)\left(\operatorname{Li}_2\left(r_3\right)-\operatorname{Li}_2\left(r_2\right)\right)\nn
&\qquad \left.-\frac{\left(1-r_3\right)\left(1-r_3+r_2\right)}{r_3} \ln \left(1-r_3\right)+\frac{1-r_2}{r_2} \ln \left(1-r_2\right)\Big]-2\right\} F^{(0)}((p+q)^2,q^2) \,, \nn
&F_{ w f c}^{(1)}((p+q)^2,q^2)= -\frac{\alpha_s C_F}{4 \pi}\left\{\frac{7-r_3}{1-r_3}\Big[\frac{1}{\epsilon}+\ln \frac{\mu^2}{m_Q^2}-\ln \left(1-r_3\right)+1\Big]-\frac{1}{1-r_3}\Big[\frac{1-7 r_3}{r_3^2} \ln \left(1-r_3\right)\right. \nn
&\qquad\left.+\frac{1}{r_3}-3\Big]\right\} F^{(0)}((p+q)^2,q^2)\,.
\end{align}
Here, we introduce the definitions $r_1 \equiv (p+q)^2 / m_Q^2$, and $r_2\equiv q^2 / m_Q^2$, with $r_3 = u r_2+\bar{u} r_1$. The tree-level amplitude $F_\mu^{(0)}$ is presented at the leading twist,
\begin{equation}
F^{(0)}((p+q)^2,q^2)=(1+{\Hat{m}_{q_2}}) f_{V}^{\perp}(\mu) \int_{0}^{1} d u\, \mathcal{C}(r_1,r_2,u)\cdot \phi_{2;V}^{\perp}(u, \mu) .
\end{equation}

The box diagram yields contributions at order $\mathcal{O}(\epsilon)$ in dimensional regularization, and therefore does not significantly contribute. This observation is consistent with previous findings \cite{Wang:2017ijn,Wang:2018wfj}, which demonstrate the vanishing of hard-collinear factorization for the one-loop box diagram in the context of hadronic photon correction. Combining the contributions from all one-loop diagrams, the results are in agreement with those presented in a previous study \cite{Li:2020rcg}:
\begin{eqnarray}
&&{\mathcal T}^{(1)}\left(r_1, r_2, \mu\right) \nn
&&=\frac{\alpha_s C_F}{4 \pi}\left\{(-2)\left[\frac{1-r_1}{r_3-r_1} \ln \frac{1-r_3}{1-r_1}+\frac{1-r_2}{r_3-r_2} \ln \frac{1-r_3}{1-r_2}+\frac{3}{1-r_3}\right]\left(\frac{1}{\epsilon}+\ln \frac{\mu^2}{m_Q^2}\right)\right. \nn
&&+2\left[\left(\frac{1-r_1}{r_3-r_1}+\frac{1-r_2}{r_3-r_2}\right) \operatorname{Li}_2\left(r_3\right)-\frac{1-r_1}{r_3-r_1} \operatorname{Li}_2\left(r_1\right)-\frac{1-r_2}{r_3-r_2} \operatorname{Li}_2\left(r_2\right)\right] \nn
&&+2\left[\left(\frac{1-r_1}{r_3-r_1}+\frac{1-r_2}{r_3-r_2}\right) \ln ^2\left(1-r_3\right)-\frac{1-r_1}{r_3-r_1} \ln ^2\left(1-r_1\right)-\frac{1-r_2}{r_3-r_2} \ln ^2\left(1-r_2\right)\right] \nn
&&+\left[\frac{1-r_3}{r_3}\left(\frac{1-r_3-2 r_1}{r_3-r_1}-\frac{2\left(1-r_3+r_2\right)}{r_3-r_2}\right)+\frac{1-6 r_3}{r_3^2}+1\right] \ln \left(1-r_3\right)+\frac{1-9 r_3}{r_3\left(1-r_3\right)} \nn
&&\left.\quad-\frac{\left(1-r_1\right)\left(1-3 r_1\right)}{r_1\left(r_3-r_1\right)} \ln \left(1-r_1\right)+\frac{2\left(1-r_2\right)}{r_2\left(r_3-r_2\right)} \ln \left(1-r_2\right)-3\right\}\,,
\end{eqnarray}

Previous work \cite{Wang:2018wfj} has demonstrated that the matrix element of the evanescent SCET operator, denoted as $\langle O_{E,\mu}\rangle^{(1)}$, vanishes at order ${\cal O}(\alpha_s)$ within the NDR scheme of $\gamma_5$ when considering hard-collinear factorization. 
This leads to the factorization formula for twist-2 contributions at next-to-leading order (NLO):
\begin{equation}
F^{(\rm NLO)}((p+q)^2,q^2)=(1+{\Hat{m}_{q_2}}) f_{V}^{\perp}(\mu)\int_{0}^{1} d u\,\mathcal{C}(r_1,r_2,u)\cdot \phi_{2;V}^{\perp}(u, \mu) \cdot {\mathcal T}^{(1)}\left(r_1, r_2, \mu\right).
\end{equation}
Now, it is imperative to advance further and obtain the NLO double spectral density:
\begin{eqnarray}
\rho^{\rm NLO}(s_1,s_2) ={1 \over \pi^2} \, {\rm Im}_{s_1} \, {\rm Im}_{s_2} F^{\rm (NLO)}((p+q)^2,q^2).
\label{}
\end{eqnarray}
For the NLO terms, we employ asymptotic vector meson DAs. This choice is justified by observing that at LO, the non-asymptotic contributions from Gegenbauer moments in the twist-2 DA (as described in Eq.(\ref{eq:confexp}) constitute only a small percentage in the LCSR, assuming typical values for the moments $a_2$ and $a_4$. Additionally, for the $K^*$ particle, terms related to $a_1$ and $a_3$ vanish due to the $u=1/2$ condition used in the calculation, effectively eliminating odd Gegenbaur terms. The additional $O(\alpha_s)$ factor significantly suppresses these contributions, keeping them well below the uncertainties inherent in the sum rule.

\section{Quark-hadron duality and sum rules \label{sec-4}}
After computing the double spectral density (\ref{eq:rhoope}), we obtain the expression,
\begin{equation}
\rho^{\rm (OPE)}(s_1,s_2)= \rho^{\rm (LO)}(s_1,s_2)+ \frac{\alpha_sC_F}{4\pi} \rho^{\rm (NLO)}(s_1,s_2)\,,
\label{eq:rhotot}
\end{equation}
where the LO contribution is provided in (\ref{eq:rhoLO}), and the NLO part consists of the twist-2 contributions, as detailed in (\ref{eq:tw2nlo}).

Following Ref.\,\cite{Khodjamirian:2020mlb}, we adopt a specific parameterization for the boundaries:
\begin{eqnarray}
\left ( {s_1 \over s_{\ast}} \right )^{\alpha} +
\left ( {s_2 \over s_{\ast}} \right )^{\alpha}   \leq 1 \,,
\qquad s_1, \, s_2 \geq  m_Q^2\,.
\label{eq:alpha}
\end{eqnarray}
We explore three different duality regions: the triangular region where $\alpha=1$ and $s_\ast=2s_0$, the concave region with $\alpha=\frac{1}{2}$ and $s_\ast=4s_0$, and the convex region where $\alpha=2$ and $s_\ast=\sqrt{2}s_0$. Our findings reveal that the terms containing $\delta(s_1-s_2)$ and its derivatives contribute equally across all duality regions. Therefore, only NLO contributions not involving the delta-function or its derivatives are affected by the choice of region. However, most of the LO and the primary part of NLO contributions come from integrating over the diagonal interval, which remains consistent across all regions. Therefore, we opt for the most practical choice: the triangular region, where $s_1+s_2 \leq 2s_0$. Returning to the LCSR Eq. (\ref{eq:SR2}), we then assume equal Borel parameter $M_1^2=M_2^2=2M^2$. This allows us to rewrite the sum rule as:
\begin{align}
f_H f_{H^*}\,g_{H^*HV}
=- \frac{1}{m_H^2 m_{H^*}}\exp
\left(\frac{m_{H}^2+m_{H^*}^2}{2M^2}\right)
\bigg[\mathcal F^{\rm (LO)}(M^2,s_0) + {\alpha_s \, C_F \over 4 \pi} \,
\mathcal F^{\rm (NLO)}(M^2,s_0)\bigg]\,.
\label{eq:SR3}
\end{align}
Here, we define the integral over the triangular region as follows:
\begin{align}
\mathcal F(M^2,s_0)\equiv &
\int^{\infty}_{-\infty} \!d s_1
\int^{\infty}_{-\infty} \!d s_2\,
\theta(2s_0-s_1-s_2)\,
\exp\left(-\frac{s_1+s_2}{2M^2}\right)
\rho(s_1,s_2) \,.
\label{eq:Fell}
\end{align}
Each DA, expressed through its Taylor expansion (\ref{eq:expansion}), decomposes the density $\rho$ into individual terms. Importantly, within the context of the triangular duality region, the resulting expressions for the integrals $\mathcal F(M^2,s_0)$ can be universally formulated for any generic DA.

To achieve this, a variable transformation is applied to the integration variables in (\ref{eq:Fell}): $s_1=s(1-v)$ and $s_2=s v$, or vice versa, $s = s_1+s_2$ and $v = {s_2}/{(s_1+s_2)}$. This transformation simplifies the difference $(s_1-s_2)\to s(1-2v)$, allowing the integration of the $\delta(s_1-s_2)= \delta(1-2v) / s$ functions and their derivatives over $v$. Simultaneously, the exponential factor in (\ref{eq:Fell}) becomes independent of $v$. 
This results in the Taylor expansion of an arbitrary DA $\phi(u)$ reducing to its value or its derivative at $u=1/2$. This yields the following simplified expression:
\begin{eqnarray}
&&\mathcal F_j(M^2,s_0)
  = \frac{1}{(j-1)!}
\bigg\{(-1)^j \big(M^2\big)^{2-j} \,
\exp \Big(-\frac{m^2_Q}{M^2}\Big)+ \delta_{j1} \, M^2 \, \exp \Big(-\frac{s_0}{M^2}\Big) \bigg\} \,\phi(u) \,\bigg|_{u=\frac{1}{2}} \,,\\
&& \widetilde{\mathcal F}_{j}^{(\phi)}(M^2,s_0)
 =  -\frac{1}{2(j-1)!}\,\frac{d^{j-1}}{d{m^2_Q}^{j-1}}
\int_{2\,m^2_Q}^{2\,s_0}ds \,\exp  \left (-\frac{s}{2\,M^2} \right )\,
\left [ u\, \left (\frac{s}{2}-m^2_Q \right ) \, \phi^{\prime}(u)
+ \frac{s}{2} \, \phi(u) \right ]
\bigg|_{u=\frac{1}{2}} \,.\label{eq:Fellphi} \hspace{0.8 cm}
\end{eqnarray}

In the subsequent analysis, $\widetilde{\mathcal F}_{3}$ stands out as the only contributor to the term in (\ref{eq:Fellphi}).
\begin{align}
\widetilde{\mathcal F}_{3}^{(\phi)}(M^2,s_0)={-}
\bigg\{\frac{m^2_Q}{2\,M^2} \,
\exp \Big(-\frac{m^2_Q}{M^2}\Big) \, \phi(u)
+\frac{1}{2}\exp \Big(-\frac{m^2_Q}{M^2}\Big)\Big[u\,\phi'(u)-\phi(u)\Big] \bigg\}
\bigg|_{u=\frac{1}{2}}\,.
\end{align}
In conclusion, the LO part of the LCSR in (\ref{eq:SR3}) is derived as a linear combination of distinct contributions from individual DAs:
\begin{eqnarray}
\label{eq:LOtot}
&&\mathcal F^{\rm (LO)}(M^2,s_0)=\mathcal F_{2p}^{\rm (LO)}(M^2,s_0)+\mathcal F_{3p}^{\rm (LO)}(M^2,s_0)
+\widetilde{\mathcal F}_{3p}^{\rm (LO)}(M^2,s_0),\\
&&\mathcal F_{2p}^{\rm (LO)}(M^2,s_0)=
- m^3_Q(1+{\Hat{m}_{q_2}})\exp\,\Big(-\frac{m^2_Q}{M^2}\Big)\,\bigg\{
\,\Big[1 - \exp\Big(\frac{m^2_Q-s_0}{M^2}\Big) \Big] f^{\perp}_V \frac{M^2}{m^2_Q}\phi^{\perp}_{2;V}(u)\nn
&&\hspace{3cm}+ 
\frac1{2}\delta_V f^{\parallel}_{V}{\psi}^{\perp}_{3;V}(u)
- \delta_V^2\Big[ {f_V^{\perp}\over 4}\big( 4 u\bar{u} \phi^{\perp}_{2;V}(u) + \phi^{\perp}_{4;V}(u)\big)+ { m^2_Q \over 4 M^2} f_V^{\perp} \phi^{\perp}_{4;V}(u)\Big]\nn
&&\hspace{3cm}-\delta_V^3\Big[\frac{ m^2_Q}{2 M^2} f_V^{\parallel} u \bar{u} {\psi}^{\perp}_{3;V}(u)+{1 \over 8}\frac{ m^4_Q}{M^4}  f_V^{\|} {\psi}^{\perp}_{5;V}(u)\Big]\nn
&&\hspace{3cm}+\delta_V^4\Big[\frac{ m^2_Q}{4M^2} f^{\perp}_V u {\bar u}\left(2u {\bar u}\phi^{\perp}_{2;V}(u)+\phi^{
\perp}_{4;V}(u)\right)+\frac{m^4_Q}{4M^4}f^{\perp}_{V}u{\bar u}\phi^{\perp}_{4;V}(u)\Big]
\bigg\}\bigg|_{u=\frac{1}{2}}\,,\label{eq:2pLO}\\
&&\mathcal F_{3p}^{\rm (LO)}(M^2,s_0)=
 m^3_Q (1+{\Hat{m}_{q_2}})\delta_V^2 f_V^\perp \,\bigg\{ \exp\Big(-\frac{m^2_Q}{M^2}\Big)\,\Big[2 \big(\overline{\Phi}^{\perp(2)}_{4;V}(u,\ell)\Big|^{\ell=0}_{\ell=1} -\overline{\Phi}^{\perp(1)}_{4;V}(u,\ell)\Big|^{\ell=0}_{\ell=1} \big) 
\nn
&&\hspace{3cm} -
\overline{\Psi}^{\perp}_{4;V}(u,\ell)\Big|_{\ell=0} -\overline{\widetilde{\Psi}}^{\perp}_{4;V}(u,\ell)\Big|_{\ell=1}+\overline{\widetilde{\Psi}}^{\perp}_{4;V}(u,\ell)\Big|^{\ell=0}_{\ell=1}\nn
&&\hspace{3cm}-
2\overline{\widehat{\Phi}}^{\perp(2134)}_{4;V}(u,\ell)\Big|_{\ell=1}+
2\overline{\widehat{\Phi}}^{\perp(2134)}_{4;V}(u,\ell)\Big|^{\ell=0}_{\ell=1}\Big]
\bigg\}\bigg|_{u=\frac{1}{2}}\,\nn
&& \hspace{3cm}-\frac{m^5_Q (1+{\Hat{m}_{q_2}})}{M^2}\delta_V^2 f_V^\perp \,\bigg\{ \exp\Big(-\frac{m^2_Q}{M^2}\Big)\,\nn
&&\hspace{3cm}\times\Big[-
\overline{\widehat{\Phi}}^{\perp(2134)}_{4;V}(u,\ell)\Big|_{\ell=1}+
\overline{\widehat{\Phi}}^{\perp(2134)}_{4;V}(u,\ell)\Big|^{\ell=0}_{\ell=1}\Big]
\bigg\}\bigg|_{u=\frac{1}{2}}\,,\\
&&\widetilde{\mathcal F}_{3p}^{\rm (LO)}(M^2,s_0)=- {\frac{1}{2}m^3_Q(1+{\Hat{m}_{q_2}})}\delta_V^2 f_V^\perp\exp \Big(-\frac{m^2_Q}{M^2}\Big) \bigg\{\left(\frac{m^2_Q}{M^2}-1\right)\nn
&&\hspace{3cm}\times\left[\overline{\widehat{\Phi}}^{\perp(2134)}_{4;V}(u,\ell)\bigg|_{\ell=1}-\overline{\widehat{\Phi}}^{\perp(2134)}_{4;V}(u,\ell)\bigg|^{\ell=0}_{\ell=1}\right]\nn
&&\hspace{3cm}+\left(u\,\frac{\partial}{\partial u}\overline{\widehat{\Phi}}^{\perp(2134)}_{4;V}(u,\ell)\bigg|_{\ell=1}-u\,\frac{\partial}{\partial u}\overline{\widehat{\Phi}}^{\perp(2134)}_{4;V}(u,\ell)\bigg|^{\ell=0}_{\ell=1}\right)\bigg|_{u=\frac{1}{2}}\bigg\},
\end{eqnarray}
where
\begin{eqnarray}
&&\overline{\widehat{\Phi}}^{\perp(2134)}_{4;V}(u,\ell)=\overline{\widehat{\Phi}}^{\perp(2)}_{4;V}(u,\ell)-\overline{\widehat{\Phi}}^{\perp(1)}_{4;V}(u,\ell)+\overline{\widehat{\Phi}}^{\perp(3)}_{4;V}(u,\ell)-\overline{\widehat{\Phi}}^{\perp(4)}_{4;V}(u,\ell)\,,\,\delta_V=\frac{m_V}{m_Q}.
\end{eqnarray}
After comparing our results with those in Ref. \cite{Wang:2020yvi}, we find agreement in the two-particle components. However, discrepancies arise in the three-particle counterparts, with numerical values displaying opposite signs. Upon careful examination, we identifiy a confusion in the simplification process observed in Eq. (25) of \cite{Wang:2020yvi} (also referred to in Eqs. (4.10-4.16) of \cite{Li:2020rcg}). This confusion stems from overlooking the introduction of additional variables, as demonstrated in Eq. \eqref{3p-redefine}, when addressing three-particle expressions. Consequently, inaccuracies occurs in integrating when applying the $1/(p\cdot x)$ replacement rule to the three-particle DAs. Our discovery emphasizes the critical importance of accurately managing higher-order terms and highlights the necessity for validation and verification in theoretical frameworks.

The subsequent stage of our analysis involves deriving the NLO contribution to Eq. \eqref{eq:SR3} by incorporating the twist-2 NLO double spectral densities, $\rho^{\rm (tw2, NLO)}(s_1, s_2)$, as explained in Eq. \eqref{eq:tw2nlo} of Appendix~\ref{appC}. The resulting expression for ${\cal F}^{\rm (NLO)}(M^2, s_0)$ within the triangular duality region is presented below, capturing the higher-order corrections to the LCSR formulation.
\begin{eqnarray}
\mathcal F^{\rm (NLO)}(M^2,s_0) = f^{\perp}_{V}  m_Q^2(1+{\Hat{m}_{q_2}}) 
\int _{2 m_Q^2}^{2 s_0} ds \exp\left(- \frac{s}{2 M^2}\right) f^{\rm (tw2)} \left ( \frac{s}{m_Q^2} -2 \right ) ,
\label{eq:NLOBorel}
\end{eqnarray}
accompanied by the NLO contributions of twist-2,
\begin{eqnarray}
f^{(\rm tw2)}(\sigma) ~&=& ~3 \, \bigg  \{ \ln \left ( {\mu^2 \over m_Q^2} \right )
+ {3 \over 4} \,   \ln \left ( {\nu^2 \over \mu^2} \right ) - {\rm Li}_2 \left ( - \sigma \right )
- {\rm Li}_2 \left ( - \sigma - 1 \right )  + 2 \,  {\rm Li}_2 \left ( - {\sigma \over 2} \right )
+ {\pi^2 \over 12} \nn
&& + \, \ln \left ( \frac{\sigma}{2} \right ) \, \ln \left ( \frac{\sigma + 2}{2} \right )
- \ln \left (\sigma + 1  \right ) \, \ln \left (\sigma + 2  \right )
+ \frac{2\,(\sigma+1)^2}{(\sigma+2)^3}\,\ln \left (\sigma+1 \right ) \nn
&& - \,  \frac{7\,\sigma^3+50\,\sigma^2+100\,\sigma+64}{4\,(\sigma+2)^3}\,
\ln \left ( \frac{\sigma}{2} \right )
- {1 \over 2} \, \ln \left ( \frac{\sigma + 2}{2} \right )
- {11\,\sigma^2+28\,\sigma+24 \over 8\,(\sigma+2)^2}  \bigg \}.
\label{eq:ftw2}
\end{eqnarray}

It's important to note that the twist-2 part of the expression precisely matches that derived in \cite{Li:2020rcg}. When transitioning to the pole-mass scheme for the heavy quark, we incorporate the terms $\Delta f^{\rm (tw2)}(\sigma)$ from (\ref{eq:deltapole}) to (\ref{eq:NLOBorel}). Additionally, we have verified the factorization-scale independence for both twist-2 terms in the LCSR (\ref{eq:SR3}) at $O(\alpha_s^2)$ in the asymptotic limit, which further enhances our confidence in the obtained results.

The derivation of the LCSR expression (\ref{eq:SR3}) for the strong $H^*HV$ coupling, applicable to both $B_{(s)}$ and $D_{(s)}$ mesons with their respective $b$ or $c$ quark masses, is now complete for the triangular duality region. This comprehensive formulation includes both LO and NLO terms, as described in (\ref{eq:LOtot}) and (\ref{eq:NLOBorel}), respectively. The formulation is now prepared for numerical analysis.

\section{Numerical analysis \label{sec-5}}
This section is dedicated to determining various input parameters and extracting the strong couplings $g_{D_{(s)}^*D_{(s)}V}$ and $g_{B_{(s)}^*B_{(s)}V}$ from the established LCSRs in Eq.(\ref{eq:SR3}). The mass of heavy-light mesons and light vector mesons are taken from the PDG \cite{ParticleDataGroup:2022pth}. The consideration of decay constants related to pseudoscalar and vector heavy-light mesons requires careful attention. We employ two distinct methods: the first utilizes LQCD values for decay constants of charmed and bottom mesons. Specifically, we incorporate $N_f=2+1+1$ results from \cite{FlavourLatticeAveragingGroupFLAG:2021npn} for heavy pseudoscalar mesons and decay constant ratios from \cite{Lubicz:2017asp}
\begin{eqnarray}
&f_D=212.0\pm 0.7 \,\mbox{MeV}, \qquad  & f_B=190.0\pm 1.3 \,\mbox{MeV} ,\nn
&f_{D_s}=249.9\pm 0.5 \,\mbox{MeV}, \qquad  & f_{B_s}=230.3\pm 1.3 \,\mbox{MeV} ,\nn
&f_{D^*}=228.5\pm 7.7\,\mbox{MeV},  \qquad & f_{B^*}=182.02\pm 4.4\,\mbox{MeV},\nn
&f_{D^*_s}=271.6\pm 5.0\,\mbox{MeV},  \qquad & f_{B^*_s}=224.3\pm 2.6\,\mbox{MeV}.
\label{eq:lattDC}
\end{eqnarray}

For comparison, the second employs two-point QCD sum rules to determine decay constants $f_H$ and $f_{H^*}$, ($H=D_{(s)},B_{(s)}$), as described in \cite{Gelhausen:2013wia,Khodjamirian:2020mlb}. This approach incorporates NLO corrections to partially cancel the perturbative effects of the LCSRs. Here, we present the decay constant values computed using the two-point QCDSRs
\begin{eqnarray}
& f_D=190.4^{+8.9}_{-7.9}~\mbox{MeV}, \qquad  & f_B=201.1^{+21.2}_{-17.6}~\mbox{MeV}
, \nn
&f_{D_s}=200.7^{+10.4}_{-8.7}~\mbox{MeV}, \qquad  &f_{B_s}=223.2^{+22.2}_{-18.0}~\mbox{MeV}, \nn
&f_{D^*}= 245.5^{+24.1}_{-21.3}~\mbox{MeV}, \qquad  &f_{B^*}=214.2^{+10.1}_{-17,8}~\mbox{MeV}\,.\nn
&f_{D^*_s}= 281.1^{+24.6}_{-20.9}~\mbox{MeV}, \qquad  &f_{B^*_s}=245.2^{+18.9}_{-25.3}~\mbox{MeV}\,.
\label{eq:SRDC}
\end{eqnarray}
Here, we use the RunDec Mathematica package \cite{Chetyrkin:2000yt} to investigate the evolution of the QCD coupling and quark mass (in the $\overline{\text{MS}}$ scheme) with four-loop precision, as outlined in Table \ref{tab:QCD}. Starting with initial conditions set at $m_Z=91.1876$ GeV, we determine the running parameters. Matching scales are identified at 4.2 GeV for the transition from $n_f=5$ to $n_f=4$ and at 1.3 GeV for the transition from $n_f=4$ to $n_f=3$.

\begin{table}[tb]
\centering
\renewcommand{\arraystretch}{1.2}
\resizebox{\columnwidth}{!}{
\begin{tabular}{|c|c|c|c|}
\hline
parameter & input value & [Ref.]& rescaled values
\\[1mm]\hline
\multirow{2}{*}{$\alpha_s(m_Z)$} &
\multirow{2}{*}{$0.1179 \pm 0.0010$} &
\multirow{6}{*}{\cite{ParticleDataGroup:2022pth}} &
$\alpha_s(1.5\,\mbox{GeV})=0.3487^{+0.0102}_{-0.0097}$
\\
~& ~& ~&
$\alpha_s(3.0\,\mbox{GeV})=0.2527^{+0.0050}_{-0.0048}$
\\[1mm]
$\overline{m}_c(\overline{m}_c)$    &  1.280 $\pm$ 0.025 \mbox{GeV} &
~ &
$\overline{m}_c(1.5\,\mbox{GeV})=1.205 \pm 0.034$ \mbox{GeV}
\\[1mm]
$\overline{m}_b(\overline{m}_b)$   &    4.18  $\pm$ 0.03 \,\mbox{GeV} &
&$\overline{m}_b(3.0\,\mbox{GeV})=4.473  \pm 0.04$  \mbox{GeV}
\\[1mm]
\multirow{2}{*}{($\overline{m}_u+\overline{m}_d$)(2 \,\mbox{GeV}) }
&
\multirow{2}{*}{$6.78 \pm 0.08 $ \mbox{MeV}}& \multirow{2}{*}{\cite{FlavourLatticeAveragingGroupFLAG:2021npn,ParticleDataGroup:2022pth}} & ($\overline{m}_u+\overline{m}_d$)(1.5 \,\mbox{MeV}) =
7.305 $\pm$ 0.09 \mbox{MeV}\\
~ & ~& ~&
($\overline{m}_u+\overline{m}_d$)(3.0 \,\mbox{GeV}) =
6.331 $\pm$ 0.07 \mbox{MeV}
\\[1mm] 
\multirow{2}{*}{($\overline{m}_s$)(2 \,\mbox{GeV}) }
&
\multirow{2}{*}{$93.1 \pm 0.6 $ \mbox{MeV}}& \multirow{2}{*}{\cite{FlavourLatticeAveragingGroupFLAG:2021npn,ParticleDataGroup:2022pth}} & ($\overline{m}_s$)(1.5 \,\mbox{MeV}) =
100.305 $\pm$ 0.65 \mbox{MeV}\\
~ & ~& ~&
($\overline{m}_s$)(3.0 \,\mbox{GeV}) =
86.936 $\pm$ 0.56 \mbox{MeV}
\\[1mm] 
\hline  
\end{tabular}}
\caption{QCD parameters used in the LCSRs and two-point QCDSRs.}
\label{tab:QCD}
\end{table}
Our discussion for the vector meson LCDAs primarily focuses on the LO part of the sum rules in Eq.(\ref{eq:LOtot}). As we explained in the last section, we need to determine the values of the DAs or their derivatives at the midpoint $u=1/2$. This relies on a complete set of coefficients in the conformal expansion, which is shaped by the inherent symmetry properties governed by the renormalization group (RG) equation controlling their scale dependence \cite{Agaev:2010aq}. It differs from the LCSRs for $B_{(s)}\to V$ and $D_{(s)}\to V$ form factors, where vector meson DAs undergo weighting by the Borel exponent and integration over a duality interval. Additionally, since we consider the NLO for asymptotic twist-2 two-particle DAs, the renormalization of Gegenbauer coefficients should extend up to NLO as well. This involves a complex multiplicative process, as illustrated by Eq.\eqref{eq:evoaNLO} in Appendix \ref{app:lcda}.

In addition to the decay constants $f^{\perp,\parallel}_V(\mu)$ and twist-2 parameters $a^{\perp,\parallel}_n(\mu)$ detailed in Table \ref{tab:numDA1}, we present the Gegenbauer moments for the twist-3 and twist-4 DAs of vector mesons in Table \ref{tab:numDA2}. These DAs are developed up to the NLO of conformal expansion in Ref.\,\cite{Ball:1998ff,Ball:1998sk}. The normalization and non-asymptotic coefficients at $\mu=1\,{\rm GeV}$, used as input parameters, are determined through computations using QCD sum rules (see Ref.\,\cite{Ball:1998ff,Ball:1998sk,Ball:2007zt} and citations therein).
For the $\omega$ meson, the decay constants $f^{\parallel}_{\omega}=197(8)\,{\rm MeV}$ and $f^{\perp}_{\omega}=148(13)\,{\rm MeV}$ are extracted from the provided reference \cite{Bharucha:2015bzk}. The remaining parameters for this meson are consistent with those employed for the $\rho$ meson.

\begin{table}[htb]
\renewcommand{\arraystretch}{1.1}
\addtolength{\arraycolsep}{3pt}
$$
\begin{array}{l || c | c ||  l | l || c | c }
\hline
& \multicolumn{2}{c||}{\rho} & \multicolumn{2}{c||}{K^*}  &  
\multicolumn{2}{c}{\phi}\\
\cline{2-7}
& \mu = 1.5\,{\rm GeV} & \mu = 3\,{\rm GeV} & \mu = 1.5\,{\rm GeV} & \mu =
3\,{\rm GeV} & \mu = 1.5\,{\rm GeV} & \mu = 3\,{\rm GeV}\\
\hline
f^\parallel_V[\rm MeV] & 216(3) & 216(3) & \phantom{-}220(5) &  \phantom{-}220(5)& 215(5) & 215(5) 
\\
f^\perp_V[\rm MeV] & 158(9) & 150(8) & \phantom{-}177(9) & \phantom{-}168(8) & 178(9) & 169(8)
\\
\hline
a_1^\parallel & 0 & 0 & \phantom{-}0.027(18) & \phantom{-}0.023(15) & 0 & 0 
\\
a_1^\perp & 0 & 0 & \phantom{-}0.036(27) & \phantom{-}0.033(25) & 0 & 0
\\
a_2^\parallel & 0.124(58) & 0.100(47) & \phantom{-}0.091(74) & 
\phantom{-}0.073(60) & 0.149(74) & 0.120(60)
\\
a_2^\perp & 0.120(51) & 0.101(43) & \phantom{-}0.086(68) &
\phantom{-}0.072(58) & 0.120(60) & 0.101(51)
\\\hline
\end{array}
$$
\renewcommand{\arraystretch}{1}
\addtolength{\arraycolsep}{-3pt}
\vspace*{-10pt}
\caption[]{\small Decay constants and twist-2 hadronic parameters, initially referenced at $\mu= 1\,{\rm GeV}$\cite{Ball:2007zt}, are then scaled to $\mu= 1.5\,{\rm GeV}$ and $\mu= 3\,{\rm GeV}$ using the evolution equations outlined in Appendix \ref{app:lcda}.}
\label{tab:numDA1}
\end{table}

\begin{table}[tb]
\renewcommand{\arraystretch}{1.1}
\addtolength{\arraycolsep}{3pt}
$$
\begin{array}{l || c | c ||  l | l || c | c }
\hline
& \multicolumn{2}{c||}{\rho} & \multicolumn{2}{c||}{K^*}  &  
\multicolumn{2}{c}{\phi}\\
\cline{2-7}
& \mu = 1.5\,{\rm GeV} & \mu = 3\,{\rm GeV} & \mu = 1.5\,{\rm GeV} & \mu =
3\,{\rm GeV} & \mu = 1.5\,{\rm GeV} & \mu = 3\,{\rm GeV}\\
\hline
\zeta_{3V}^\parallel 
& 0.022(7) & 0.016(5) & \phantom{-}0.017(6) & \phantom{-}0.012(4) & 
0.018(6) & 0.013(4)
\\
\widetilde\lambda_{3V}^\parallel 
& 0 & 0 & \phantom{-}0.025(11)& \phantom{-}0.017(7) & 0 & 0 
\\
\widetilde\omega_{3V}^\parallel 
& -0.059(19) & -0.037(12) & -0.046(19) &  -0.028(12) & -0.030(10) & -0.019(6)
\\
\kappa_{3V}^\parallel 
& 0 & 0 & \phantom{-}0.000(1) & \phantom{-}0.000(1) & 0 & 0
\\
\omega_{3V}^\parallel 
& 0.110(36) & 0.078(24) & \phantom{-}0.074(28) & \phantom{-}0.052(19) &
0.066(21) & 0.046(15)
\\
\lambda_{3V}^\parallel 
& 0 & 0 & -0.005(3) & -0.003(2) & 0 & 0
\\
\kappa_{3V}^\perp 
& 0 & 0 & \phantom{-}0.003(3) & \phantom{-}0.002(2) & 0 & 0 
\\
\omega_{3V}^\perp 
& 0.435(198) & 0.335(152) & \phantom{-}0.237(79) & \phantom{-}0.183(61) &  
0.158(63) & 0.122(49)
\\
\lambda_{3V}^\perp 
& 0 & 0 & -0.018(14) & -0.012(9) & 0 & 0\\
\hline 
\zeta_{4V}^\parallel 
& 0.062(27) & 0.054(23) & \phantom{-}0.018(18) & \phantom{-}0.015(15) & 0.000(18) & 0.000(15)\\
\widetilde\omega_{4V}^\parallel 
& -0.021(7) & -0.014(5) & -0.014(7) & -0.010(5) & -0.014(7) & -0.010(5)\\
\zeta_{4V}^\perp
& -0.026(41) & -0.022(32) & -0.009(24) & -0.008(19) & -0.009(24) & -0.008(19)\\
\widetilde\zeta_{4V}^\perp
& -0.065(41) & -0.052(32) & -0.041(32) & -0.032(26) & -0.025(32) & -0.020(26)\\
\kappa_{4V}^\parallel
& 0 & 0 & -0.022(5) & -0.019(5) & 0 & 0\\
\kappa_{4V}^\perp
& 0 & 0 & \phantom{-}0.012(5) & \phantom{-}0.011(5) & 0 & 0\\
\hline 
\end{array}
$$
\renewcommand{\arraystretch}{1}
\addtolength{\arraycolsep}{-3pt}
\vspace*{-10pt}
\caption[]{\small The twist-3 and twist-4 hadronic parameters, initially referenced at $\mu= 1\,{\rm GeV}$ \cite{Ball:2007zt}, are then scaled to $\mu= 1.5\,{\rm GeV}$ and $\mu= 3\,{\rm GeV}$ using the evolution equations outlined in Appendix \ref{app:lcda}.}\label{tab:numDA2}
\end{table}

Subsequently, we specify the renormalization scale of the quark-gluon coupling and quark masses, the factorization scale, the Borel parameter, and the quark-hadron duality thresholds of the LCSRs in Eq.\eqref{eq:SR3}. These parameters play crucial role in determining the behavior of the sum rules and extracting the strong couplings. 
We choose these parameters to ensure that the sum rules exhibit strong convergence and stability, while also minimizing theoretical uncertainties in the final outcomes.
Since we use the finite heavy quark mass, these parameters are naturally different for the couplings with charmed and bottom mesons. However, the heavy-quark spin symmetry enables certain scale and the Borel parameter to be aligned in the $H$ and $H^*$ channels.
Detailed default values and intervals for these parameters are provided in Table \ref{tab:scales}, which offers a comprehensive overview of the specific values and intervals for the parameters mentioned above. 
\begin{table}[tb]
\centering
\begin{tabular}{|c|c|c|c|c|c|}
\hline
Parameter & default value (interval) & [Ref.]
&Parameter & default value (interval) & [Ref.] \\[1mm]
\hline
\multicolumn{3}{|c|}{ charmed meson sum rules}
&
\multicolumn{3}{|c|}{bottom meson sum rules}\\
\hline
$\mu$ (GeV) & 1.5~(1.0\,-\,3.0)&
\multirow{3}{*}{\cite{Khodjamirian:2009ys}}
&$\mu$  (GeV) & {3.0~(2.5\,-\,4.5)}&
\multirow{3}{*}{\cite{Khodjamirian:2011ub}} \\[1mm]
$M^2$ (GeV$^2$)  & 4.5~(3.5\,-\,5.5) &
& $M^2$  (GeV$^2$)  & {16.0~(12.0\,-\,20.0)} &\\[1mm]
$ s_0$  (GeV$^2$) & 7.0~(6.5\,-\,7.5) &
&$ s_0$  (GeV$^2$) &  {37.5\,(35.0\,-\,40.0)} &\\[1mm]\hline
$\bar{M}^2$(GeV$^2$) & 2.0~(1.5\,-\,2.5)&
\multirow{5}{*}{\cite{Gelhausen:2013wia}}
&$\bar{M}^2$ (GeV$^2$)& {5.5~(4.5\,-\,6.5)} &
\multirow{5}{*}{\cite{Gelhausen:2013wia}}\\[1mm]
\multirow{2}{*}{$\bar{s}_{0}$ (GeV$^2$)} & 5.6 ($f_D$) &
&\multirow{2}{*}{$\bar{s}_{0}$(GeV$^2$)}  & 33.9 ($f_B$) & 
\\
& 6.3 ($f_{D_s}$) & & & 35.6($f_{B_s}$) & \\
\multirow{2}{*}{$\bar{s}_{0}^{\,*}$(GeV$^2$)} & {6.2}($f_{D^*}$) & 
&\multirow{2}{*}{$\bar{s}_{0}^{*}$(GeV$^2$)} & 34.1 ($f_{B^*}$)&  \\
& 7.4 ($f_{D^*_s}$) & & & 36.3 ($f_{B^*_s}$) & \\
\hline
\end{tabular}
\caption{
The renormalization scale $\mu$, Borel parameter $M^2$ and $\bar{M}^2$, along with duality threshold $s_0$ and $\bar{s}_0$ ($\bar{s_0}^{*}$), are employed in the LCSR and the two-point QCDSRs to determine the decay constants of $H$ ($H^*$) mesons for both charmed and bottom particles.}
\label{tab:scales}
\end{table}
The determination of the Borel parameter $M^2$ and the effective threshold $s_0$ in LCSRs follows established criteria aimed at ensuring the smallness of subleading power contributions in the light-cone OPE while simultaneously suppressing contributions from excited states. These criteria typically involve considerations of convergence and stability in the sum rule framework, and in our analysis, we adhere to them to ensure the reliability and accuracy of our results.  These parameters, presented in Table \ref{tab:scales}, are consistent with prior choices in LCSRs analyses of heavy-to-light decay form factors and couplings \cite{Khodjamirian:2020mlb,Khodjamirian:2009ys, Khodjamirian:2011ub}. 
Furthermore, we choose the renormalization scale $\mu$ equal to the factorization scale used in the OPE, which is approximately given by $\mu\sim\sqrt{m^2_H-m^2_Q}\sim \sqrt{2m_Q \bar{\Lambda}}$ ($\bar{\Lambda}=m_H-m_Q$). To maintain the perturbative expansion of the correlation function in $\alpha_s$ within reasonable bounds, the NLO twist-2 and higher power terms are constrained to be no more than $30\%$ of LO counterparts. Consequently, we adopt default values of $\mu_c = 1.5\, {\rm GeV}$  for charm and $\mu_b = 3\, {\rm GeV}$ for bottom and  investigate the range of variation within specified intervals $(1.0\sim3.0){\rm GeV}$  for charm and $(2.5\sim4.5){\rm GeV}$  for bottom to quantify the associated uncertainties. This choice is motivated by considerations of consistency within the theoretical framework and has been adopted in previous analyses within the field \cite{Khodjamirian:1997ub,Bagan:1997bp,Khodjamirian:2000ds,Ball:2004ye,Duplancic:2008ix,Khodjamirian:2009ys,Khodjamirian:2011ub}. By aligning these scales, we ensure coherence in our calculations and facilitate comparisons with prior studies.

\begin{figure}[!tb]
\centering
\includegraphics[width=0.49 \columnwidth]{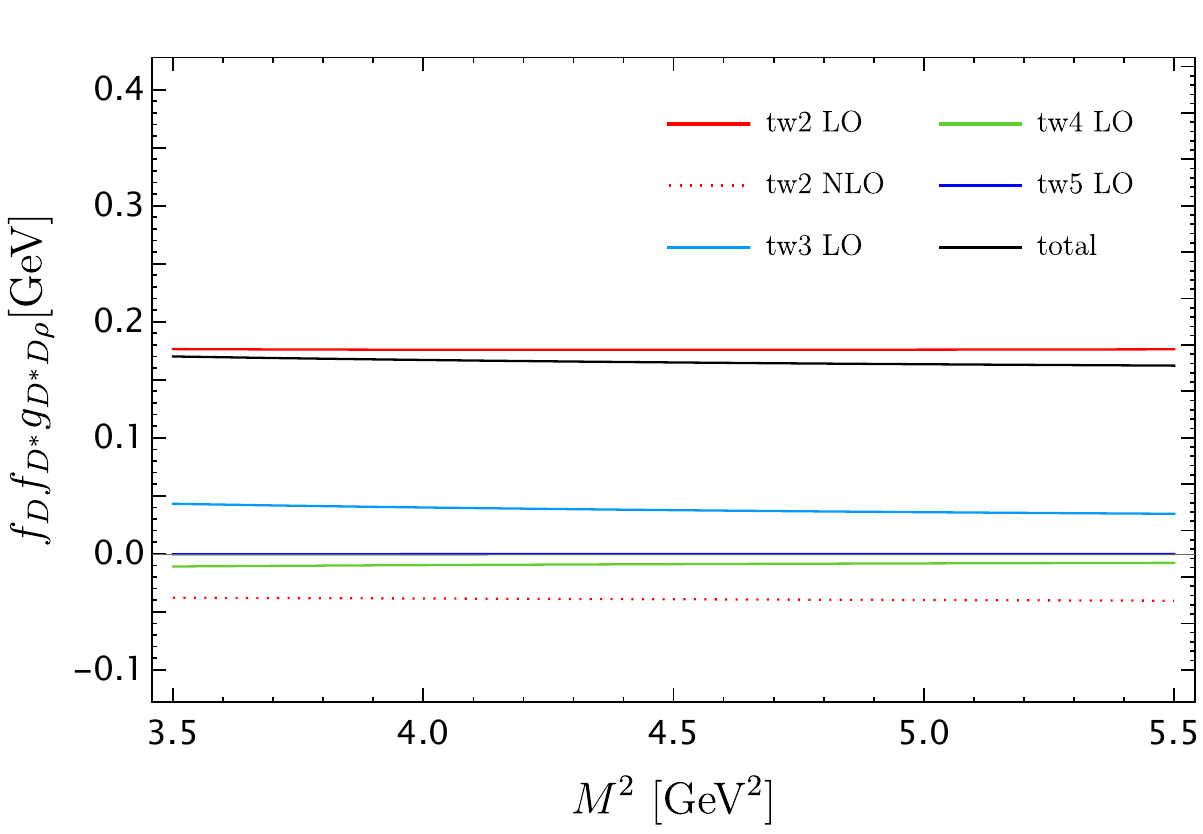}
\includegraphics[width=0.49 \columnwidth]{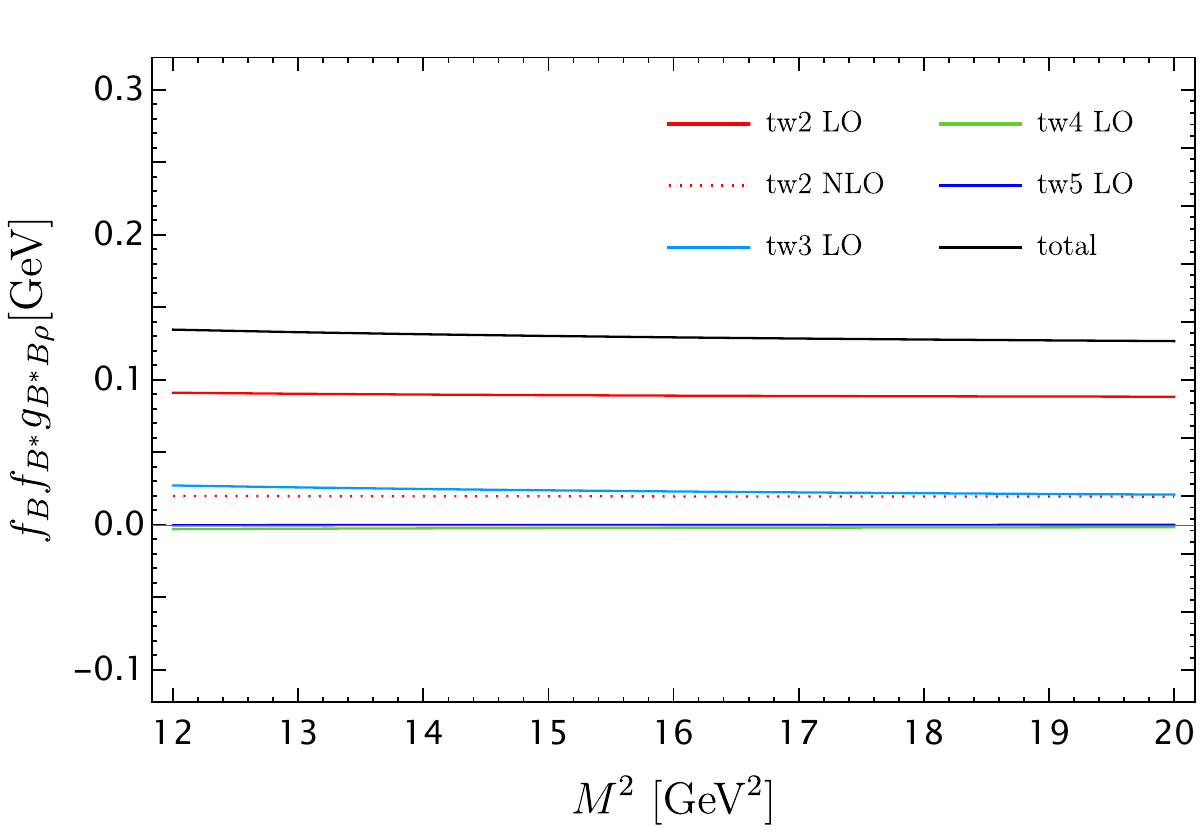}
\includegraphics[width=0.49 \columnwidth]{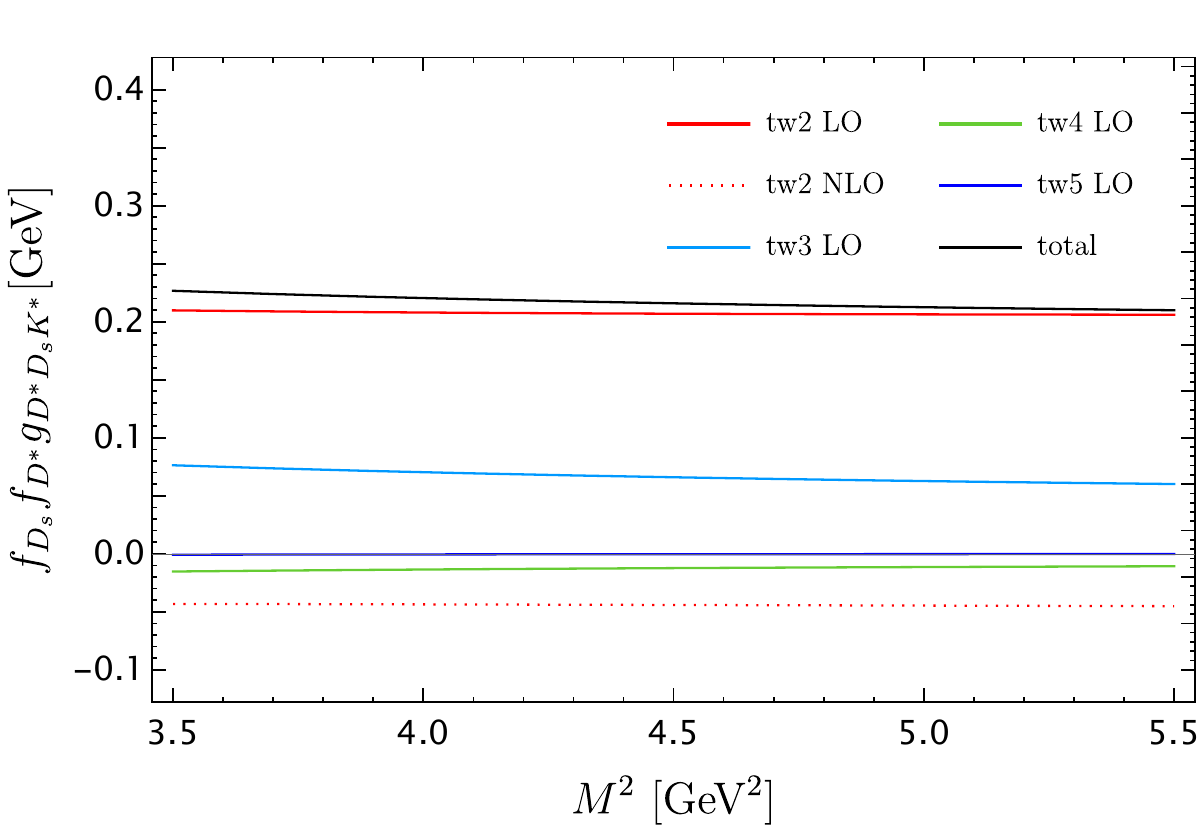}
\includegraphics[width=0.49 \columnwidth]{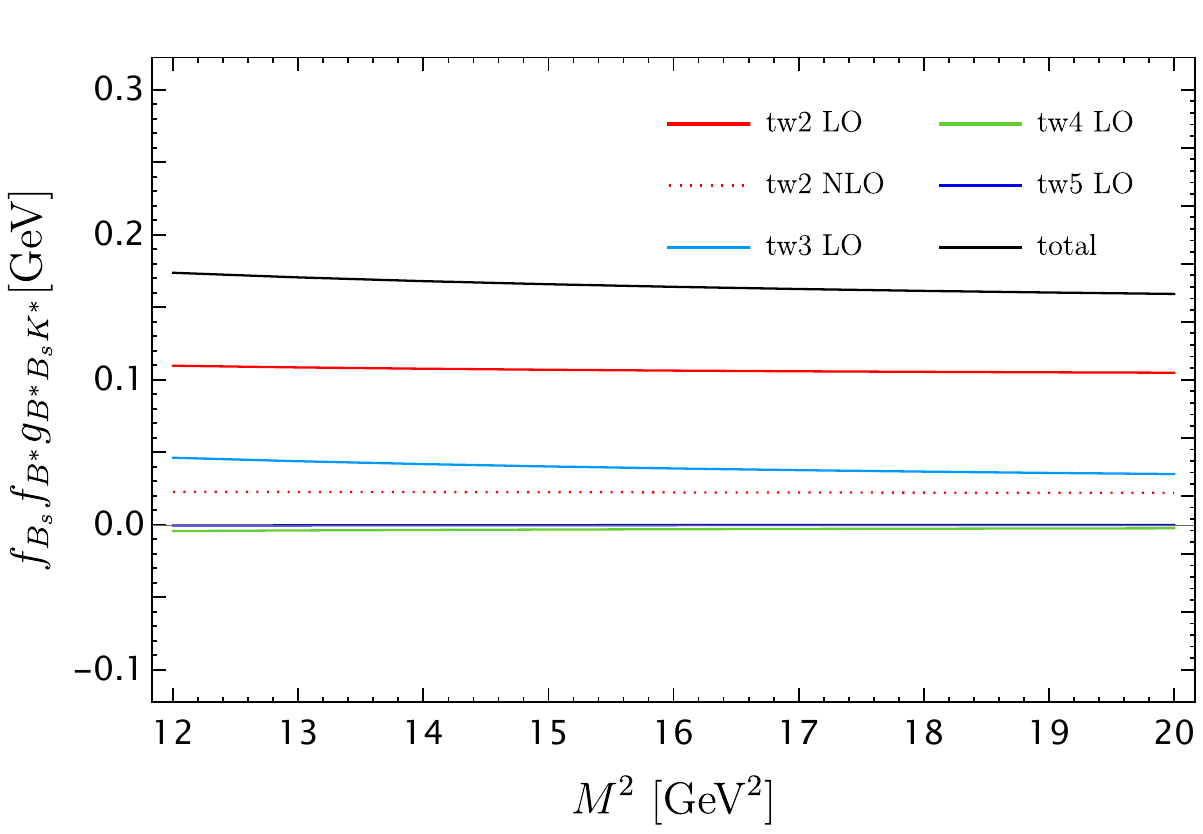}
\includegraphics[width=0.49 \columnwidth]{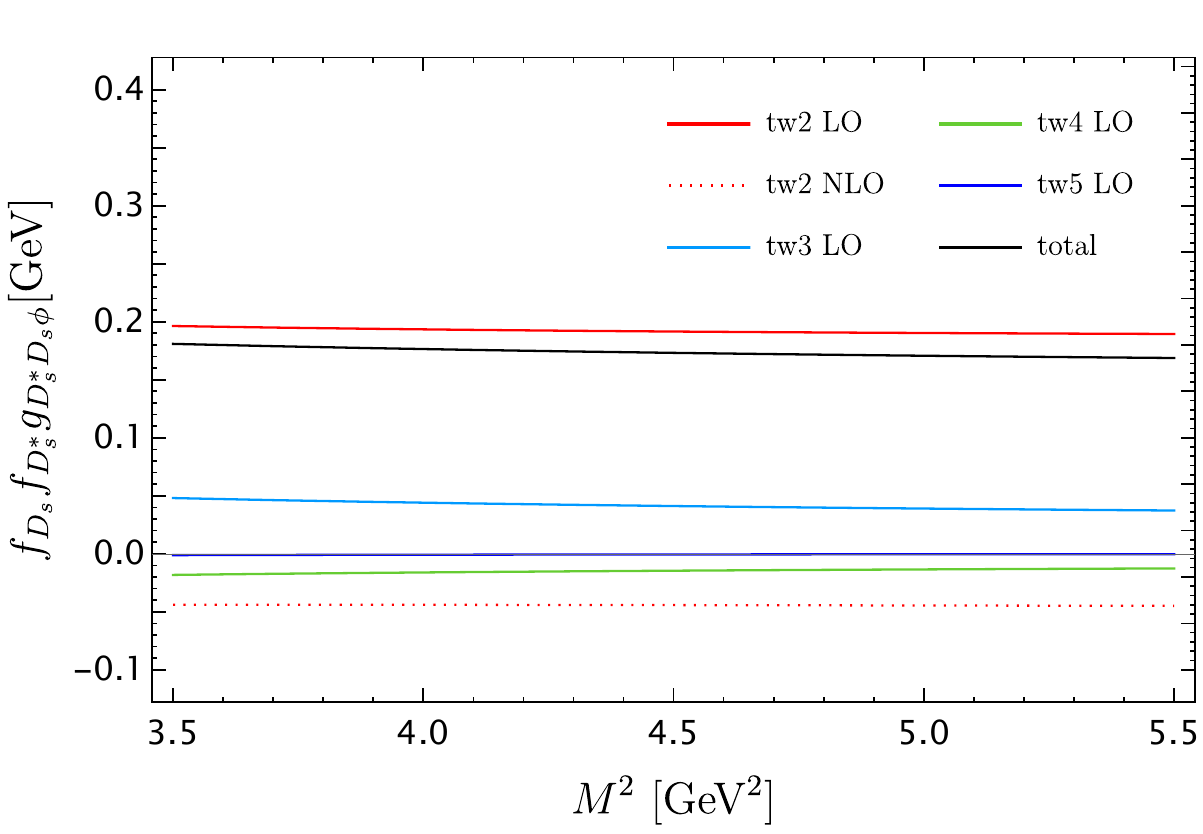}
\includegraphics[width=0.49 \columnwidth]{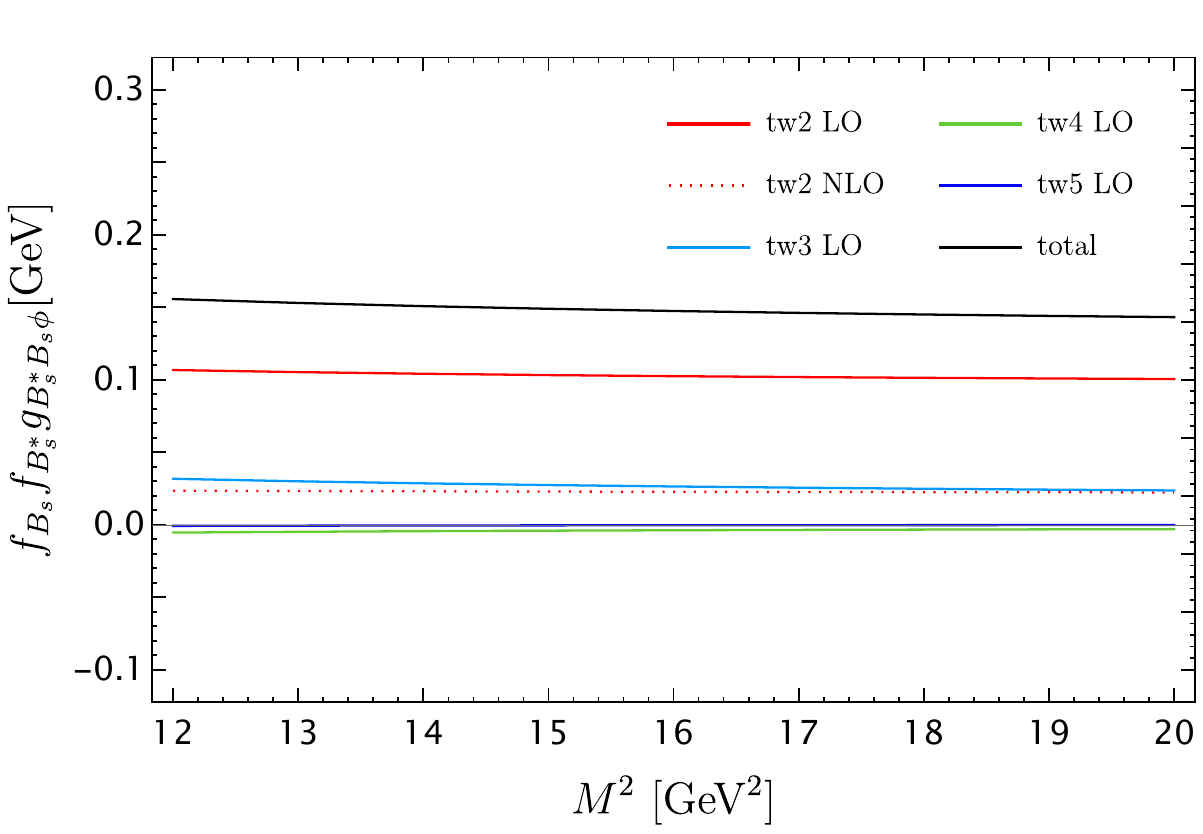}
\caption{Borel parameter dependence of the products
$ f_{D^*_{(s)}} f_{D_{(s)}} g_{D^*_{(s)}D_{(s)}V}$ and $ f_{B^*_{(s)}}f_{B_{(s)}} g_{B^*_{(s)}B_{(s)}V}$ calculated from LCSR. Displayed are the total values and separate  contributions from  twist-2 to twist-5 
 of the vector meson ($\rho,K^*,\phi$) DAs, considering central values for all the other input parameters.}
\label{fig:M2dep}
\end{figure}

\begin{figure}[!tb]
\centering
\includegraphics[width=0.49 \columnwidth]{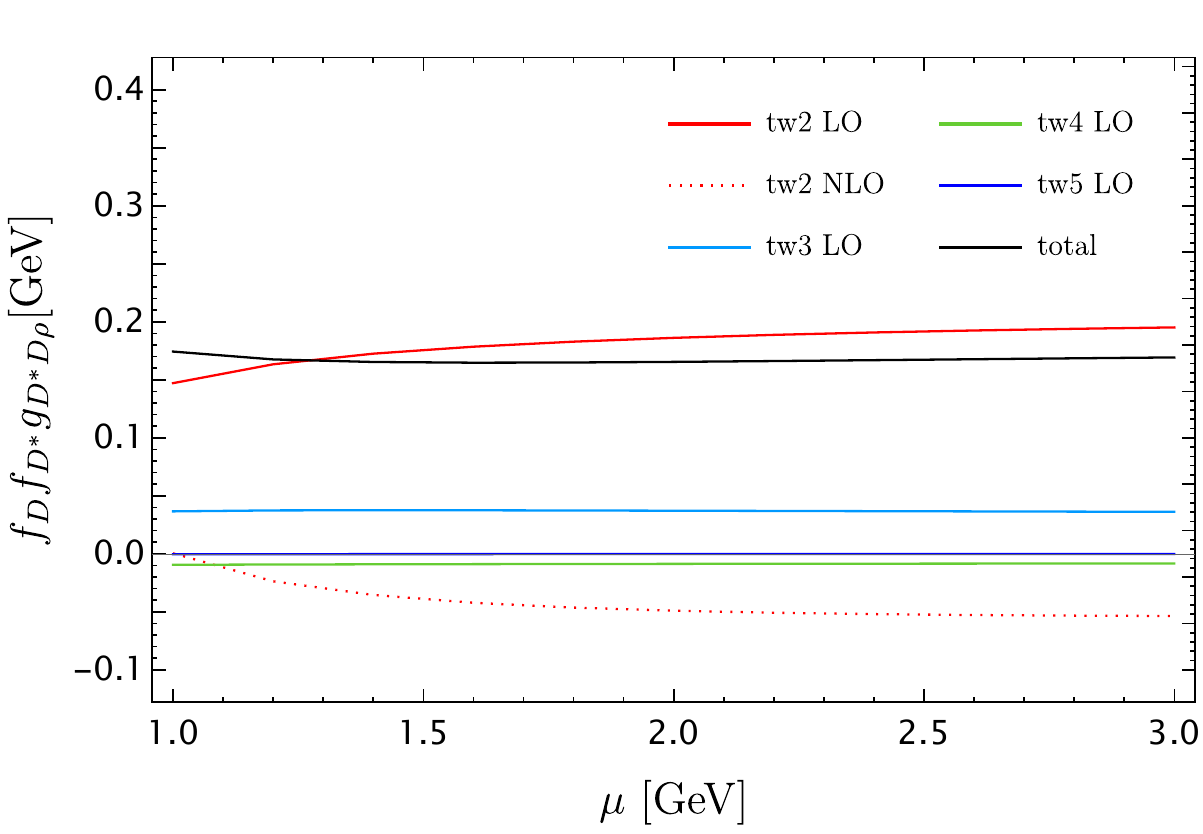}
\includegraphics[width=0.49 \columnwidth]{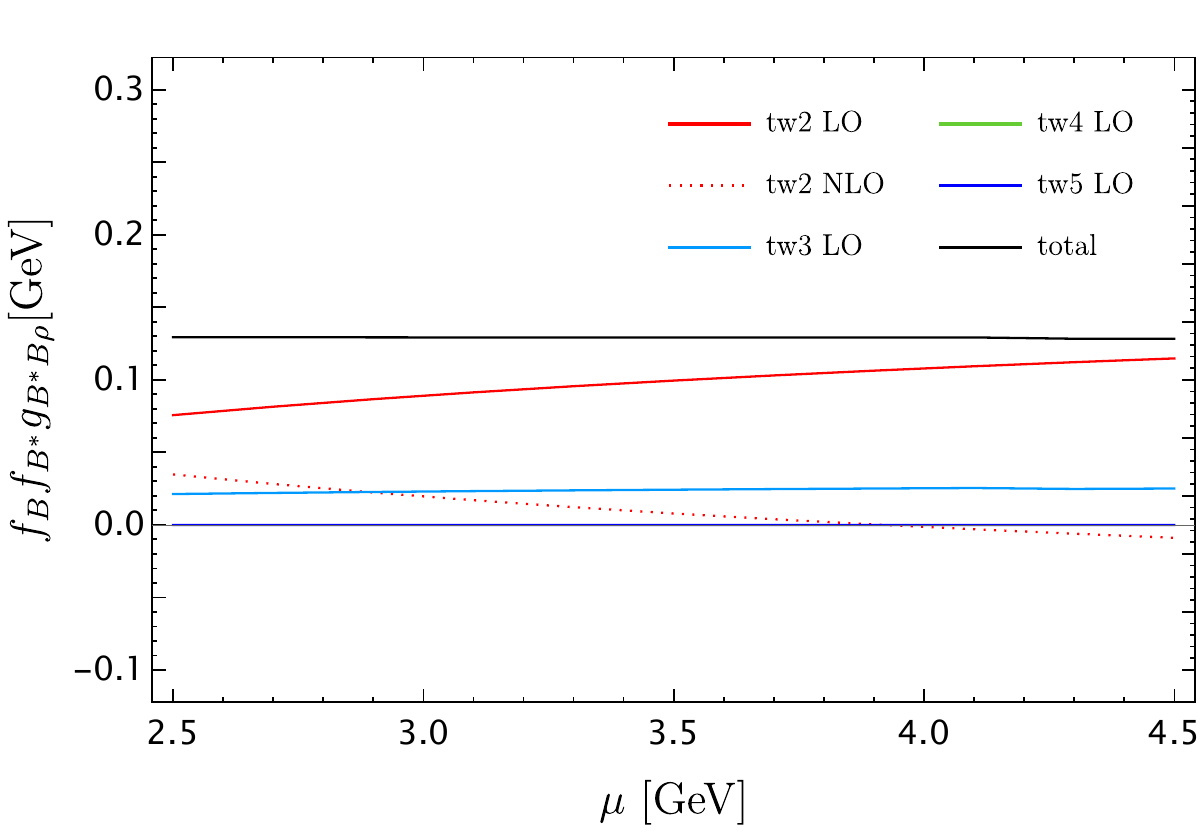}
\includegraphics[width=0.49 \columnwidth]{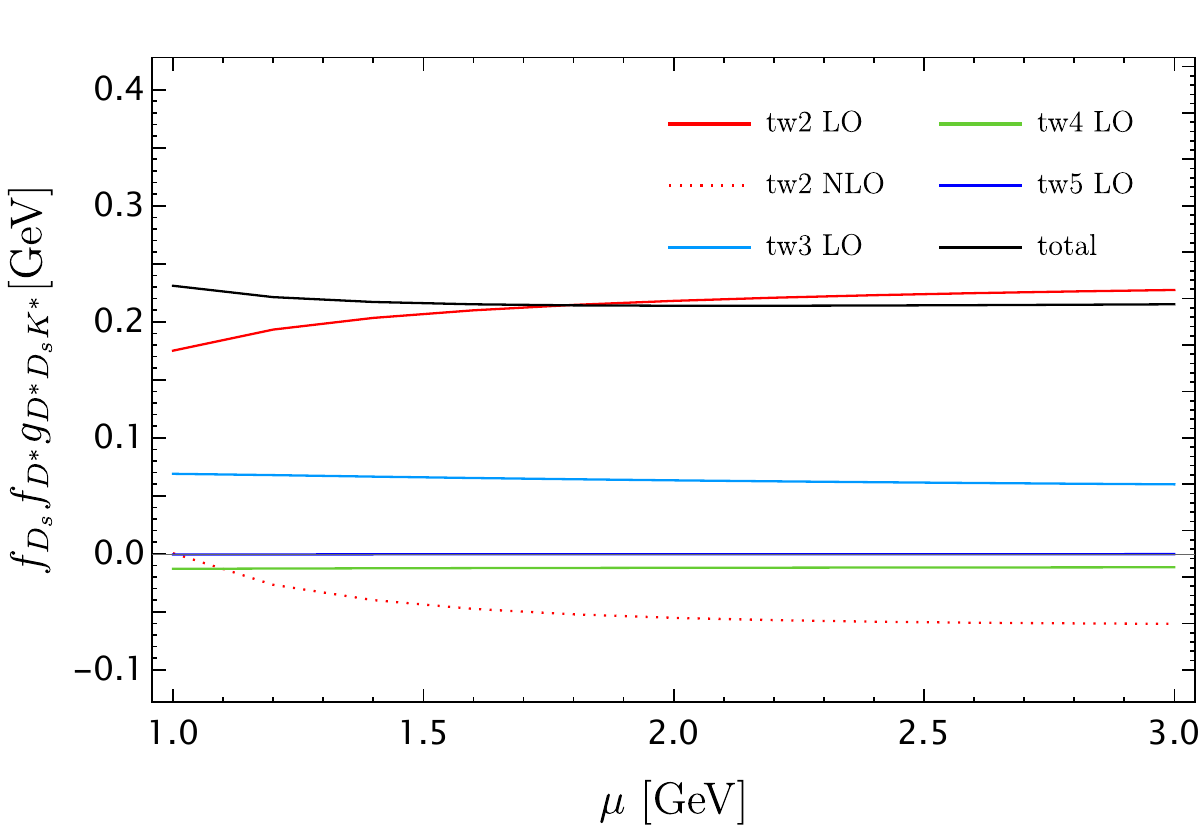}
\includegraphics[width=0.49 \columnwidth]{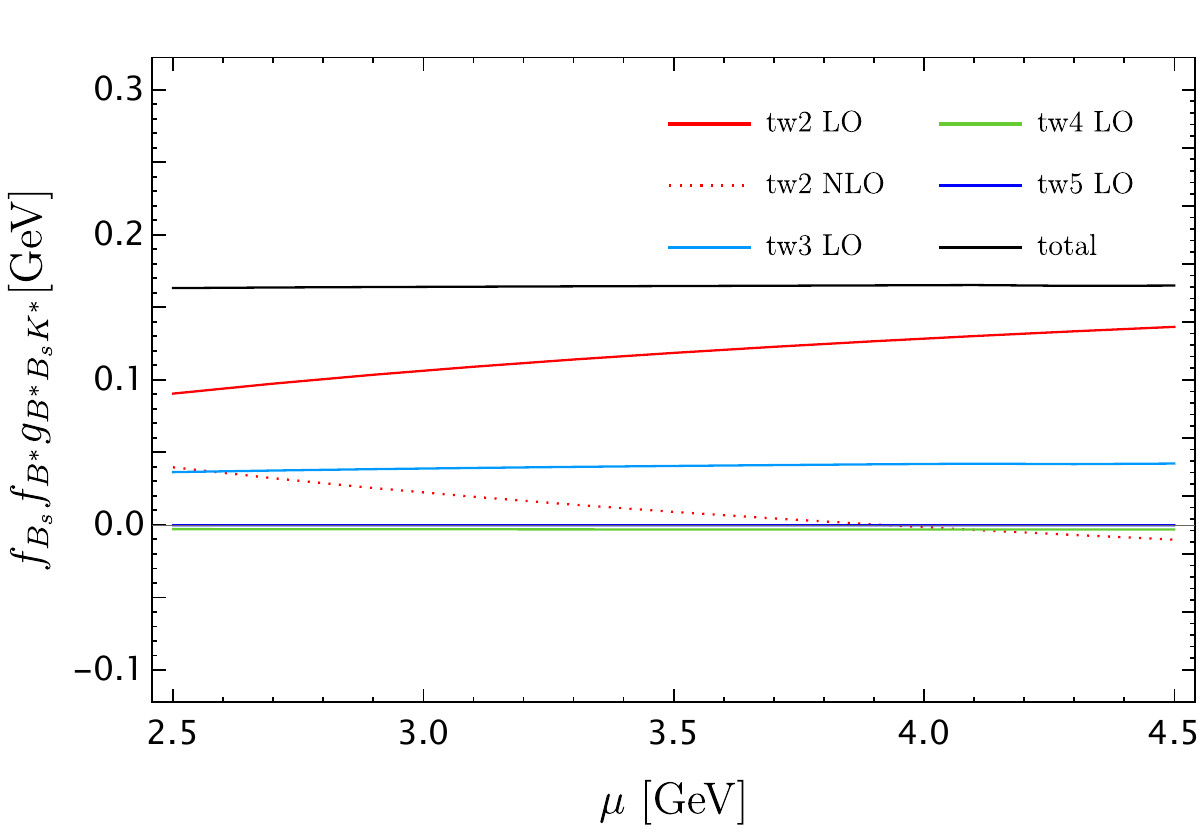}
\includegraphics[width=0.49 \columnwidth]{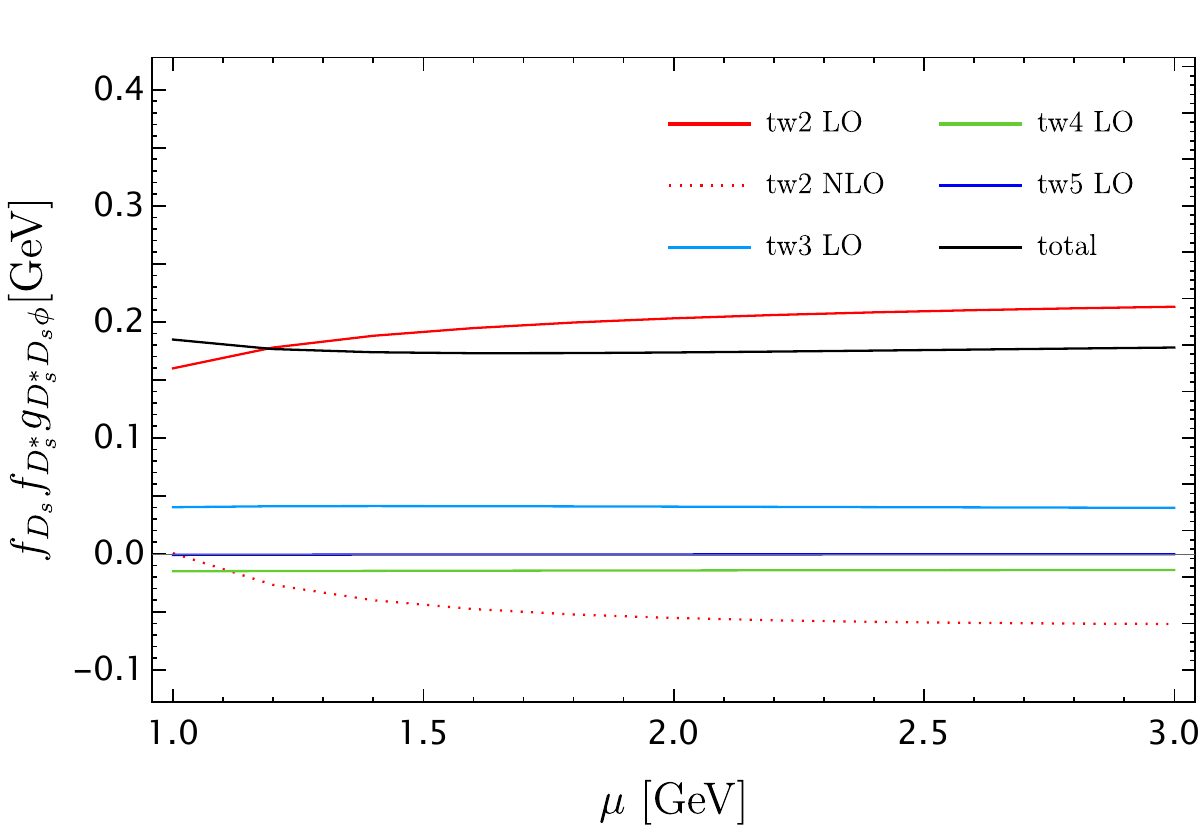}
\includegraphics[width=0.49 \columnwidth]{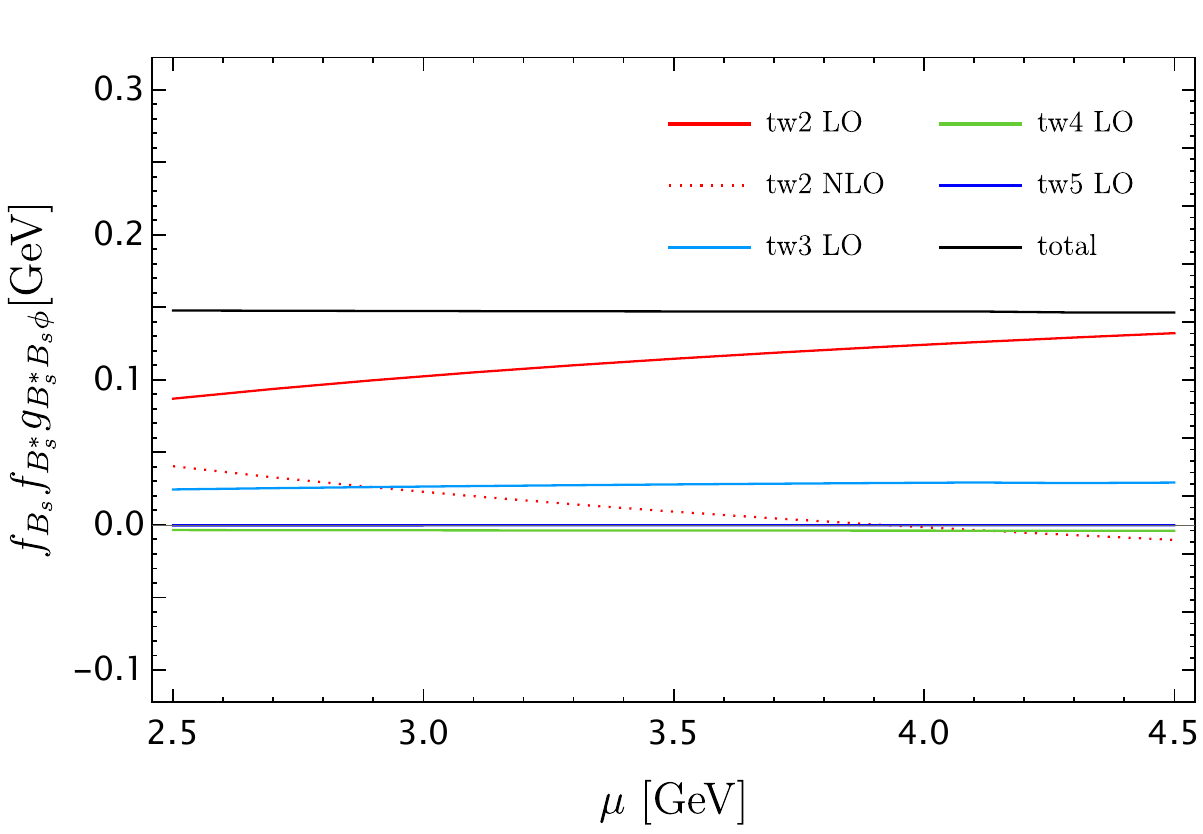}
\caption{Scale dependence of the products
$ f_{D^*_{(s)}} f_{D_{(s)}} g_{D^*_{(s)}D_{(s)}V}$ and $ f_{B^*_{(s)}}f_{B_{(s)}} g_{B^*_{(s)}B_{(s)}V}$ calculated from LCSR.
Displayed are the total values and separate  contributions from  twist-2 to twist-5
 of the vector meson($\rho,K^*,\phi$) DAs, considering central values for all other input parameters.}
\label{fig:mudep}
\end{figure}

Moreover, it's valuable to analyze how LCSRs for $D^*_{(s)}D_{(s)}V$ and $B^*_{(s)}B_{(s)}V$ couplings, along with decay constant products, respond to changes in the Borel parameter $M^2$ and the scale $\mu$.
Figure \ref{fig:M2dep} depicts the dependence on the Borel parameter $M^2$, illustrating a nearly flat behavior within the range considered, indicating minimal sensitivity to variations in $M^2$. This observation suggests that the extracted quantities are robust against changes in the Borel parameter, contributing to the reliability of our results.
In Figure \ref{fig:mudep}, we observe the dependence of individual contributions on the scale $\mu$. Remarkably, contributions from twist-2 at both LO and NLO display significant but opposite dependencies on $\mu$, almost canceling each other. While other higher-twist contributions exhibit minimal dependence, the overall outcome indicates only slight sensitivity to the scale $\mu$. This behavior suggests a delicate balance between different contributions in the LCSRs framework, highlighting the importance of considering various factors when interpreting the results.

\begin{table}[tb]
\centering
\setlength\tabcolsep{5pt}
\def\arraystretch{1.4}
\begin{tabular}{|c||c|c|c|c|c|c||}
\hline
 Power  & $\delta^0_{\phi}$ & $\delta^1_{\phi}$ & $\delta^2_{\phi}$ & $\delta^3_{\phi}$ &  $\delta^4_{\phi}$ & total\\
\hline
$g_{D^*_s D_s\phi}$
& 2.41
& 0.63 
& $-0.48$
& $-0.045$
& 0.020
& 2.54
\\
$g_{B^*_s B_s\phi} $
& 2.47
& 0.53
& $-0.48$
& $-0.016$
& 0.020
& 2.86
\\
\hline
\end{tabular}
\caption{Power corrections for the strong couplings $g_{H^*H\phi}$ using the decay constats from LQCD, in units of $[{\rm GeV}^{-1}]$. The expansion extends to the power of $\delta_{\phi}$ ranging from 0 to 4, where $\delta_{\phi}=\delta_{\phi}^{(Q)}=m_{\phi}/m_Q$, for the coupling with $D_s^{(\ast)}$ meson $\delta_{\phi}^{(c)}=0.85$ and for the coupling $B_s^{(\ast)}$ meson $\delta_{\phi}^{(b)}=0.23$. }
\label{tab:power}
\end{table}

To ensure the efficiency of power corrections, we illustrate the contributions of each term in the power correction series, taking the heaviest $\phi$ meson as an example. From the Table \ref{tab:power}, it is  clear that for the $D$ meson, despite the significant value of $\delta_{\phi}$, the contributions generally decrease in order of power. Contributions at $\delta^1_{\phi}$ and $\delta^2_{\phi}$, compared to $\delta^0_{\phi}$, are reduced by an order of magnitude, although $\delta^1_{\phi}$ and $\delta^2_{\phi}$ exhibit similar magnitudes.  Likewise,  $\delta^3_{\phi}$ and $\delta^4_{\phi}$, relative to $\delta^2_{\phi}$ and $\delta^3_{\phi}$, are also suppressed by an order of magnitude, with $\delta^2_{\phi}$ and $\delta^3_{\phi}$ displaying comparable magnitudes. This pattern can be explained by noting that in Eq.\eqref{eq:2pLO}, higher-power contributions are accompanied by a reduction in the ratio $m^2_c/M^2$. 
In contrast to the $D$ meson case, although the $\delta_{\phi}$ for the $B$ channel is much smaller, the ratio $m_b^2/M^2$ played a compensatory role, this leads to a similar power expansion behaviour for the $B_s^\ast B_s\phi$ coupling as the $D_s^\ast D_s\phi$. 

\begin{table}
\centering
\setlength\tabcolsep{3.5pt}
\def\arraystretch{1.1}
\begin{tabular}{|c|c|c|c|c|c|c|c|}
\hline
LCSR result  & tw~2 LO  & tw~2 NLO & tw~3 LO & tw~4 LO
&tw~5 LO & total \\
\hline
$f_{D^*}\,f_D\,g_{D^*D\rho}$
& $0.175^{+0.028}_{-0.036}$ & $-0.039^{+0.041}_{-0.018}$
& $0.038^{+0.006}_{-0.004}$ & $-0.009^{+0.003}_{-0.003}$
& $-0.0002^{+0.0001}_{-0.0003}$ & $0.165^{+0.021}_{-0.022}$ 
\\
\hline
$f_{D^*}\,f_D\,g_{D^*D\omega}$
& $0.111^{+0.020}_{-0.024}$ & $-0.025^{+0.026}_{-0.011}$
& $0.024^{+0.004}_{-0.003}$ & $-0.006^{+0.002}_{-0.002}$
& $-0.0002^{+0.0001}_{-0.0002}$ & $0.105^{+0.015}_{-0.015}$ 
\\
\hline
$f_{D^*_s}\,f_D\,g_{D^*_s D K^*}$
& $0.206^{+0.035}_{-0.043}$ & $-0.044^{+0.046}_{-0.020}$
&$0.041^{+0.007}_{-0.004}$& $-0.012^{+0.003}_{-0.004}$
& $-0.0004^{+0.0002}_{-0.0004}$  & $0.190^{+0.028}_{-0.029}$
\\
\hline
$f_{D^*}\,f_{D_s}\,g_{D^* D_s K^*}$
& $0.207^{+0.035}_{-0.043}$ & $-0.044^{+0.046}_{-0.020}$
&$0.041^{+0.007}_{-0.004}$& $-0.012^{+0.003}_{-0.004}$
& $-0.0004^{+0.0002}_{-0.0005}$  & $0.191^{+0.028}_{-0.029}$
\\
\hline
$f_{D^*_s}\,f_{D_s}\,g_{D^*_s D_s\phi}$
& $0.192^{+0.033}_{-0.041}$  & $-0.044^{+0.046}_{-0.020}$
& $0.041^{+0.007}_{-0.005}$ & $-0.015^{+0.004}_{-0.005}$
& $-0.0006^{+0.0003}_{-0.0007}$  & $0.173^{+0.026}_{-0.027}$
\\
\hline
\hline
$f_{B^*}\,f_B\,g_{B^*B\rho}$
& $0.089^{+0.028}_{-0.018}$ & $0.020^{+0.015}_{-0.029}$
& $0.023^{+0.005}_{-0.003}$ & $-0.002^{+0.001}_{-0.001}$
& $-0.0002^{+0.0001}_{-0.0002}$ & $0.129^{+0.013}_{-0.013}$ 
\\
\hline
$f_{B^*}\,f_B\,g_{B^*B\omega}$
& $0.056^{+0.018}_{-0.012}$ & $0.012^{+0.010}_{-0.020}$
& $0.015^{+0.003}_{-0.002}$ & $-0.001^{+0.001}_{-0.001}$
& $-0.0001^{+0.0001}_{-0.0001}$ & $0.082^{+0.009}_{-0.009}$ 
\\
\hline
$f_{B^*_s}\,f_B\,g_{B^*_s B K^*}$
& $0.106^{+0.033}_{-0.022}$  & $0.022^{+0.018}_{-0.033}$
& $0.026^{+0.006}_{-0.003}$ & $-0.003^{+0.001}_{-0.001}$
& $-0.0002^{+0.0001}_{-0.0003}$  & $0.151^{+0.017}_{-0.017}$
\\
\hline
$f_{B^*}\,f_{B_s}\,g_{B^* B_s K^*}$
& $0.106^{+0.034}_{-0.022}$  & $0.022^{+0.018}_{-0.033}$
& $0.026^{+0.006}_{-0.003}$ & $-0.003^{+0.001}_{-0.002}$
& $-0.0003^{+0.0001}_{-0.0003}$  & $0.151^{+0.017}_{-0.017}$
\\
\hline
$f_{B^*_s}\,f_{B_s}\,g_{B^*_s B_s\phi}$
& $0.102^{+0.033}_{-0.021}$  & $0.023^{+0.018}_{-0.033}$
& $0.026^{+0.006}_{-0.004}$ & $-0.004^{+0.001}_{-0.002}$
& $-0.0004^{+0.0002}_{-0.0004}$  & $0.147^{+0.016}_{-0.015}$
\\
\hline
\end{tabular}
\caption{Summary of the contribution from individual twist and NLO effects on the products of the strong couplings $g_{H^*HV}$ and decay constants presented with both central values and associated data errors in units of $[{\rm GeV}]$.}
\label{tab:products}
\end{table}

We now explore the numerical effects of perturbative QCD and distinct higher-twist corrections on the products of strong couplings and decay constants computed in Eq.\eqref{eq:SR3}. The outcomes of individual twist and NLO effects are summarized in Table \ref{tab:products}, along with relevant data uncertainties. 
As observed in Table \ref{tab:products}, the NLO corrections of twist-2 magnitude exhibit numerical equivalence with the LO contributions of twist-3, constituting roughly $20\%$ of the leading twist for both charm and bottom scenarios. However, it's noteworthy that a sign difference is observed in NLO corrections of twist-2 between charm and bottom. This sign difference may arise from the difference in the scale ratio squared $\mu^2/m_Q^2$ between charm and bottom systems, where $\mu^2_c/m_c^2 \textgreater 1$ for charm and $\mu^2_b/m_b^2 \textless 1$ for bottom, leading to distinct behaviors in the higher-order corrections. Additionally, our analysis indicates that the twist-5 contribution in the OPE of the correlation function has negligible impact, as validated through numerical assessments. This observation suggests that higher-order contributions beyond twist-4 have minimal influence on the computed results. Furthermore, the total uncertainty reported in Table \ref{tab:products} should involve variations stemming from changes in the duality region within the NLO aspect. Specifically, for the charm scenario, this variation amounts to around $20\%$, while for the bottom scenario, it stands at approximately $15\%$. These variations are reflected in the parameter $\Delta \alpha$, as depicted in Tables \ref{tab:paraerr1} and \ref{tab:paraerr2}, where $\alpha$ is defined in Eq. \eqref{eq:alpha}. It's suggested that these uncertainties, to some extent, address the ``systematic error'' inherent in the quark-hadron duality ansatz applied in LCSRs.

In our final analysis, we normalize the couplings using heavy meson decay constants by two methods: the two-point QCDSRs in Eq.\eqref{eq:SRDC} and LQCD results  in Eq.\eqref{eq:lattDC}. Table  \ref{tab:res} highlights the main findings, while Tables \ref{tab:paraerr1} and \ref{tab:paraerr2} provide more details on the parameters for each set of decay constants. 
In the $D$ meson sector, the main sources of uncertainty lie in the Gegenbauer moment $a^{\perp}_2$ and tensor decay constants $f^{\perp}_V$ derived from LQCD. 
When employing two-point sum rules to calculate decay constants, an additional uncertainty arises from the decay constant itself, which is comparable in magnitude. In the $B$ meson sector, the uncertainties related to the Borel parameter and threshold somehow increase relative to those in the $D$ meson sector.

\begin{table}[tb]
\centering
\setlength\tabcolsep{5pt}
\def\arraystretch{1.2}
\begin{tabular}{|c|c||c|c|c|c|c||}
\hline
\multirow{3}{*}{Charm}&decay constants & $g_{D^*D\rho}$ & $g_{D^*D\omega}$ & $g_{D^*_s D K^*}$ & $g_{D^* D_s K^*}$ &  $g_{D^*_s D_s \phi}$\\
\hline
&  two-point QCDSRs
& $3.53^{+0.61}_{-0.57}$ 
& $2.25^{+0.42}_{-0.41}$ 
& $3.55^{+0.65}_{-0.61}$
& $3.87^{+0.73}_{-0.68}$
& $2.69^{+0.55}_{-0.52}$
\\
& LQCD 
& $3.40^{+0.49}_{-0.43}$
& $2.17^{+0.34}_{-0.32}$
& $3.30^{+0.52}_{-0.48}$
& $3.34^{+0.54}_{-0.53}$
& $2.54^{+0.40}_{-0.39}$
\\
\hline
\hline
\multirow{3}{*}{bottom}&decay constants & $g_{B^*B\rho}$ &$g_{B^*B\omega}$ & $g_{B^*_s B K^*}$ &
$g_{B^* B_s K^*}$ &$g_{B^*_s B_s \phi}$\\
\hline
&  two-point QCDSRs
& $3.00^{+0.49}_{-0.44}$
& $1.88^{+0.34}_{-0.30}$
& $3.12^{+0.60}_{-0.50}$
& $3.16^{+0.52}_{-0.47}$
& $2.70^{+0.49}_{-0.41}$
\\
& LQCD 
& $3.74^{+0.38}_{-0.38}$
& $2.38^{+0.28}_{-0.28}$
& $3.61^{+0.40}_{-0.39}$
& $3.61^{+0.41}_{-0.40}$
& $2.86^{+0.32}_{-0.30}$\\
\hline
\end{tabular}
\caption{LCSR  results for the strong couplings $g_{H^*HV}$ for the two
methods of dividing out the decay constants in units of $[{\rm GeV}^{-1}]$.}
\label{tab:res}
\end{table}

\begin{table}[tb]
\centering
\setlength\tabcolsep{4pt}
\def\arraystretch{1.3}
\begin{tabular}{|c|c||c|c|c|c|c|c|c|c|c|c|}
\hline
 &{ central value} & $\Delta f_{H}$ & $\Delta f_{H^*}$ & $\Delta \mu$ &  $\Delta m_Q$ &  $\Delta s_0$ & $\Delta M^2$ &  $\Delta f^{\perp}_V$ &
 $\Delta a^{\perp}_{2}$ & $\Delta \alpha$   & $\Delta_{\rm tot}$\\
\hline
\multirow{2}{*}{$g_{D^*D\rho}$} 
& $3.53$ 
&$^{+0.15}_{-0.15}$
&$^{+0.33}_{-0.32}$
& $^{+0.20}_{-0.01}$
& $^{+0.07}_{-0.08}$
& $^{+0.06}_{-0.06}$
&$^{+0.11}_{-0.06}$
& $\pm{0.14}$
& $\pm{0.35}$
& $\pm 0.17$
&$^{+0.61}_{-0.57}$
\\\cline{2-12}
&$3.40$
&$^{+0.01}_{-0.01}$
&$^{+0.11}_{-0.11}$
&$^{+0.19}_{-0.01}$
& $^{+0.07}_{-0.07}$
& $^{+0.06}_{-0.08}$
& $^{+0.10}_{-0.06}$
& $\pm 0.14$
& $\pm 0.34$
& $\pm 0.16$
&$^{+0.49}_{-0.43}$
\\
\hline
\multirow{2}{*}{$g_{D^*D\omega}$} 
& $2.25$ 
&$^{+0.10}_{-0.10}$
&$^{+0.21}_{-0.20}$
& $^{+0.13}_{-0.01}$
& $^{+0.04}_{-0.05}$
& $^{+0.04}_{-0.05}$
&$^{+0.07}_{-0.04}$
& $\pm{0.15}$
& $\pm{0.22}$
& $\pm 0.15$
&$^{+0.42}_{-0.41}$
\\\cline{2-12}
&$2.17$
&$^{+0.01}_{-0.01}$
&$^{+0.08}_{-0.07}$
&$^{+0.12}_{-0.01}$
& $^{+0.04}_{-0.05}$
& $^{+0.04}_{-0.05}$
& $^{+0.07}_{-0.04}$
& $\pm 0.14$
& $\pm 0.21$
& $\pm 0.15$
&$^{+0.34}_{-0.32}$
\\
\hline
\multirow{2}{*}{$g_{D^*_s D K^*}$} 
&$3.55$ 
&$^{+0.15}_{-0.16}$
&$^{+0.29}_{-0.29}$
& $^{+0.20}_{-0.03}$
& $^{+0.06}_{-0.07}$
& $^{+0.07}_{-0.09}$
&$^{+0.13}_{-0.07}$
& $\pm 0.13$
& $\pm{0.45}$
& $\pm 0.16$
&$^{+0.65}_{-0.61}$ 
\\\cline{2-12}
&$3.30$
&$^{+0.01}_{-0.01}$
&$^{+0.06}_{-0.06}$
&$^{+0.18}_{-0.01}$
& $^{+0.06}_{-0.07}$
& $^{+0.07}_{-0.08}$
& $^{+0.12}_{-0.07}$
& $\pm 0.12$
& $\pm 0.42$
& $\pm 0.15$
&$^{+0.52}_{-0.48}$
\\
\hline
\multirow{2}{*}{$g_{D^* D_s K^*}$} 
&$3.87$ 
&$^{+0.18}_{-0.19}$
&$^{+0.37}_{-0.35}$
& $^{+0.21}_{-0.03}$
& $^{+0.07}_{-0.07}$
& $^{+0.08}_{-0.10}$
&$^{+0.14}_{-0.08}$
& $\pm 0.14$
& $\pm{0.49}$
& $\pm 0.18$
&$^{+0.73}_{-0.68}$ 
\\\cline{2-12}
&$3.34$
&$^{+0.01}_{-0.01}$
&$^{+0.12}_{-0.11}$
&$^{+0.19}_{-0.03}$
& $^{+0.06}_{-0.06}$
& $^{+0.07}_{-0.09}$
& $^{+0.12}_{-0.07}$
& $\pm 0.12$
& $\pm 0.42$
& $\pm 0.15$
&$^{+0.54}_{-0.53}$
\\
\hline
\multirow{2}{*}{$g_{D^*_s D_s \phi}$}
& $2.69$
& $^{+0.14}_{-0.15}$
& $^{+0.25}_{-0.25}$                        &$^{+0.17}_{-0.01}$
& $^{+0.06}_{-0.06}$
&$^{+0.06}_{-0.08}$
& $^{+0.13}_{-0.08}$
& $\pm 0.11$
& $\pm 0.37$
& $\pm 0.16$
& $^{+0.55}_{-0.52}$
\\\cline{2-12}
&$2.54$
&$^{+0.01}_{-0.01}$
&$^{+0.05}_{-0.05}$
&$^{+0.14}_{-0.01}$
& $^{+0.05}_{-0.05}$
& $^{+0.05}_{-0.06}$
& $^{+0.11}_{-0.06}$
& $\pm 0.09$
& $\pm 0.31$
& $\pm 0.13$
&$^{+0.40}_{-0.39}$
\\
\hline
\end{tabular}
\caption{Summary of the individual uncertainties for the strong couplings $g_{H^*HV}$
predicted from the LCSR (\ref{eq:SR3}) in units of $[{\rm GeV}^{-1}]$.
The negligible uncertainties arising from variations in the remaining input parameters have been considered but are not explicitly enumerated here, but are already taken into account in the determinations
of the total errors.}
\label{tab:paraerr1}
\end{table}

\begin{table}[tb]
\centering
\setlength\tabcolsep{4pt}
\def\arraystretch{1.3}
\begin{tabular}{|c|c||c|c|c|c|c|c|c|c|c|c|}
\hline
 &{ central value} & $\Delta f_{H}$ & $\Delta f_{H^*}$ & $\Delta \mu$ &  $\Delta m_Q$ &  $\Delta s_0$ & $\Delta M^2$ &  $\Delta f^{\perp}_V$ &
 $\Delta a^{\perp}_{2}$ & $\Delta \alpha$   & $\Delta_{\rm tot}$\\
\hline
\multirow{2}{*}{$g_{B^* B \rho}$} 
&$3.00$
& $\pm 0.29$
& $^{+0.27}_{-0.14}$
&$^{+0.01}_{-0.02}$
& $^{+0.04}_{-0.04}$
&$^{+0.15}_{-0.18}$
& $^{+0.12}_{-0.06}$
& $\pm 0.13$
&$\pm 0.16$
& $\pm 0.07$
&$^{+0.49}_{-0.44}$
\\\cline{2-12}
&$3.74$
&$\pm 0.03$
&$^{+0.09}_{-0.09}$
& $^{+0.01}_{-0.02}$
& $^{+0.05}_{-0.06}$
& $^{+0.19}_{-0.23}$
& $^{+0.15}_{-0.07}$
&$\pm 0.17$
& $\pm 0.20$
& $\pm 0.08$
&$^{+0.38}_{-0.38}$
\\
\hline
\multirow{2}{*}{$g_{B^*B\omega}$} 
&$1.88$
& $\pm 0.18$
& $^{+0.17}_{-0.09}$
&$^{+0.00}_{-0.01}$
& $^{+0.03}_{-0.03}$
&$^{+0.10}_{-0.12}$
& $^{+0.08}_{-0.04}$
& $\pm 0.14$
&$\pm 0.10$
& $\pm 0.06$
&$^{+0.34}_{-0.30}$
\\\cline{2-12}
&$2.38$
&$\pm 0.02$
&$^{+0.06}_{-0.06}$
& $^{+0.01}_{-0.02}$
& $^{+0.03}_{-0.04}$
& $^{+0.12}_{-0.15}$
& $^{+0.10}_{-0.05}$
&$\pm 0.17$
& $\pm 0.13$
& $\pm 0.08$
&$^{+0.28}_{-0.28}$
\\
\hline
\multirow{2}{*}{$g_{B^*_s B K^*}$} 
& $3.12$
& $\pm{0.30}$
& $^{+0.36}_{-0.22}$
&$^{+0.01}_{-0.01}$
& $^{+0.05}_{-0.05}$
&$^{+0.16}_{-0.20}$
& $^{+0.15}_{-0.08}$
& $\pm 0.12$
& $\pm 0.21$
& $\pm 0.08$
& $^{+0.60}_{-0.50}$
\\\cline{2-12}
&$3.61$
&$\pm 0.02$
&$^{+0.04}_{-0.04}$
& $^{+0.01}_{-0.01}$
& $^{+0.05}_{-0.06}$
& $^{+0.19}_{-0.23}$
& $^{+0.17}_{-0.09}$
& $\pm 0.14$
& $\pm 0.25$
& $\pm 0.08$
&$^{+0.40}_{-0.39}$
\\
\hline
\multirow{2}{*}{$g_{B^* B_s K^*}$} 
&$3.16$
& $\pm 0.29$
& $^{+0.29}_{-0.14}$
&$^{+0.01}_{-0.01}$
& $^{+0.05}_{-0.05}$
&$^{+0.16}_{-0.20}$
& $^{+0.15}_{-0.08}$
& $\pm 0.12$
& $\pm 0.22$
& $\pm 0.07$
&$^{+0.52}_{-0.47}$
\\\cline{2-12}
&$3.61$
&$\pm 0.02$
&$^{+0.09}_{-0.09}$
& $^{+0.01}_{-0.01}$
& $^{+0.05}_{-0.06}$
& $^{+0.18}_{-0.23}$
& $^{+0.17}_{-0.09}$
& $\pm 0.14$
& $\pm 0.25$
& $\pm 0.08$
&$^{+0.41}_{-0.40}$
\\
\hline
\multirow{2}{*}{$g_{B^*_s B_s \phi}$} 
& $2.70$
& $\pm 0.24$
& $^{+0.31}_{-0.19}$
& $^{+0.01}_{-0.01}$
& $^{+0.04}_{-0.04}$
& $^{+0.14}_{-0.17}$
& $^{+0.15}_{-0.08}$
& $\pm 0.10$
& $\pm 0.17$
& $\pm 0.06$
& $^{+0.49}_{-0.41}$
\\\cline{2-12}
& $2.86$
&$\pm 0.02$
&$^{+0.03}_{-0.03}$
& $^{+0.01}_{-0.01}$
& $^{+0.04}_{-0.04}$
& $^{+0.15}_{-0.18}$
& $^{+0.16}_{-0.08}$
&$\pm 0.11$
& $\pm 0.18$
& $\pm 0.07$
& $^{+0.32}_{-0.30}$
\\\hline
\end{tabular}
\caption{Summary of the individual theory uncertainties for the strong couplings $g_{H^*HV}$
predicted from the LCSR (\ref{eq:SR3}) in units of $[{\rm GeV}^{-1}]$.
The negligible uncertainties arising from variations in the remaining input parameters have been considered but are not explicitly enumerated here, but are already taken into account in the determinations
of the total errors.}
\label{tab:paraerr2}
\end{table}

\begin{table}[htb]
\centering
\setlength\tabcolsep{5pt}
\def\arraystretch{1.3}
\begin{tabular}{|c|c||c|c|c|c||}
\hline
$\Delta r_{D^*_{(s)}D_{(s)}V}$ & decay constants & $\Delta r_{D^* D \omega}$& $\Delta r_{D^*_s D K^*}$ & $\Delta r_{D^* D_s K^*}$ &  $\Delta r_{D^*_s D_s \phi}$\\
\hline
charm&  two-point QCDSRs
& $-10\%$
& $7.6\%$
& $9.8\%$
& $-13\%$
\\
& LQCD 
& $-10\%$
& $-3\%$
& $-1.8\%$
& $-25\%$
\\
\hline
\hline
$\Delta r_{B^*_{(s)}B_{(s)}V}$ & decay constants &  $\Delta r_{B^* B \omega}$&  $\Delta r_{B^*_s B K^*}$ &
$\Delta r_{B^* B_s K^*}$ &$\Delta r_{B^*_s B_s \phi}$\\
\hline
bottom&  two-point QCDSRs
& $-10\%$
& $3.9\%$
& $5.4\%$
& $-9.9\%$
\\
& LQCD 
& $-10\%$
& $-3.5\%$
& $-3.5\%$
& $-23\%$\\
\hline
\end{tabular}
\caption{Summary of the extent of $SU(3)$ breaking effects as determined by LCSR, with $\Delta r_{H^*HV}$ serving as a measurement parameter.}
\label{tab:SU3}
\end{table}
The estimation for the $D^*_{(s)}D_{(s)}V$ and $B^*_{(s)}B_{(s)}V$ strong couplings from LCSRs yields uncertainty ranges of $13\%$ to $18\%$ and $10\%$ to $18\%$, respectively, when using heavy-meson decay constants from the two-point QCDSRs. These uncertainty intervals are calculated based on variations in the input parameters and methodologies used in the LCSR analysis. 
While the uncertainty intervals for $g_{D^*_{(s)}D_{(s)}V}$ remain consistent regardless of the decay constant choice, only slight agreement is observed in the intervals for $g_{B^*_{(s)}B_{(s)}V}$.
The difference in the central values of the $B^*_{(s)}B_{(s)}V$ couplings mainly arise from a $20\%$ discrepancy between the lattice-QCD value of $f_{B^*_{(s)}}$ and the central value predicted by the two-point sum rule. For consistency, we use the two-point QCDSRs at NLO, with more precise sum rules at NNLO \cite{Gelhausen:2013wia}, yielding results consistent with LQCD calculations within uncertainties. This ensures that our analysis incorporates the latest advancements in theoretical frameworks and maintains consistency with the most accurate calculations available.

To investigate the $SU(3)_F$ symmetry breaking effects, we introduce $\Delta r_{H^*HV}$ as the parameter characterizing the size of the breaking effects, 
\begin{eqnarray}
\Delta r_{D^*_{(s)}D_{(s)}V}= \frac{C_V\cdot  g_{D^*_{(s)}D_{(s)}V}-g_{D^*D\rho}}{g_{D^*D\rho}},\quad \Delta r_{B^*_{(s)}B_{(s)}V}=\frac{C_V\cdot   g_{B^*_{(s)}B_{(s)}V}-g_{B^*B\rho}}{g_{B^*B\rho}},
\end{eqnarray}
where $g_{D^*D\rho}$ and $g_{B^*B\rho}$ are defined in Eq.\eqref{eq:coupling-def1} and \eqref{eq:coupling-def2}, respectively. Additionally, $C_V=\sqrt{2}$ for $V$ takes the $\omega$ meson, otherwise, it's set to 1.


In Table \ref{tab:SU3}, we detail the relative magnitudes of $SU(3)_F$ symmetry breaking effects for the various strong couplings. Interestingly, both $\Delta r_{D^* D \omega }$ and $\Delta r_{B^* B \omega }$ remain constant at $-10\%$. This stability is attributed to the interplay of specific factors such as $f^{\perp}_{\omega}$, $f^{\parallel}_{\omega}$, and $m_{\omega}$, which decrease at rates of $8\%$, $1.9\%$, and $0.1\%$, respectively. The other parameters remain consistent with those associated with the $\rho$ meson, resulting in a cancellation of terms in both the numerator and denominator. Shifting focus to the $9.8\%$ breaking observed in $\Delta{r_{D^* D_s K^*}}$, analysis using the decay constants from two-point QCDSRs reveals four primary sources of this deviation. Firstly, a substantial $12\%$ deviation arises from the LCDAs of the $K^*$ meson ($8\%$) and the effect of the light quark mass ($4\%$). Secondly, the decay constants of the $D_s$ meson contribute approximately $-5\%$. Thirdly, the decay constant $f^{\perp}_{K^*}$ explains $9\%$ of the observed breaking. Lastly, the mass of the $D_s$ meson contributes roughly $-6\%$. As for $\Delta r_{D^*_s D K^*}$, the contribution from the $D_s$ meson's decay constant $f_{D^*_s}$ is around $-12.9\%$, larger than that of $D_s$. Additionally, a portion of this difference can be attributed to the mass of the $D^*_s$ meson, resulting in a decrease of about $0.27\%$. This results in an overall breaking of $7.6\%$. Substituting decay constants calculated via LQCD for those obtained from two-point QCDSRs notably alters the quantification of symmetry breaking effects, primarily reducing the attributed deviation from decay constants. It's crucial to highlight that the breaking effects observed in $\phi$, as determined through LQCD, tends to exceed that calculated using  two-point QCDSRs (taking their absolute magnitudes into account). This discrepancy arises from the substantial difference in the decay constant $f_{D_s}$ obtained from the two methods, which causes the negative breaking effects up to $15\%$ from LQCD  , larger than $5\%$ from the two-point QCDSRs. However, for $f_{D^*_s}$, the situation is reversed but less difference, $-13\%$ from  two-point QCDSRs compared to $-15\%$ from LQCD. Consequently, the discrepancy attributed to LQCD calculated decay constants amounts to $-25\%$, compared to $-13\%$ from  two-point QCDSRs. Finally, after accounting for additional factors contributing around $5\%$ to the observed symmetry breaking effects, we arrive at our comprehensive analysis of $SU(3)$ flavour symmetry deviations. The situation with $SU(3)$ symmetry breaking in the $B$ sector resembles that in the $D$ sector, albeit with milder decrease.

When comparing our numerical outcomes with those of Ref. \cite{Wang:2020yvi}, depicted in Tables \ref{tab:compar1} and \ref{tab:compar2}, we observed an enhancement of the reported values $3.80^{+0.59}_{-0.45}\,{\rm GeV^{-1}}$ and $3.89^{+0.52}_{-0.48}\,{\rm GeV^{-1}}$ for $g_{D^*D\rho}$ and $g_{B^*B\rho}$ respectively. 
The primary reason for this difference lies in variations in our chosen input parameters such as scale $\mu$, Borel parameter $M^2$, and effective threshold $s_0$. The impact of the scale $\mu$ is especially pronounced in the twist-2 LO and NLO aspects, although its overall effect decreases. 
Shifting the scale in the bottom sector from $\mu=3\,{\rm GeV}$ to $\mu=m_b$ leads to an opposite sign for the NLO terms. Moreover, their specified values of $M^2$ and $s_0$ differ from our using.\footnote{The authors in \cite{Wang:2020yvi} specified $M^2=4.5\pm1.0\,{\rm GeV^2}$ for charm and $M^2=18\pm3.0\,{\rm GeV^2}$ for bottom sector, with $M^2_1=M^2_2=M^2$, whereas we used $M^2_1=M^2_2=2M^2$\, differing by a factor of 2. Their choices of $s_0=6.0\pm 0.5\,{\rm GeV^2}$ for charm and $s_0=34.0\pm 1.0\,{\rm GeV^2}$ for bottom sector contrast with our selections of $7.0\pm 0.5\,{\rm GeV^2}$ and $37.5\pm 2.5\,{\rm GeV^2}$, respectively. } We also extend our analysis to include sub-subleading power corrections ranging from twist-3 to twist-5 (considering $\delta^3_V$ and $\delta^4_V$) in Eq.\eqref{eq:2pLO}, which were absent in \cite{Wang:2020yvi}. 
Additionally, our treatment of the three-particle contributions deviates from that in \cite{Wang:2020yvi} due to our modifications to the three-particle DAs, resulting in a reversal of signs, although their influence remains negligible compared to the leading twist contribution. 
Furthermore, it is noteworthy that the LCSR studies conducted in \cite{Wang:2007mc,Li:2002pp} only focused on leading-order calculations, while the significant compensatory effects from the twist-2 NLO calculations were absent. As mentioned previously, in the case of charm mesons, the NLO effects at the leading twist effectively counterbalance the corrections from higher-twist contributions. Consequently, we expect our results to align closely with those from earlier sum rule investigations, within the errors. However, in the case of bottom mesons, the opposite sign of the NLO effects enhance the impact of the leading twist.

\begin{table}[tb]
\centering
\begin{tabular}{|c||c|c|c|c|c|}
\hline
Method  &$g_{D^*D\rho}$  & $g_{D^*D\omega}$ & $g_{D^*_s D K^*}$& $g_{D^* D_s K^*}$ &  $g_{D^*_s D_s \phi}$ \\
\hline
LCSR\cite{Wang:2020yvi}& $3.80^{+0.59}_{-0.45}$&-- & --& --&--\\
\hline
LCSR\cite{Wang:2007mc}  &$3.56\pm 0.60$ &-- &-- &$4.04\pm 0.8$ &$3.28\pm 0.64$ \\
\hline
LCSR\cite{Li:2002pp}&$4.17\pm1.04$&-- & --& --&--\\
\hline
QCDSR\cite{Azizi:2010jj}  &-- &--  & $3.74\pm1.38$&--&-- \\[1mm]
\hline
\hline
\multirow{2}{*}{LCSR (this work)}& $3.53^{+0.61}_{-0.57}$ 
& $2.25^{+0.42}_{-0.41}$ 
& $3.55^{+0.65}_{-0.61}$
& $3.87^{+0.73}_{-0.68}$
& $2.69^{+0.55}_{-0.52}$\\[1mm]
\cline{2-6}
& $3.40^{+0.49}_{-0.43}$
& $2.17^{+0.34}_{-0.32}$
& $3.30^{+0.52}_{-0.48}$
& $3.34^{+0.54}_{-0.53}$
& $2.54^{+0.40}_{-0.39}$ \\[1mm]
\hline
\end{tabular}
\caption{Numerical values of coupling  $g_{D^*_{(s)} D_{(s)}V}({\rm GeV^{-1}})$ from several methods.}
\label{tab:compar1}
\end{table}

\begin{table}[tb]
\centering
\begin{tabular}{|c||c|c|c|c|c|}
\hline
Method  &$g_{B^*B\rho}$  & $g_{B^*B\omega}$& $g_{B^*_s B K^*}$& $g_{B^* B_s K^*}$ &  $g_{B^*_s B_s \phi}$ \\
\hline
LCSR\cite{Wang:2020yvi}& $3.89^{+0.52}_{-0.48}$&-- & --& --&--\\
\hline
LCSR\cite{Li:2002pp}& $5.70\pm1.43$&-- & --& --&--\\
\hline
QCDSR\cite{Azizi:2010jj}  &-- &-- & $3.24\pm 1.08$&-- &--\\[1mm]
\hline
\multirow{2}{*}{LCSR (this work)} & 
 $3.00^{+0.49}_{-0.44}$
& $1.88^{+0.34}_{-0.30}$
& $3.12^{+0.60}_{-0.50}$
& $3.16^{+0.52}_{-0.47}$
& $2.70^{+0.49}_{-0.41}$
\\[1mm]
\cline{2-6}
& $3.74^{+0.38}_{-0.38}$
& $2.38^{+0.28}_{-0.28}$
& $3.61^{+0.40}_{-0.39}$
& $3.61^{+0.41}_{-0.40}$
& $2.86^{+0.32}_{-0.30}$\\
\hline
\end{tabular}
\caption{Numerical values of coupling $g_{B^*_{(s)} B_{(s)}V}({\rm GeV^{-1}})$ from several methods.}
\label{tab:compar2}
\end{table}

In addition, we  introduce an alternative method to determine the $B^*_{(s)} B_{(s)} V$ coupling. This method, as extensively described in \cite{Wang:2020yvi,Khodjamirian:2020mlb}, utilizes the hadronic dispersion relation for the $B_{(s)}\to V$ vector(tensor) form factor as a foundational element:
\begin{eqnarray}
V(q^2)\,\,&=&\,\,\frac{(m_{B_{(s)}}+m_V)g_{B^*_{(s)}B_{(s)}V} f_{B^*_{(s)}}}{2m_{B^*_{(s)}}(1-q^2/m^2_{B^*_{(s)}})}+
\frac{1}{\pi}\!\!\!\!\!\!\int\limits_{(m_{B_{(s)}}+m_V)^2}^\infty \!\!\!\!\!dt\,
 \frac{\mbox{Im} V(t)}{t-q^2},\\
T_1(q^2)\,\,&=&\,\,\frac{g_{B^*_{(s)}B_{(s)}V} f^{\perp}_{B^*_{(s)}}}{2(1-q^2/m^2_{B^*_{(s)}})}+
\frac{1}{\pi}\!\!\!\!\!\!\int\limits_{(m_{B_{(s)}}+m_V)^2}^\infty \!\!\!\!\!dt\,
 \frac{\mbox{Im} T_1(t)}{t-q^2}.
\label{eq:HVdisp}  
\end{eqnarray}

This relation features the vector-meson $B^*_{(s)}$ pole, just below the kinematic threshold $q^2=(m_{B_{(s)}}+ m_V)^2$. The residue of the $B^*$ pole represents the product of the $B^*_{(s)}B_{(s)}V$ coupling and the $B^{*(\perp)}_{(s)}$ decay constant. Multiplying both sides of Eq.\eqref{eq:HVdisp} by the denominator of the pole term and taking the limit $q^2\to m_{B^*_{(s)}}^2$, we avoid the integral over the hadronic spectral density, leading to:
\begin{eqnarray}
 g_{B^*_{(s)}B_{(s)}V}\,\,&=&\,\, \frac{2m_{B^*_{(s)}}}{(m_{B_{(s)}}+m_V)f_{B^*_{(s)}}}\lim_{q^2\to m_{B^*_{(s)}}^2}
\Big[\left(1-q^2/m^2_{B^*_{(s)}}\right)V(q^2)\Big] \,,\\
 g_{B^*_{(s)}B_{(s)}V}\,\,&=&\,\,\frac{2}{f^{\perp}_{B^*_{(s)}}}\lim_{q^2\to m_{B^*_{(s)}}^2}
\Big[\left(1-q^2/m^2_{B^*_{(s)}}\right)T_1(q^2)\Big] \,.
\label{eq:BstarBVlimit}   
\end{eqnarray}
Employing Eq.\eqref{eq:BstarBVlimit} and assuming $f^T_{B^*_{(s)}}=f_{B^*_{(s)}}$, we obtain various numerical values for the coupling $g_{B^*_{(s)}B_{(s)}V}$ using the $B_{(s)}\to V$ form factors from two distinct LCSRs \cite{Bharucha:2015bzk,Gao:2019lta}. These results are summarized in Table \ref{tab:FFcompare}.

\begin{table}[tb]
\centering
\begin{tabular}{|c||c|c|c|c|c|}
\hline
Method  & This work& $V(q^2)$\cite{Bharucha:2015bzk} & $V(q^2)$\cite{Gao:2019lta} &  $T_1(q^2)$\cite{Bharucha:2015bzk}&  $T_1(q^2)$\cite{Gao:2019lta} \\
\hline
\multirow{2}{*}{$g_{B^*B\rho}$} &  $3.00^{+0.49}_{-0.44}$
& $7.24^{+1.38}_{-1.26}$
& $5.51^{+3.69}_{-2.34}$
& $6.83^{+1.26}_{-1.15}$
& $6.17^{+4.14}_{-2.61}$
\\[1mm]
\cline{2-6}
& $3.74^{+0.38}_{-0.38}$
& $8.53^{+1.44}_{-1.44}$
& $6.49^{+4.31}_{-2.74}$
& $8.04^{+1.31}_{-1.31}$
& $7.25^{+4.83}_{-3.06}$
\\
\hline
\multirow{2}{*}{$g_{B^*B\omega}$} &$1.88^{+0.34}_{-0.30}$
& $6.78^{+1.73}_{-1.65}$
& $6.00^{+4.00}_{-2.55}$
& $6.50^{+1.58}_{-1.5}$
& $6.69^{+4.52}_{-2.83}$
\\[1mm]
\cline{2-6}
&$2.38^{+0.28}_{-0.28}$
& $7.98^{+1.92}_{-1.92}$
& $7.06^{+4.67}_{-2.99}$
& $7.65^{+1.74}_{-1.74}$
& $7.88^{+5.28}_{-3.31}$
\\
\hline
\multirow{2}{*}{$g_{B^*_s B K^*}$}
& $3.12^{+0.60}_{-0.50}$
& $7.34^{+1.69}_{-1.55}$
& $6.22^{+3.92}_{-2.39}$
& $6.96^{+1.54}_{-1.41}$
& $6.95^{+4.38}_{-2.67}$
\\[1mm]
\cline{2-6}
& $3.61^{+0.40}_{-0.39}$
& $8.02^{+1.60}_{-1.60}$
& $6.80^{+4.21}_{-2.57}$
& $7.61^{+1.44}_{-1.44}$
& $7.60^{+4.71}_{-2.87}$
\\
\hline
\multirow{2}{*}{$g_{B^* B_s K^*}$} & $3.16^{+0.52}_{-0.47}$
& $6.75^{+1.42}_{-1.32}$
& $--$
& $6.43^{+1.26}_{-1.16}$
& $--$
\\[1mm]
\cline{2-6}
& $3.61^{+0.41}_{-0.40}$
& $7.95^{+1.52}_{-1.52}$
& $--$
& $7.57^{+1.33}_{-1.33}$
& $--$
\\
\hline
\multirow{2}{*}{$g_{B^*_s B_s \phi}$} & $2.70^{+0.49}_{-0.41}$
& $6.88^{+1.33}_{-1.18}$
& $--$
& $6.54^{+1.24}_{-1.09}$
& $--$
\\[1mm]
\cline{2-6}
& $2.86^{+0.32}_{-0.30}$
& $7.51^{+1.18}_{-1.18}$
& $--$
& $7.15^{+1.08}_{-1.08}$
& $--$
\\
\hline
\end{tabular}
\caption{Coupling constant $g_{B^*_{(s)}B_{(s)}V}$ in units of GeV$^{-1}$ extracted from the residue of the transition form factors $V$ and $T_1$ in the framework of LCSR fit, compared with our obtained values.}
\label{tab:FFcompare}
\end{table}

Our study reveals a discrepancy between our central value and the projections from the transition form factors, which can be attributed to two main factors. Firstly,  different theoretical calculations of form factors yield different results, leading to inconsistent conclusions \cite{Bharucha:2015bzk,Gao:2019lta}. In Ref.\,\cite{ Gao:2019lta}, our results are consistent with their theoretical predictions when considering uncertainties. This consistency is attributed to the significant magnitude of the uncertainties. However, in Ref.\,\cite{ Bharucha:2015bzk}, notable deviations persist even after accounting for uncertainties. Moreover, there might be an underestimation of uncertainties stemming from the parameterizations of form factors to address unphysical singularities. 
Secondly, assuming the validity of these findings, the Borel suppression in the double dispersion relation may not effectively eliminate contributions from isolated excitations, as discussed in Ref.\,\cite{Becirevic:2002vp}. 
In sum rule calculations, it is common practice to include contributions from higher states in the perturbative estimation of a continuum, often without explicit consideration of isolated excited states. This approach is supported by the stability criterion being met adequately without requiring substantial excitation contributions. However, Ref.\,\cite{Becirevic:2002vp} proposes a different viewpoint, suggesting that neglecting explicit radial excitation contributions could be the root of the discrepancy.  While acknowledging the Borel suppression effect, they argue that in certain cases, such as the one under consideration, the suppression might be insufficient, leading to the need for additional considerations. Incorporating explicit radial excitation contributions to the hadronic side of the LCSRs provides a plausible explanation for the failure of the LCSR prediction. The relatively weak exponential suppression due to the Borel transformation in the LCSR framework significantly amplifies the impact of radial excitations without necessitating excessively large couplings. This suppression is significantly weaker compared to standard sum rules, primarily due to the additional factor in the exponent and the larger mean value of $M^2$ within the allowed range. Overall, according to Ref.\,\cite{Becirevic:2002vp}, assuming partial cancellation between contributions of excited and ground states in the dispersion relation may lead to an increased LCSR coupling.

When extending Eq.\eqref{eq:BstarBVlimit} to include transition form factors from $D$ to $V$, we can derive the coupling for $D^*DV$ as well. By employing the $D$ to $\rho$ form factor detailed in the Ref.\,\cite{Fu:2018yin}, our calculations yielded couplings $g_{D^*D\rho}$ of $8.91^{+0.77}_{-0.55}\,{\rm GeV}^{-1}$ using the heavy meson decay constants from LQCD, and $8.29^{+0.71}_{-0.51}\,{\rm GeV}^{-1}$ using the decay constants from two-point QCDSRs , which exceed the values of $3.40^{+0.49}_{-0.43}\,{\rm GeV}^{-1}$ and $3.53^{+0.61}_{-0.57}\,{\rm GeV}^{-1}$ obtained from LCSRs a lot, mirroring the trend observed in $B \to\rho$ transitions. Nevertheless, since the Ref. \cite{Fu:2018yin} only provides uncertainty estimates for $V(0)$ and not for the fitting parameters $a_1$ and $a_2$, the resulting uncertainty remains relatively small.

Finally, we also compare our results with other model-dependent studies concerning chiral and heavy quark limits. The coupling parameter $\lambda$ characterizes the $H^*H\rho$ interaction, and it's related to $g_{H^*H\rho}$ through the expression,
\begin{align}
   \lambda=\frac{\sqrt{2}}{4}\frac{g_{H^*H\rho}}{{g_V}} \,,
\end{align} 
where the parameter $g_V = m_\rho/f_\pi$ ($m_\rho$ is the $\rho$ meson mass and $f_\pi$ denotes the pion decay constant) is determined from the first and second KSRF relation \cite{Casalbuoni:1996pg,Kawarabayashi:1966kd,Riazuddin:1966sw}.
Our calculations yield $\lambda=0.21\pm0.03\,{\rm GeV}^{-1}$ with  decay constants obtained from two-point sum rule and $\lambda=0.20\pm0.03\,{\rm GeV}^{-1}$ from LQCD-calculated decay constants at the leading power, based on our determination of $g_{B^*B\rho}$. These values are compatible with those obtained from Ref.\,\cite{Li:2007dv,Wang:2007mc,Wang:2020yvi} but significantly smaller than those predicted by the VMD model \cite{Isola:2003fh}, where $\lambda$ is estimated to be $0.56\,{\rm GeV}^{-1}$, and the constituent quark model \cite{Melikhov:2000yu}, which suggests a value of $0.47\,{\rm GeV}^{-1}$. One possible reason for this discrepancy could be the likelihood of larger uncertainties in the model predictions, favoring the reliability of the LCSRs method. However, uncertainties in the determination of NLO corrections to higher-twists (such as twist-3 NLO) and sub-sub-leading power contributions remain unknown factors contributing to the observed deviation. Advancements in computational techniques, including refinements in sum rule methods and modeling, may offer valuable insights into resolving these discrepancies in the future.

\section{Conclusion}
In this paper, we have conducted a comprehensive investigation into the improved calculation of the strong couplings $g_{D^\ast_{(s)} D_{(s)}V}$ and $g_{B^\ast_{(s)} B_{(s)} V}$, where $V=\rho, \omega, K^*$ and $\phi$, within the framework of LCSR. This framework is originally developed in Refs.\,\cite{Khodjamirian:2020mlb,Li:2020rcg,Wang:2020yvi}, which primarily focus on the calculation of $g_{D^\ast D\rho}$ and $g_{B^\ast B \rho}$. 

Our study built upon the OPE method, focusing on the correlation function Eq.\,(\ref{eq:corr}) and incorporating vector meson LCDAs with increasing twist. We analytically derived the double spectral representation of the QCD factorization formula, employing the parton-hadron duality ansatz and the double Borel transformation. This allows us to establish LCSRs for both leading and sub-leading contributions, encompassing two-particle DAs up to twist-5 and three-particle DAs up to twist-4. Exploring the newly determined double spectral densities for subleading twist contributions allows us to perform conduct analytical continuum subtractions. This facilitates the development of higher-twist sum rules on the light-cone, improving the precision and scope of our analysis. Furthermore, we derived concise analytical expressions for NLO terms within the $\overline{\text{MS}}$ scheme for heavy quark mass. This ensures the applicability of our findings to both charmed and bottom mesons. Additionally, we have updated all input parameters for the LCSRs, focusing particularly on the Borel parameter $M^2$, scale $\mu$, and effective threshold $s_0$. These modifications are based on newly cited data from Ref.\,\cite{Khodjamirian:2020mlb}, reflecting improvements in our understanding compared to the findings presented in Ref.\,\cite{Wang:2020yvi}.

Examining the numerically calculated LCSRs for the strong couplings $g_{H^*HV}$, we observed that the dominant contribution arises from twist-2. In addition, we found that the NLO corrections closely match the LO contributions at twist-three, constituting nearly $20\%$ of the leading twist. Notably, our analysis revealed opposite signs for the NLO corrections between charm-meson and bottom-meson couplings. Our analysis also indicated that the contributions from subleading twists to charm-meson couplings, particularly from two-particle and three-particle LCDAs, are more significant than those for bottom-meson couplings.

When comparing our optimized LCSRs predictions with the previous studies \cite{Wang:2020yvi,Wang:2007mc,Li:2002pp,Azizi:2010jj}, we found satisfactory agreement in the values obtained for the $H^*HV$ couplings within theoretical uncertainties. Have these couplings in hand, we extracted the effective coupling $\lambda$ in the HM$\chi$PT Lagrangian, it is compatible with those obtained in Ref.\,\cite{Li:2007dv,Wang:2007mc,Wang:2020yvi} but significantly smaller than those predicted by the VMD model \cite{Isola:2003fh} and the constituent quark model \cite{Melikhov:2000yu}.
Furthermore, we fulfilled a detailed exploration to the $SU(3)$ flavour symmetry breaking effects by comparing the $H^*HV$ (for $V$ is $K^\ast$, $\omega$ and $\phi$) couplings with $g_{D^\ast D \rho}$ and $g_{B^\ast B \rho}$. 
The analysis revealed breaking effects ranging from $-25\%$ to $-1.8\%$ when employing the decay constants of heavy mesons calculated from LQCD, and from $-13\%$ to $+9.8\%$ when utilizing the decay constants obtained from two-point QCDSRs. 
Where, both $g_{H^*H\omega}$ and $g_{H_s^*H_s\phi}$ demonstrate negative symmetry breaking effects, however the breaking effects of $g_{H^\ast H_s K^*}$ and $g_{H_s^* H K^*}$ exhibit opposite signs for the heavy meson decay constants computed from the two methods. 
The most obvious $SU(3)_F$ breaking effects were observed in the coupling $g_{D^*_s D_s\phi}$ which reach to $-25\%$, this is primarily due to the involvement of the maximum number of strange quarks in this channel.
Moreover, we performed a comprehensive investigation of $g_{B^*_{(s)}B_{(s)}V}$ and $g_{D^\ast D\rho}$ derived from $B\to V$ \cite{Bharucha:2015bzk,Gao:2019lta} and $D\to\rho$ transition form factors by using dispersion relation. 
Which demonstrated an obvious deviation from our findings in LCSRs and the results obtained in the previous studies \cite{Wang:2020yvi,Wang:2007mc,Li:2002pp,Azizi:2010jj}. 


Developing the LCSRs for the strong coupling $g_{H^*HV}$ beyond our current investigation offers several promising directions. Firstly, it's crucial to calculate NLO QCD corrections to the higher-twist contributions of the vector meson DAs, this is important for clarifying the discrepancies observed in $g_{H^*HV}$ between the direct LCSRs calculations and the indirect predictions from the $H\to V$ transition form factors. Secondly, updating the non-perturbative parameters in the conformal expansion of vector meson DAs holds significant phenomenological importance and can enhance the accuracy of our LCSRs predictions. 
As noticed in \cite{Khodjamirian:2020mlb}, a twist-2 model for pion LCDA was proposed from the LQCD data \cite{RQCD:2019osh}, which has led to a notable improvement in predictions for $B^*B\pi$. 
Another potential solution to this problem is to incorporate high excited heavy-meson states in the hadronic components of the sum rules, as suggested in the case of $D^* D\pi$ \cite{Becirevic:2002vp}. 


\section*{Acknowledgments}
\addcontentsline{toc}{section}{Acknowledgements}
We appreciate the useful discussions with Chao Wang and Yu-Ming Wang.
This work is supported by the National Natural Science Foundation of China with Grant No. 12075125 and 12247162. 
H. Y. Jiang gratefully acknowledges financial support from China Scholarship Council.

\appendix

\section {The definitions of the vector meson LCDAs}\label{appA}
Vector meson LCDAs are defined through expansions based on matrix elements, specifically involving the DAs of quark-antiquark (or quark-antiquark-gluon) within the vector meson. These DAs are characterized by increasing twists.
The widely accepted standard definition is applied for the two-particle LCDAs of vector mesons from twist-$2$ to twist-$5$:
\begin{align} \label{eq:DAvec}
&\left\langle V(p,\eta^*)\left|\bar{q_1}(x) \gamma_{\mu} q_2(0)\right| 0\right\rangle=  f^{\parallel}_{V} m_{V} \int_{0}^{1} d u e^{i u p \cdot x}   \nn
&\hspace{1.5cm} \times\left\{ p_{\mu}\frac{\eta^* \cdot x}{p \cdot x}  \left[\phi^{\|}_{2;V}(u) -\phi^{\perp}_{3;V}(u) + \frac{m_{V}^{2} x^{2}}{16} (\phi^{\|}_{4;V}(u)-\phi^{\perp}_{5;V}(u))\right]\right. \nn
&\hspace{1.5cm}+\eta^*_{\mu}\left(\phi^{\perp}_{3;V}(u) +\frac{m_{V}^{2} x^{2}}{16}\phi^{\perp}_{5;V}(u)\right) \left. -  \frac{\eta^* \cdot x}{2(p \cdot x)^{2}} x_{\mu}m_{V}^{2} \left(\psi^{\|}_{4;V}(u)-2\phi^{\perp}_{3;V}(u)+\phi^{\|}_{2;V}(u)\right)\right\}, \\
&\left\langle V(p,\eta^*)\left|\bar{q_1}(x) \sigma_{\mu \nu} q_2(0)\right| 0\right\rangle
 =-i f_{V}^{\perp} \int_{0}^{1} d u e^{i u p \cdot x}\left\{\left(\eta^*_{\mu} p_{\nu}-\eta^*_{\nu} p_{\mu}\right) \left[\phi^{\perp}_{2;V}(u)+\frac{m_{V}^{2} x^{2}}{16} \phi^{\perp}_{4;V}(u)\right]\right. \nn
&\hspace{1.5cm}+ \left(p_{\mu} x_{\nu} -p_{\nu} x_{\mu}\right)\frac{\eta^* \cdot x}{(p \cdot x)^{2}}  m_{V}^{2} \left(\phi^{\|}_{3;V}(u)-\frac{1}{2}\phi^{\perp}_{2;V}(u)-\frac{1}{2}\psi^{\perp}_{4;V}(u)\right)\nn
&\hspace{1.5cm}+\left. \frac{\eta^*_{\mu} x_{\nu}-\eta^*_{\nu} x_{\mu}}{2 \,p \cdot x} m_{V}^{2} \left(\psi^{\perp}_{4;V}(u)-\phi^{\perp}_{2;V}(u)\right)\right\},\,\\
&\left\langle V(p,\eta^*)\left|\bar{q_1}(x) \gamma_{\mu} \gamma_{5} q_2(0)\right| 0\right\rangle  = \frac{1}{4} f^{\parallel}_{V} m_{V} \epsilon_{\mu \nu \rho \sigma} \eta^{*\nu}p^{\rho}x^{\sigma} \int_{0}^{1} d u e^{i u p \cdot x}\left(\psi^{\perp}_{3,V}(u)
 +\frac{m_{V}^{2} x^{2}}{16}\psi^{\perp}_{5,V}(u) \right),\\
&\left\langle V(p,\eta^*)|\bar{q_1}(x) q_2(0)| 0\right\rangle  = -\frac{i}{2} f_{V}^{\perp}\left(\eta \cdot x\right) m_V^2 \int_0^1 d u e^{i u p \cdot x} \psi^{\parallel}_{3;V}(u).
\end{align}

The definitions of $f^{\parallel}_V$ and $f^\perp_V$ are given by:
\begin{eqnarray}
 \left\langle V(p,\eta^*)\left|\bar{q}_1(0) \gamma_{\mu} q_2(0)\right| 0\right\rangle= f^{\|}_V m_V \eta^*_\mu,\quad \left\langle V(p,\eta^*)\left|\bar{q}_1(0) \sigma_{\mu \nu} q_2(0)\right| 0\right\rangle=-if^{\perp}_V\left(\eta^*_\mu p_\nu-\eta^*_\nu p_\mu\right).\qquad\,.
\end{eqnarray}

The three-particle chiral-even DAs \cite{Ball:1998ff} are defined by the following matrix elements
\begin{align}
& \left\langle V(p,\eta^*)\left|\bar{q}_1(x)g_s G_{\mu\nu}(v x) \gamma_\alpha q_2(-x)\right| 0\right\rangle = i f_V^{\|} m_V P_\alpha \left[ P_\nu \eta_{\perp\mu}^\ast - P_\mu \eta_{\perp\nu}^\ast \right]  \Phi_{3;V}^{\|}(v, Px) \nn
& \hspace{6.3cm} + i f_V^{\|} m_V^3 {\eta^\ast \cdot x \over P\cdot x} \left[ P_\mu g_{\alpha\nu}^{\perp} - P_\nu g_{\alpha\mu}^\perp \right] \Phi^{\|}_{4;V}(v, Px) \nn
& \hspace{6.3cm} + i f_V^{\|} m_V^3 {\eta^\ast \cdot x \over (P\cdot x)^2} P_\alpha \left[ P_\mu x_{\nu} - P_\nu x_{\mu} \right] \Psi^{\|}_{4;V}(v, Px)\,,  \\
& \big\langle V(p,\eta^*)\big|\bar{q}_1(x) g_s \widetilde{G}_{\mu\nu}(v x) \gamma_\alpha \gamma_5 q_2(-x)\big| 0 \big\rangle = - f_V^{\|} m_V P_\alpha \left[ P_\nu \eta_{\perp\mu}^\ast - P_\mu \eta_{\perp\nu}^\ast \right] \widetilde\Phi_{3;V}^{\|}(v, Px) \nn
& \hspace{6.3cm} - f_V^{\|} m_V^3 {\eta^\ast \cdot x \over P\cdot x} \left[ P_\mu g_{\alpha\nu}^{\perp} - P_\nu g_{\alpha\mu}^\perp \right] \widetilde{\Phi}^{\|}_{4;V}(v, Px) \nn
& \hspace{6.3cm} - f_V^{\|} m_V^3 {\eta^\ast \cdot x \over (P\cdot x)^2} P_\alpha \left[ P_\mu x_{\nu} - P_\nu x_{\mu} \right] \widetilde{\Psi}^{\|}_{4;V}(v, Px)\,,
\end{align}
and the three-particle chiral-odd DAs\cite{Ball:1998ff} can be defined as
\begin{align}
& \left\langle V(p,\eta^*)\left|\bar{q}_1(x)g_s G_{\mu\nu}(v x) \sigma_{\alpha\beta} q_2(-x)\right| 0\right\rangle \nn
& = f_V^{\perp} m_V^2 \frac{\eta^*\cdot x}{2(P\cdot x)} \left[P_\alpha P_\mu g_{\beta \nu}^{\perp} - P_\beta P_\mu g_{\alpha \nu}^{\perp} - P_\alpha P_\nu g_{\beta \mu}^{\perp} + P_\beta P_\nu g_{\alpha \mu}^{\perp}\right] \Phi_{3;V}^{\perp}(v, P x) \nn
& \quad + f_V^{\perp} m_V^2\left[P_\alpha \eta_{\perp \mu}^{\ast} g_{\beta \nu}^{\perp} - P_\beta \eta_{\perp \mu}^{\ast} g_{\alpha \nu}^{\perp} - P_\alpha \eta_{\perp \nu}^{\ast} g_{\beta \mu}^{\perp} + P_\beta \eta_{\perp \nu}^{\ast} g_{\alpha \mu}^{\perp}\right]\Phi_{4;V}^{\perp(1)}(v, P x) \nn
& \quad + f_V^{\perp} m_V^2\left[P_\mu \eta_{\perp \alpha}^{\ast} g_{\beta \nu}^{\perp} - P_\mu \eta_{\perp \beta}^{\ast} g_{\alpha \nu}^{\perp} - P_\nu \eta_{\perp \alpha}^{\ast} g_{\beta \mu}^{\perp} + P_\nu \eta_{\perp \beta}^{\ast} g_{\alpha \mu}^{\perp}\right] \Phi_{4;V}^{\perp(2)}(v, P x) \nn
& \quad +\frac{f_V^{\perp} m_V^2}{P\cdot x} \left[P_\alpha P_\mu \eta_{\perp \beta}^{\ast} x_\nu - P_\beta P_\mu \eta_{\perp \alpha}^{\ast} x_\nu - P_\alpha P_\nu \eta_{\perp \beta}^{\ast} x_\mu + P_\beta P_\nu \eta_{\perp \alpha}^{\ast} x_\mu\right] \Phi_{4;V}^{\perp(3)}(v, P x) \nn
& \quad +\frac{f_V^{\perp} m_V^2}{P\cdot x}\left[P_\alpha P_\mu \eta_{\perp \nu}^{\ast} x_\beta - P_\beta P_\mu \eta_{\perp \nu}^{\ast} x_\alpha - P_\alpha P_\nu \eta_{\perp \mu}^{\ast} x_\beta + P_\beta P_\nu \eta_{\perp \mu}^{\ast} x_\alpha\right] \Phi_{4;V}^{\perp(4)}(v, P x)\,, \\
& \left\langle V(p,\eta^*)\left|\bar{q}_1(x)g_s G_{\mu\nu}(v x) q_2(-x)\right| 0\right\rangle =  - i f_V^{\perp} m_V^2\left[ \eta_{\perp \mu}^{\ast} P_\nu - \eta_{\perp \nu}^{\ast} P_\mu\right]\Psi^{\perp}_{4;V}(v, P x)\,, \\
& \big\langle V(p,\eta^*)\big|\bar{q}_1(x)g_s \widetilde{G}_{\mu\nu}(v x) \gamma_5 q_2(-x) \big| 0 \big\rangle = f_V^{\perp} m_V^2\left[ \eta_{\perp \mu}^{\ast} P_\nu - \eta_{\perp \nu}^{\ast} P_\mu\right] \widetilde{\Psi}^{\perp}_{4;V}(v, P x)\,,
\end{align}
where
\begin{align}
& P_\mu = p_\mu - {1\over 2} x_\mu {m_V^2\over P\cdot x}, \\
& \eta_\mu^\ast = {\eta^\ast \cdot x \over P\cdot x} \Big( P_\mu - {m_V^2 \over 2 P\cdot x} x_\mu \Big) + \eta_{\perp \mu}^\ast, \\
& g_{\mu\nu}^\perp = g_{\mu\nu}-\frac1{P\cdot x}(P_\mu x_\nu + P_\nu x_\mu), \\
& \Phi_{3p}(v,Px) = \int \mathcal{D}\underline{\alpha} \, e^{iP\cdot x (\alpha_1 -\alpha_2+ v\alpha_3)} \Phi_{3p}(\alpha_1,\alpha_2,\alpha_3),
\end{align}
$\underline{\alpha}$ is the set of three momentum fractions
$\underline{\alpha}=\{\alpha_1,\alpha_2,\alpha_3\}$.
 The integration measure is defined as 
\begin{equation}
 \int {\cal D}\underline{\alpha} \equiv \int_0^1 d\alpha_1
  \int_0^1 d\alpha_2\int_0^1 d\alpha_3 \,\delta\left(1-\sum \alpha_i\right).
\label{eq:measure}
\end{equation}
$P$ and $x$ are two introduced light-like vectors and the dual gluon field strength tensor is defined as $\widetilde{G}_{\mu\nu} = {1\over 2} \epsilon_{\mu\nu\rho\sigma} G^{\rho\sigma}$, {with $\epsilon_{0123} = -1$}.

\section{Parameterizations for the vector meson DAs }\label{app:lcda}
 The parameterizations for the vector meson DAs  in the numerical analysis are collected below:
\begin{itemize}
\item  the twist-2 DA:
\vspace{-2mm}
\begin{align}
\label{eq:confexp}
\phi_{2;V}^{\parallel,\perp}(u,\mu) = 6 u \bar u \left\{ 1 + \sum_{n=1}^\infty
a_n^{\parallel,\perp}(\mu) C_n^{3/2}(u-{\bar u})\right\},
\end{align}

in terms of the (non-perturbative) Gegenbauer moments $a_n^{\parallel,\perp}$ and the
Gegenbauer polynomials $C_n^{3/2}$.
To leading-logarithmic accuracy, the  
$a^{\parallel,\perp}_n$ renormalise multiplicatively as
\begin{equation}
a^{\parallel,\perp}_n(\mu) = E_{n,\rm LO}^{\parallel,\perp}(\mu, \mu_0)\, a^{\parallel,\perp}_n(\mu_0).
\end{equation}
To next-to-leading order accuracy, the scale dependence of the
Gegenbauer moments is more complicated and reads \cite{Wang:2017ijn,Agaev:2010aq}
\begin{eqnarray}
\left[ f^{\perp}_V(\mu) a^{\perp}_{n,{\rm NLO}}(\mu)\right]\,& =&\,  E_{n,\rm NLO}^{\perp} (\mu,\mu_0)
\left[f^{\perp}_V(\mu_0)  a^{\perp}_n(\mu_0)\right]+\frac{\alpha_s}{4\pi}\sum_{k=0}^{n-2}   
E_{n,\rm LO}^{\perp}(\mu,\mu_0)\, d^{k}_{n}(\mu,\mu_0)\left[f^{\perp}_V(\mu_0) a^{\perp}_k(\mu_0)\right],\nn
a^{\parallel}_{n,{\rm NLO}}(\mu) \,& =& \,  E_{n,\rm NLO}^{\parallel} (\mu,\mu_0)\,
a^{\parallel}_n(\mu_0)+\frac{\alpha_s}{4\pi}\sum_{k=0}^{n-2} \,  
E_{n,\rm LO}^{\parallel}(\mu,\mu_0)\, d^{k}_{n}(\mu,\mu_0)\,a^{\parallel}_k(\mu_0), 
\label{eq:evoaNLO}
\end{eqnarray}
where  the formulas for the RG factors $E^{\parallel,\perp}_{n,{\rm (N)LO}}(\mu,\mu_0)$ and off-diagonal mixing coefficients $d^k_n(\mu,\mu_0)$ within the $\overline {\text{MS}}$ scheme are explicitly provided in references \cite{Agaev:2010aq,Wang:2017ijn}. The evolution of $a^{\perp}_n(\mu)$ is closely linked to $f^{\perp}_V(\mu)$ because of the definition of the conformal operator.

\item the twist-3 two-particle DAs\cite{Ball:2007rt}:
\begin{eqnarray}
&&\psi_{3;V}^\perp(u,\mu)  = 6 u\bar u \left\{ 1 + \left(
\frac{1}{3} a_1^\parallel(\mu) + \frac{20}{9} \kappa_{3V}^\parallel(\mu)\right)
C_1^{3/2}(u-{\bar u})\right.\nn
&& \hspace{2cm}\left. + \left( \frac{1}{6} a_2^\parallel(\mu) +
\frac{10}{9}\zeta_{3V}^\parallel(\mu) + \frac{5}{12}\,
\omega_{3V}^\parallel(\mu)-\frac{5}{24}\,
\widetilde\omega_{3V}^\parallel(\mu)\!\right)
C_2^{3/2}(u-{\bar u})\right.\nn
&&\hspace{2cm} \left. + \left(\frac{1}{4}\widetilde\lambda_{3V}^\parallel(\mu) - \frac{1}{8}
\lambda_{3V}^\parallel(\mu)\right) C_3^{3/2}(u-{\bar u})\right\}\nn
&& \hspace{2cm}+ 6\, \frac{m_{q_2}(\mu)+m_{q_1}(\mu)}{m_{V}}\,\frac{f_{V}^\perp(\mu)}{f_{V}^\parallel} \left\{
u\bar u \left(2 + 3u-{\bar u} a_1^\perp(\mu) + 2 (11-10u\bar u) a_2^\perp(\mu)\right)\right.\nn
&&\hspace{2cm}\left.+ \bar u \ln \bar u \left(1+3 a_1^\perp(\mu) + 6 a_2^\perp(\mu)\right) + u \ln u 
\left(1-3 a_1^\perp(\mu) + 6 a_2^\perp(\mu)\right)\right\} \nn
&& \hspace{2cm}- 6\,\frac{m_{q_2}(\mu)-m_{q_1}(\mu)}{m_{V}}\,\frac{f_{V}^\perp(\mu)}{f_{V}^\parallel}\left\{
u\bar u \left(9 a_1^\perp(\mu) + 10(u-{\bar u}) a_2^\perp(\mu)\right) \right.\nn
&&\hspace{2cm}\left. + \bar u \ln \bar u \left(1+3 a_1^\perp(\mu) + 6 a_2^\perp(\mu)\right)- u \ln u 
\left(1-3 a_1^\perp(\mu) + 6 a_2^\perp(\mu)\right)\right\}.
\end{eqnarray}

The scale dependence of twist-3 parameters are as followed 
\begin{eqnarray}
&&\kappa_{3V}^\perp(\mu) = L^{ (C_A + \frac{4}{3}\,C_F)/\beta_0} 
\kappa_{3V}^\perp(\mu_0)\,,\kappa_{3V}^\parallel(\mu) = L^{(3\,C_A - \frac{1}{3}\,C_F)/\beta_0} 
\kappa_{3V}^\parallel(\mu_0)\,,\nn
&&\omega_{3V}^\perp(\mu) =  L^{( C_A + \frac{17}{6}\, C_F)/\beta_0} 
\omega_{3V}^\perp(\mu_0)\,,\lambda_{3V}^\perp(\mu) = L^{(\frac{10}{3}\,C_A + \frac{1}{6}\,C_F))/\beta_0} 
\lambda_{3V}^\perp(\mu_0),\,\nn
&&\zeta_{3V}^\parallel(\mu) = L^{(3\,C_A - \frac{1}{3}\,C_F)/\beta_0} 
\zeta_{3V}^\parallel(\mu_0),\,
\end{eqnarray}
and
\begin{eqnarray}
\left(\begin{array}{c}
\frac{2}{3}\omega_{3V}^\parallel(\mu)-\widetilde\omega_{3V}^\parallel(\mu) \\
\frac{2}{3}\omega_{3V}^\parallel(\mu)+
\widetilde\omega_{3V}^\parallel(\mu)
\end{array}\right)=\left((L^{\overrightarrow{\Gamma_3^+}) / \beta_0}\right)_D\left(\begin{array}{c}
\frac{2}{3}\omega_{3V}^\parallel(\mu_0)-\widetilde\omega_{3V}^\parallel(\mu_0) \\
\frac{2}{3}\omega_{3V}^\parallel(\mu_0)+
\widetilde\omega_{3V}^\parallel(\mu_0))
\end{array}\right),
\end{eqnarray}

\begin{eqnarray}
\left(\begin{array}{c}
\frac{2}{3}\widetilde\lambda_{3V}^\parallel(\mu)+
\lambda_{3V}^\parallel(\mu) \\
\frac{2}{3}\widetilde\lambda_{3V}^\parallel(\mu)-
\lambda_{3V}^\parallel(\mu)
\end{array}\right)=\left((L^{\overrightarrow{\Gamma_3^-}) / \beta_0}\right)_D\left(\begin{array}{c}
\frac{2}{3}\widetilde\lambda_{3V}^\parallel(\mu_0)+
\lambda_{3V}^\parallel(\mu_0) \\
\frac{2}{3}\widetilde\lambda_{3V}^\parallel(\mu_0)-
\lambda_{3V}^\parallel(\mu_0)
\end{array}\right),
\end{eqnarray}
where $D$ represent diagnoal and $\overrightarrow{\Gamma_3}$ is the vector containing the diagonal elements of the diagonal matrix $(\Gamma_3)_D$, with
\begin{eqnarray}
\Gamma_3^+ &\, =\, & \left(\begin{array}{l@{\quad}l}
\phantom{-}\frac{7}{3}\, C_A + \frac{8}{3}\,C_F & 
 -\frac{2}{3}\,C_A + \frac{2}{3}\,C_F\\[10pt]
 -\frac{4}{3}\,C_A + \frac{5}{3}\,C_F & 
\phantom{-}4C_A +\frac{1}{6}\,C_F
\end{array}\right),\nn
\Gamma_3^- &\, =\, & \left(\begin{array}{l@{\quad}l}
\phantom{-}4\, C_A + \frac{1}{6}\,C_F & 
 -\frac{4}{3}\,C_A + \frac{5}{3}\,C_F\\[10pt]
 -\frac{2}{3}\,C_A + \frac{2}{3}\,C_F & 
\phantom{-}\frac{7}{3}\,C_A +\frac{8}{3}\,C_F
\end{array}\right).
\end{eqnarray}

\item the twist-4  two-particle DAs\cite{Ball:2007zt}:
\begin{eqnarray}
&&\phi^\perp_{4;V}(u,\mu) =
30 u^2\bar u^2 \left\{ \left(\frac{4}{3}\, \zeta^\perp_{4V}(\mu) -
\frac{8}{3}\, \widetilde{\zeta}^\perp_{4V}(\mu) + \frac{2}{5} +
\frac{4}{35}\, a_2^\perp(\mu)\right) \right.
\nn
&&{}\hspace*{1.5cm} + \left( \frac{3}{25}\,a_1^\perp(\mu) +
\frac{1}{3}\,\kappa_{3V}^\perp(\mu) - \frac{1}{45}\, \lambda_{3V}^\perp(\mu)
- \frac{1}{15}\,\theta_1^\perp(\mu)  + \frac{7}{30}\,\theta_2^\perp(\mu)
+\frac{1}{5}\, \widetilde{\theta}_1^\perp(\mu) - 
\frac{3}{10}\,\widetilde{\theta}_2^\perp(\mu)\right) 
\nn
&&\hspace*{1.5cm}\,\,\times C_1^{5/2}(u-{\bar u})+\left.\left(\frac{3}{35}\,a_2^\perp(\mu) +
\frac{1}{60}\,{\omega}_{3V}^\perp(\mu) \right) C_2^{5/2}(u-{\bar u})
-\frac{4}{1575}\,\lambda_{3V}^\perp(\mu) C_3^{5/2}(u-{\bar u})\right\}
\nn
&&\hspace*{1.5cm}+ \left( 5 \kappa_{3V}^\perp(\mu) - a_1^\perp(\mu) - 
20 \widetilde{\phi}_2^\perp(\mu)\right)\nn
&&\hspace*{1.5cm}\times\left\{-4 u^3 (2-u) \ln u + 4 \bar u^3 (2-\bar u) \ln \bar u +
\frac{1}{2}\,u \bar u (u-{\bar u}) \left(3(u-{\bar u})^2-11\right)\right\}
\nn
&&\hspace*{1.5cm}+  \left( 2 \omega_{3V}^\perp(\mu) - \frac{36}{11}\, a_2^\perp(\mu) 
-\frac{252}{55}\, \langle\!\langle Q^{(1)}\rangle\!\rangle(\mu) -
\frac{140}{11}\, \langle\!\langle Q^{(3)}\rangle\!\rangle(\mu)\right)
\nn
&&\hspace*{1.5cm}\times\left\{ u^3
(6 u^2-15u+10)\ln u + \bar u^3 (6 \bar u^2-15\bar u+10)\ln\bar u -
\frac{1}{8}\,u\bar u \left( 13(u-{\bar u})^2-21\right)\right\}.
\label{4.24}
\end{eqnarray}

\item the twist-4  three-particle DAs \footnote{
We adopt a convention denoted as $\underline{\alpha}=\{\alpha_1,\alpha_2,\alpha_3\}=\{\alpha_{\bar q_1},\alpha_{q_2},\alpha_g\}$. It's worth mentioning that this differs from the convention used in Ref.\cite{Ball:2007zt} where the only distinction is the exchange of $\alpha_1$ for $\alpha_2$.}
\begin{eqnarray}
{\Psi}_{4;V}^\perp\left(\alpha_i,\mu\right)= &&\quad 30  \alpha_3^2\left\{ \psi^\perp_{0}(\mu)(1-\alpha_3)
     +\psi^\perp_{1}(\mu)\left[\alpha_3(1-\alpha_3)-6\alpha_1\alpha_2\right]\right.\nn
 &&   \hspace*{-2cm}   \left. {} +\psi^\perp_{2}(\mu)\left[\alpha_3(1-\alpha_3)-\frac{3}{2}(\alpha_1^2 +\alpha_2^2)\right]+(\alpha_1-\alpha_2)\left[ \theta^\perp_0(\mu) + \alpha_3\theta^\perp_1(\mu) + \frac{1}{2}\,(5 \alpha_3-3)\theta^\perp_2(\mu)\right]\right\} , \nn
\widetilde{\Psi}_{4;V}^{\perp}\left(\alpha_i,\mu\right) =&&\quad30 \alpha_3^2\left\{ \widetilde\psi^\perp_{0}(\mu)(1-\alpha_3)
     +\widetilde\psi^\perp_{1}(\mu)\left[\alpha_3(1-\alpha_3)-6\alpha_1\alpha_2\right]\right.\nn
  &&  \hspace*{-2cm} \left.  {} +\widetilde\psi^\perp_{2}(\mu)\left[\alpha_3(1-\alpha_3)-
\frac{3}{2}(\alpha_1^2  +\alpha_2^2)\right]+ (\alpha_1-\alpha_2)\left[ \widetilde\theta^\perp_0(\mu) + \alpha_3
\widetilde\theta^\perp_1(\mu) + \frac{1}{2}\,(5 \alpha_3-3) \widetilde\theta^\perp_2(\mu)\right]\right\},\nn
\Phi_{4;V}^{\perp(1)}\left(\alpha_i,\mu\right)=&&\quad 120\alpha_1\alpha_2\alpha_3 [ \phi_0^\perp(\mu) -
\phi_1^\perp(\mu)(\alpha_1-\alpha_2) +\phi_2^\perp(\mu) (3\alpha_3-1)]\,,\nn
 {\Phi}^{\perp(2)}_{4;V}\left(\alpha_i,\mu\right) =&&\quad -30 \alpha_3^2\left\{ \widetilde\theta^\perp_{0}(\mu)(1-\alpha_3)
 +\widetilde\theta^\perp_{1}(\mu)\left[\alpha_3(1-\alpha_3)-6\alpha_1\alpha_2\right]\right. \nn
 &&\hspace*{-2cm} \left. {}  +\widetilde\theta^\perp_{2}(\mu)\left[\alpha_3(1-\alpha_3)-\frac{3}{2}(\alpha_1^2
 +\alpha_2^2)\right]+(\alpha_1-\alpha_2)\left[ \widetilde\psi^\perp_0(\mu) + \alpha_3\widetilde\psi^\perp_1(\mu) + \frac{1}{2}\,(5 \alpha_3-3)\widetilde \psi^\perp_2(\mu)\right]\right\}\,,\nn
 \Phi^{\perp(3)}_{4;V}\left(\alpha_i,\mu\right)=&&\quad -120 \alpha_1\alpha_2\alpha_3 [ \widetilde\phi_0^\perp(\mu) -
  \widetilde\phi_1^\perp(\mu)(\alpha_1-\alpha_2) +
\widetilde\phi_2^\perp(\mu) (3\alpha_3-1)],\,\nn
{\Phi}^{\perp(4)}_{4;V}\left(\alpha_i,\mu\right) =&&\quad 30 \alpha_3^2\left\{ \theta^\perp_{0}(\mu)(1-\alpha_3)
+\theta^\perp_{1}(\mu)\left[\alpha_3(1-\alpha_3)-6\alpha_1\alpha_2\right]\right. \nn
&& \hspace*{-2cm}  \left. {}  +\theta^\perp_{2}(\mu)\left[\alpha_3(1-\alpha_3)-\frac{3}{2}(\alpha_1^2
+\alpha_2^2)\right]+(\alpha_1-\alpha_2)\left[ \psi^\perp_0(\mu) + \alpha_3\psi^\perp_1(\mu) + \frac{1}{2}\,(5 \alpha_3-3)
\psi^\perp_2(\mu)\right]\right\}. \nn
\end{eqnarray}
The parameters associated with the three particle amplitudes are as followed
\begin{eqnarray}
\psi_0^\perp(\mu)\, &=&\, \phantom{-}\zeta_{4V}^\perp(\mu) \,,\hspace{3cm}
 \widetilde{\psi}_0^\perp(\mu) = \widetilde\zeta_{4V}^\perp(\mu) \,,\nn
 \phi_0^\perp(\mu)  \, &=&\,\phantom{-}\frac{1}{6}\kappa_{3V}^\perp(\mu) +
\frac{1}{3}\kappa_{4V}^\perp(\mu) \,,\qquad \widetilde\phi_0^\perp(\mu) \, =\,\frac{1}{6}\kappa_{3V}^\perp(\mu) -\frac{1}{3}\kappa_{4V}^\perp(\mu)\,,\nn
\theta_0^\perp(\mu) \, &=&\,-\frac{1}{6}\kappa_{3V}^\perp(\mu) -
\frac{1}{3}\kappa_{4V}^\perp(\mu) \,,\qquad\,\widetilde\theta_0^\perp(\mu) =-\frac{1}{6}\kappa_{3V}^\perp(\mu) +\frac{1}{3}\kappa_{4V}^\perp(\mu)\,.\nn
\phi_1^\perp(\mu)  \, &=&\,\phantom{-}\frac{9}{44}\, a_2^\perp(\mu) + \frac{1}{8}\,
\omega_{3V}^\perp(\mu) +\frac{63}{220}\,\langle\!\langle Q^{(1)}\rangle\!\rangle(\mu) -
\frac{119}{44}\, \langle\!\langle Q^{(3)}\rangle\!\rangle(\mu)\,,\nn
\widetilde\phi_1^\perp(\mu) \, &=&\, -\frac{9}{44}\, a_2^\perp(\mu) + \frac{1}{8}\,
\omega_{3V}^\perp(\mu) -\frac{63}{220}\, \langle\!\langle Q^{(1)}\rangle\!\rangle(\mu) -\frac{35}{44}\, \langle\!\langle Q^{(3)}\rangle\!\rangle(\mu)\,,\nn
\psi_1^\perp(\mu) \, &=&\,\phantom{-}\frac{3}{44}\, a_2^\perp(\mu) + \frac{1}{12}\,
\omega_{3V}^\perp(\mu) + \frac{49}{110}\,\langle\!\langle Q^{(1)}\rangle\!\rangle(\mu) -
\frac{7}{22}\, \langle\!\langle Q^{(3)}\rangle\!\rangle(\mu)
+ \frac{7}{3}\, \langle\!\langle Q^{(5)}\rangle\!\rangle(\mu)\,,\nn
\widetilde\psi_1^\perp(\mu)\, &=&\,-\frac{3}{44}\, a_2^\perp(\mu) + \frac{1}{12}\,
\omega_{3V}^\perp(\mu) - \frac{49}{110}\, \langle\!\langle Q^{(1)}\rangle\!\rangle(\mu) +
\frac{7}{22}\, \langle\!\langle Q^{(3)}\rangle\!\rangle(\mu)
 + \frac{7}{3}\, \langle\!\langle Q^{(5)}\rangle\!\rangle(\mu)\,,\nn
\psi_2^\perp(\mu)\, &=&\, -\frac{3}{22}\, a_2^\perp(\mu) - \frac{1}{12}\,
\omega_{3V}^\perp(\mu) + \frac{28}{55}\,
\langle\!\langle Q^{(1)}\rangle\!\rangle(\mu) +
\frac{7}{11}\, \langle\!\langle Q^{(3)}\rangle\!\rangle(\mu) +
\frac{14}{3}\, \langle\!\langle Q^{(5)}\rangle\!\rangle(\mu)\,,
\nn
\widetilde\psi_2^\perp(\mu)\, &=&\, \phantom{-}\frac{3}{22}\, a_2^\perp(\mu) - \frac{1}{12}\,
\omega_{3V}^\perp(\mu) - \frac{28}{55}\,
\langle\!\langle Q^{(1)}\rangle\!\rangle(\mu) -
\frac{7}{11}\, \langle\!\langle Q^{(3)}\rangle\!\rangle(\mu)
+ \frac{14}{3}\, \langle\!\langle Q^{(5)}\rangle\!\rangle(\mu)\,.
\end{eqnarray}
and
\begin{eqnarray}
\left(\zeta_{4V}^{\perp}+ \widetilde{\zeta}_{4V}^{\perp}\right)(\mu) = L^{\gamma^+/{\beta_0}}
\left(\zeta_{4V}^{\perp} + \widetilde{\zeta}_{4V}^{\perp}\right)(\mu_0),& \qquad & \gamma_+ = 3 C_A -
\frac{8}{3}\, C_F,\nn
\left(\zeta_{4V}^{\perp} - \widetilde{\zeta}_{4V}^{\perp}\right)(\mu) = L^{\gamma^-/{\beta_0}}
\left(\zeta_{4V}^{\perp} - \widetilde{\zeta}_{4V}^{\perp}\right)(\mu_0),& \qquad & \gamma_- = 4 C_A -
4 C_F.
\end{eqnarray}

The $\kappa_{4V}^\perp(\mu)$
depends on quark masses and $a_1^\perp(\mu)$  and was obtained in Ref.~\cite{Ball:2006fz}.
Like with $\kappa_{4V}^\parallel(\mu)$, 
the scale-dependence of $\kappa_{4V}^\perp$ follows from that of the
parameters on the right-hand side and is given by
\begin{eqnarray}
\kappa^\perp_{4V}(\mu) 
\,&=& \,
\kappa^\perp_{4V}(\mu_0)
+ \left( L^{8/(3\beta_0)} - 1\right) \frac{1}{10}\,a_1^{\perp}(\mu_0^2)
+ \left( L^{8/(3\beta_0)} - 1\right) 
\frac{f_{V}^\parallel}{f_{V}^\perp(\mu_0)}\,
\frac{[m_{q_2}-m_{q_1}](\mu_0^2)}{12 m_{V}} 
\nn
&&{}
- \left( L^{8/\beta_0} - 1\right)\frac{[m_{q_2}^2-m_{q_1}^2](\mu_0)}{4 m_{V}^2},\,
\end{eqnarray}
with $L=\alpha_s(\mu^2)/\alpha_s(\mu_0^2)$.

The precise definition of $\langle\!\langle Q^{(i)}\rangle\!\rangle$ is
given in Ref.\cite{Ball:1998ff}. $\omega_{3K^*}^\perp$ is a twist-3
parameter.
Existing numerical determinations of these parameters from QCD sum
rules are far from
being precise, so we decide to estimate them using the renormalon
model of Ref.\cite{Braun:2004bu} instead.

The renormalon model implies the following estimates for the G-conserving twist-4 parameters:
\begin{equation}
\begin{array}[b]{
r@{\ =\ }l@{\quad}r@{\ =\ }l@{\quad}r@{\ =\ }l@{\quad}r@{\ =\ }l}
\displaystyle\langle\!\langle Q^{(1)}\rangle\!\rangle^{\rm R}
& \displaystyle-\frac{10}{3}\zeta_{4K^*}^\perp\,, & 
\langle\!\langle Q^{(3)}\rangle\!\rangle^{\rm R}
& -\zeta_{4K^*}^\perp\,, &
\langle\!\langle Q^{(5)}\rangle\!\rangle^{\rm R}
& 0\,.
\end{array}
\end{equation}

\item the twist-5 two-particle DAs(asymptotic form)\cite{Bharucha:2015bzk}:
\begin{eqnarray}
   {\psi}^{\perp}_{5;V}(u)= 12u^2 {\bar u}^2.
\end{eqnarray}
\end{itemize}
Recall that $\bar u = 1-u$ . The expression discussed is applicable to a $V=(q_2\bar q_1)$ meson. {In the case of $\bar V = (q_1\bar q_2)$, one must replace $u$ with $1-u$ and exchange the values of $m_{q_2}$ and $m_{q_1}$}.

\section{Double spectral density and integral at NLO}\label{appC}

The expression for the twist-2 component of the Next-to-Leading Order (NLO) double spectral density in the sum rule (\ref{eq:SR3}) is obtained from the calculations presented in Ref.\cite{Li:2020rcg}. The analytical expression for the double spectral density can be collected:
\begin{eqnarray}
\rho^{\rm NLO}(s_1, s_2)\,  &=&\, \frac1{m_Q}
\big \{  \left [\rho_{\rm I}(r, \sigma)  + \rho_{\rm II}(r, \sigma) \, \ln r
+ \rho_{\rm III}(r, \sigma) \, (\ln^2 r - \pi^2) \right ] \,
\delta^{(2)}(r-1) \nn
&& +  \left [ \rho_{\rm II}(r, \sigma) + 2 \, \rho_{\rm III}(r, \sigma)  \, \ln r  \right ] \,
{d^3 \over d r^3} \, \ln|1-r|  \big \}  \, \theta(s_1 - m_Q^2)  \, \theta(s_2 - m_Q^2) \,,
\label{eq:tw2nlo}
\end{eqnarray}
where the invariant functions have been introduced,
\begin{eqnarray}
\rho_{\rm I}(r, \sigma)\,  &=&\,  \frac{(r+1)}{\sigma^3}
\Bigg\{6 \,r\,\sigma^2\, \left [ 2 \, \ln \left ( \frac{\mu^2}{m_Q^2} \right )
+ {3 \over 2} \, \ln \left ( \frac{\nu^2}{\mu^2} \right )
+ 3\,{\rm Li}_2 \left (- \frac{\sigma }{r+1} \right )
+ {\rm Li}_2 \left (- \frac{r\, \sigma }{r+1} \right )  \right ]  \nn
&&  + \, 6 \, r\,\sigma^2\,  \ln \left ( \frac{\sigma}{r+1} \right ) \,
\left [ \ln \left ( \frac{\sigma+r+1 }{r+1} \right ) + \ln \left ( \frac{r \sigma + r + 1 }{r+1} \right ) \right ]  \nn
&& + \, 3\,(r+1)\,\big[2\, r \,(\sigma+1)+5 \,\sigma+2\big]\, \ln \left ( \frac{\sigma+r+1 }{r+1} \right ) \nn
&& - \, \frac{3\,r\,\sigma^2}{(r+\sigma+1)(r\sigma+r+1)}\left [9\,r\,\sigma^2+(7\,r+10)\,(r+1)\,\sigma+8\,(r+1)^2 \right ] \,
\ln \left ( \frac{\sigma }{r+1} \right )  \nn
&& + \, \frac{\sigma}{r+\sigma+1}\,\left [ \big((3+2\,\pi^2)\, r+6\big)\, \sigma^2
+\big((9+2\,\pi^2)\, r-6\big)\, (r+1)\, \sigma-6\, (r+1)^2 \right ] \Bigg\}, \nn
\rho_{\rm II}(r, \sigma)\, &=& \,  3 \,r(r+1)\,\left[ \frac{r}{r\,\sigma+r+1}
+ \frac{2}{r+\sigma+1} - \frac{2}{\sigma}\, \ln \left( \frac{\sigma+r+1}{r\,\sigma+r+1} \right ) \right ] \,,
\nn
\rho_{\rm III}(r, \sigma)\,  &=&\,  - \frac{6 \,r\,(r+1)}{\sigma} \,,
\end{eqnarray}
with the introduction of two new dimensionless variables carried out through the following definitions:
\begin{eqnarray}
r= {s_1 - m_Q^2 \over s_2 - m_Q^2 } \,, \qquad
\sigma = {s_1 \over m_Q^2} + {s_2 \over m_Q^2} - 2 \,.
\label{Jacobi transformation r and sigma}
\end{eqnarray}

Switching to the pole-mass scheme of the heavy quark,
one needs to add the following expression
\begin{align}
\Delta \rho^{\rm (tw2,NLO)}_{\rm pole}(r,\sigma)  = -6 \, m_Q^4\, \left [ 4+3\,\ln \left ( \frac{\mu^2}{m_Q^2} \right )\right ]\,
\frac{d}{dm_Q^2}\,\left [ \frac{1}{m_Q^2}\,\frac{r\,(r+1)}{\sigma} \right ] \,.  
\label{eq:poeladd}
\end{align}

Finally, for completeness we present the pole-mass scheme
additions to the final NLO expressions in (\ref{eq:NLOBorel}):
\begin{align}
\Delta f^{\rm (tw2)}_{\rm pole}(\sigma) =-3 (4+3\ln\frac{\mu^2}{m_Q^2})\exp\left(- \frac{m_Q^2}{M^2}\right).
 \label{eq:deltapole}
\end{align}

\medskip

\bibliographystyle{elsarticle-num-names}
\bibliography{paper}


\end{document}